\documentclass[twocolumn,superscriptaddress,secnumarabic,
amssymb,amsmath,nobibnotes,aps,prd,showpacs,nofootinbib]{revtex4}
\usepackage{graphicx}
\usepackage{epsf}
\usepackage{bm}
\usepackage{amsmath}
\usepackage{amsfonts}
\usepackage{amssymb}
\usepackage{epstopdf}
\usepackage{color}
\begin{document}

%%%%%%%%%%%%%%%%%%%%%%%%%%%%%%%%%%%%%%%%%%%%%%%%%%%%%%%%%%%%%%%%%%%%%%%%%%%%%%%%%%%%
\title{Large-scale (in-) stability Analysis of an Exactly Solved Coupled Dark-Energy Model}

\author{Weiqiang Yang}
\email{d11102004@163.com}
\affiliation{Department of Physics, Liaoning Normal University, Dalian, 116029, P. R. China}

\author{Supriya Pan}
\email{span@research.jdvu.ac.in}
\affiliation{Department of Mathematics, Raiganj Surendranath Mahavidyalaya, Sudarsharpur, Raiganj, Uttar Dinajpur, West Bengal 733134, India}

\author{Ram\'{o}n Herrera}
\email{ramon.herrera@pucv.cl} \affiliation{Instituto de
F\'{\i}sica, Pontificia Universidad Cat\'{o}lica de
Valpara\'{\i}so, Casilla 4059, Valpara\'{\i}so, Chile.}

\author{Subenoy Chakraborty}
\email{schakraborty@math.jdvu.ac.in}
\affiliation{Department of Mathematics, Jadavpur University, Kolkata 700032, West Bengal, India}

%%%%%%%%%%%%%%%%%%%%%%%%%%%%%%%%%%%%%%%%%%%%%%%%%%%%%%%%%%%%%%%%%%%%%%%%%%%%%%%%%%%%%%%%
%\keywords{Cosmological parameters; Dark energy; Dark matter; Interaction.}
%%%%%%%%%%%%%%%%%%%%%%%%%%%%%%%%%%%%%%%%%%%%%%%%%%%%%%%%%%%%%%%%%%%%%%%%%%%%%%%%%%%%%%%%%%%%%%%%%%%%%
%%%%%%%%%%%%%%%%%%%%%%%%%%%%%%%%%%%%%%%%%%%%%%%%%%%%%%%%%%%%%%%%%%%%%%%%%%%%%%%%%%%%%%%%%%%%%%%%%%%%%
%%%%%%%%%%%%%%%%%%%%%%%%%%%%%%%%%%%%%%%%%%%%%%%%%%%%%%%%%%%%%%%%%%%%%%%%%%%%%%%%%%%%%%%%%%%%%%%%%%%
\begin{abstract}
Assuming a non-gravitational interaction amongst the dark fluids
of our universe namely, the dark matter and dark energy, we study
a specific interaction model in the background of a spatially flat
Friedmann-Lema\^itre-Robertson-Walker geometry. The
interaction model, as we found, solves the background evolution in
an analytic way when the dark energy takes a constant
barotropic equation of state, $w_x$. In particular, we analyze two
separate interaction scenarios, namely, when the dark energy is a fluid
other than the vacuum energy (i.e., $w_x \neq -1$) and when
it is vacuum energy itself (i.e., $w_x = -1$). We found that the
interacting model with $w_x \neq -1$ produces stable perturbation
at large scales for $w_x < -1$ with the coupling strength $\xi
<0$.  Both the scenarios have been constrained with the latest
astronomical data having distinct origin. The analyses show that a very small
interaction with coupling strength is allowed and within 68.3\% confidence-region,
$\xi =0$ is recovered. The analyses further
show that a large coupling strength significantly affects the
large scale dynamics of the universe while according to the
observational data the interaction models are very well
consistent with the $\Lambda$-cosmology. Furthermore, we observe
that for the vacuum interaction scenario, the tension on $H_0$ is
not released while for the interacting dark energy scenario 
with $w_x < -1$, the tension on $H_0$ seems to be
released partially because of the high error bars in $H_0$.
Finally, we close the work with the Bayesian evidence 
which shows that the $\Lambda$CDM cosmology is favored over the 
two interacting scenarios.  
\end{abstract}

%%%%%%%%%%%%%%%%%%%%%%%%%%%%%%%%%%%%%%%%%%%%%%%%%%%%%%%%%%%%%%%%%%%%%%%%%%%%%%%%%%%%%%%%%%%%%%%%%%%%%
\maketitle
%%%%%%%%%%%%%%%%%%%%%%%%%%%%%%%%%%%%%%%%%%%%%%%%%%%%%%%%%%%%%%%%%%%%%%%%%%%%%%%%%%%%%%%%%%%%%%%%%%%%%%%%%%%%%%%%%%
% \myclassification{98.80.Cq, 98.80.-k}\\\\
%%%%%%%%%%%%%%%%%%%%%%%%%%%%%%%%%%%%%%%%%%%%%%%%%%%%%%%%%%%%%%%%%%%%%%%%%%%%%%%%%%%%%%%%%%%%%%%%%%%%%%%%%%%%%%%%%%
\section{Introduction}
\label{Intro}

The theory of non-gravitational interactions between dark matter and dark energy is the main concern of this work.  The origin of such interacting theory did not appear suddenly in the cosmological scheme. It has a well motivated history that we shall discuss here. However, before that, we need a basic introduction about the dark matter and dark energy.
According to latest observational suggestions \cite{Ade:2015xua} dark matter (DM) and dark energy (DE) are the main influential sources of the total energy budget of the universe where the dark matter contributes around 26\% of its total energy density, is pressureless and unseen while the dark energy, a hypothetical fluid occupying 68\% of the total energy density of the universe is accelerating the expansion history of the universe. The best description for such observational information is the $\Lambda$CDM cosmology where $\Lambda$ acts as the dark energy fluid and the CDM is the cold dark matter that is pressureless. This is a non-interacting scenario in the sense that both $\Lambda$ and CDM are conserved separately. Despite of great success of $\Lambda$CDM cosmology, the cosmological constant problem \cite{Weinberg} still lacks a satisfactory explanation. The cosmological constant problem is basically confined with the mismatched value of the cosmological constant predicted from the high and low energy scales of the universe. In the following we shall discuss how the interacting dynamics is closely related to the cosmological constant problem. In fact this coupling mechanics was originated because of the cosmological constant problem and finally it became very useful to explain some other issues. Let us move to the next section for an elaborative discussion on the origin of interacting dark matter and dark energy.

In the late eighties, there was no concept of dark energy but the discrepancy in $\Lambda$ was remaining to be a serious issue for modern cosmology. To account of such issue, one of the attempts was to consider a toy model where scalar field is coupled to gravity  \cite{Wetterich:1994bg}. The energy-momentum tensor of such coupled scalar field introduces a time dependent cosmological constant and consequently, it became a possible solution to the cosmological constant problem since the objection on the time-independent cosmological constant is naturally solved due to having a variable nature of the cosmological constant.  After the official introduction of dynamical dark energy models in several forms (see \cite{Copeland:2006wr, AT, Bamba:2012cp} for a detailed survey on them), it was found that they automatically induce coincidence problem \cite{Zlatev:1998tr}. We note that the cosmological constant being time-independent cannot escape from the same problem. Quite interestingly, it was reported in \cite{Amendola:1999er} that if dark energy and dark matter are allowed to interact non-gravitationally with each other, the coincidence problem can be solved. Following this, a series of works with coupled dark matter and dark energy had the same conclusions \cite{Chimento:2003iea,Cai:2004dk, Hu:2006ar, delCampo:2008sr,delCampo:2008jx}.  However, some recent results fueled the investigations of coupled dark matter and dark energy with the claim that the observational data favor an interaction in the dark sector \cite{Salvatelli:2014zta,Nunes:2016dlj,Kumar:2016zpg,Yang:2016evp,vandeBruck:2016hpz,Yang:2017yme,Kumar:2017dnp,Yang:2017zjs,Kumar:2017bpv,Yang:2017ccc}. Additionally, again some very recent investigations in this direction strongly claim that the tension on the local Hubble constant can be released if the interaction in the dark sector is allowed.

However, the most important question in the coupling dynamics is, what should be the
energy transfer rate between the dark sectors? Before we look for an appropriate transfer rate we recall that the nature of both dark matter and dark energy is unknown. On this ground the sensible approach is to consider some well motivated phenomenological transfer rates, or interaction functions and test the expansion history with the available astronomical data.  A number of different interaction rates between dark matter and dark energy have been studied in the last several years 
\cite{Billyard:2000bh,Gumjudpai:2005ry,Barrow:2006hia,%
Zimdahl:2006yq,Amendola:2006dg,CalderaCabral:2008bx,%
Chimento:2009hj,Quartin:2008px,Valiviita:2009nu,%
Clemson:2011an,Thorsrud:2012mu,Pan:2013rha,%
Yang:2014hea,Faraoni:2014vra,Yang:2014gza,Nunes:2014qoa,%
Pan:2014afa,Chen:2011cy,G:2014mea,Pan:2012ki,Li:2013bya,Duniya:2015nva,%
Valiviita:2015dfa,Sola:2016ecz,Mukherjee:2016shl,%
Pan:2016ngu,Dutta:2017kch,Cai:2017yww,Odintsov:2017icc,Pan:2017ent,%
Yang:2018pej, Yang:2018ubt, Yang:2018euj}. 
For a comprehensive review on different interaction rates, we refer to 
\cite{Bolotin:2013jpa, Wang:2016lxa}. We also note that the interaction between the dark sectors has also been examined in a more general framework where the geometry of the universe is inhomogeneous \cite{Izquierdo:2017igb,Izquierdo:2017pnp}.

In this work we concentrate on the spatially flat
Friedmann-Lema\^itre-Robertson-Walker universe where we introduce
an interaction between dark energy and pressureless dark matter
that exactly solves the background evolution. That means the
evolution equations for dark matter and dark energy are
analytically solved. The appearance of analytic structure of the
background evolution makes the cosmological model quite
interesting because the cosmological parameters associated with
this model take analytic forms too. However, this is not new
because the analytic structure for such interaction model has
already been reported by some of the authors in a previous work
\cite{Sharov:2017iue}. But the motivation of the present work is
completely different. Here we aim to test the large-scale
stability of the model which is very important because without
stable perturbations there will be no such structure formation of
the universe.

The analysis  of structure formation in models of DE and DM,
from the point of view of the cosmological perturbations theory,
plays an essential  role when the different  models are confronted
with the observations-data \cite{Hwang:2009zj}. As it is well
known, these dark scenarios  imprint a signature on the cosmic
microwave background (CMB) power spectrum
\cite{Bean:2003fb,Weller:2003hw}. Thus, the study of the
cosmological perturbations is important and also need to be
well-analyzed. In particular, for  models with interaction between
DE and DM,  with adiabatic initial conditions and the perturbation
theory were studied  in Ref. \cite{Bean:2007ny}, see also
\cite{Herrera:2016uci, delCampo:2013hka, Wang:2016lxa}. 
Also an analysis in models with an
interacting DE-DM  with a constant equation of state, was analyzed
in \cite{Valiviita:2008iv}. Here, the authors demonstrated  that
perturbations were unstable together with a rapid growth of DE
fluctuations. In this sense the test the large-scale stability is
fundamental,  since without stable perturbations there will be no
such structure formation.

We organize the work in the following way.
In section \ref{sec-eqns} we describe the basic equations for the interacting
model both at background and perturbative levels. The analytical solutions are
discussed in section \ref{sec-solutions}. The section \ref{sec-results} details the results
of the analysis following the observational data used in this work. Finally, we close our work with a brief summary in section \ref{sec-conclu}.

\section{Interacting Dynamics in flat FLRW}
\label{sec-eqns}

We consider a spatially flat Friedmann-Lema\^itre-Robertson-Walker
(FLRW) universe where pressureless dark matter, also known as
cold-dark-matter (CDM), interacts with a dark energy fluid. The
interaction is non-gravitational, that means gravity does not play
any role in their interaction. Additionally, we consider the
existence of baryons and radiation in the universe sector. To
avoid any kinds of inflexible constraints like ``fifth force'', we
assume that neither baryons nor radiation takes part in the
interaction. In other words, they are conserved separately. Since
the interaction exists between CDM and DE, the  total conservation
of this (CDM$+$DE) sector is,

\begin{eqnarray}\label{total-cons}
\dot{\rho}_c+ 3 H  (1+w_c) \rho_c= -\dot{\rho}_x  - 3 H (1+w_x )\rho_x \, ,
\end{eqnarray}
where ($\rho_c$, $\rho_x$) are respectively the energy density of
CDM and DE. The parameter $H \equiv \dot{a}/a$,
corresponds to the Hubble rate, $a$ being the FLRW scale factor.
The quantity  $w_x = p_x/\rho_x$, corresponds to
the equation of state for DE and  $p_x$ is the pressure of the DE
fluid. Also, we note that $w_c = p_c/\rho_c$ since $p_c$, the
pressure of CDM is zero, hence $w_c= 0$. The total conservation
equation (\ref{total-cons}) can be decoupled into the following
equations

\begin{eqnarray}
&&\dot{\rho}_{c} + 3  H \rho_{c} =  - Q \, ,\label{conservation1}\\
&&\dot{\rho}_{x} + 3 H (1+w_x )\rho_x =  Q\, ,\label{conservation2}
\end{eqnarray}
where an overhead dot represents the cosmic time differentiation.
The parameter $Q$ denotes the energy transfer between the
dark sectors. In this sense, the sign of $Q$ determines the
 direction of energy transfer. For instance, $Q < 0$ indicates the energy transfer from
 dark energy to CDM  while $Q > 0$ means the energy flow occurs from CDM  to  DE. In terms of the Hubble parameter, $H$, we  have the following constrain or Friedmann equation

\begin{eqnarray}
H^2  &=& \frac{8 \pi G}{3} (\rho_r+\rho_b+\rho_{c}+ \rho_{x}).\label{friedmann1}
\end{eqnarray}
Thus, the dynamical evolution of the universe can be determined from eqns. (\ref{conservation1}), (\ref{conservation2}) and (\ref{friedmann1}) once the
interaction rate $Q$ is specified.

Introducing, $\rho_t= \rho_{c}+ \rho_{x}$, as the total energy density of the dark sector,  one can express the energy densities for dark energy and dark matter respectively as
%%%%%%%%%%%%%%%%%%%%%%%%%%%%%%%%%%%%%%%%%%%%%%%%%%%%%%%%%%%%%%%%%%%%%%%%%%%%%%%%%%%%
%%%%%%%%%%%%%%%%%%%%%%%%%%%%%%%%%%%%%%%%%%%%%%%%%%%%%%%%%%%%%%%%%%%%%%%%%%%%%%%%%%%
\begin{eqnarray}
\rho_{x} =-\left(\frac{\rho^\prime _{t}+ 3 \rho_{t}}{3 w_{x}}\right),
\label{DE-density}\\
\rho_{c}=\left(\frac{\rho^\prime _{t}+ 3 (1+ w_{x})\rho_{t}}{3 w_{x}}\right),
\label{DM-density}
\end{eqnarray}
%%%%%%%%%%%%%%%%%%%%%%%%%%%%%%%%%%%%%%%%%%%%%%%%%%%%%%%%%%%%%%%%%%%%%%%%%%%%%
where the prime stands for the differentiation with respect to the
lapse function $N = \ln (a/a_0) = \ln a$ (Here, we set $a_0$, the
present day value of the scale factor to be unity). Now,
inserting (\ref{DE-density}) into (\ref{conservation2}) or
(\ref{DM-density}) into (\ref{conservation1}), we find that the
 differential equation by the total energy density of the dark
 sector is given by 
\begin{eqnarray}
\rho^{\prime\prime}_{t}+ 3 \Bigg[2+ w_{x}- \frac{w^\prime_{x}}{3w_{x}}\Bigg]\rho^\prime_{t}+ 9 \Bigg[(1+w_{x})-\frac{w^\prime_{x}}{3w_{x}}    \Bigg]\rho_{t} \nonumber\\ =  -\, 3 \bar{Q} w_{x},\label{ode12}
\end{eqnarray}
%%%%%%%%%%%%%%%%%%%%%%%%%%%%%%%%%%%%%%%%%%%%%%%%%%%%%%%%%%%%%%%%%%%%%%%%%
where $\bar{Q} = Q/H$. Giving the interaction $Q$ and the
equation of state $w_x$, the differential equation (\ref{ode12}),
if solved, can determine the evolution of each dark sector
separately which can be obtained from  equations
(\ref{DE-density}) and (\ref{DM-density}).
%%%%%%%%%%%%%%%%%%%%%%%%%%%%%%%%%%%%%%%%%%%%%%%%%%%%%%%%%%%%%%%%%%%%%%%%%
In this context because both dark components are assumed to
interact with each other, we must define the interaction rate $Q$ in
order to obtain analytical solutions. As we mentioned before
different  expressions  have been considered in the literature for
the interaction rate $Q$.
The most commonly studied energy transfer between the
dark sectors depends on the energy densities ($\rho_x$, $\rho_c$,
$\rho_t$) or some combinations of these, multiplied by a quantity with
units of the inverse of time, that could  be  a rate or a
differentiation with respect to the time. Commonly,  this rate
corresponds  to the Hubble rate. In particular, in the scenario of
the reheating, this rate was considered as
a constant \cite{Turner:1983he} and an analogous situation for the
curvaton field case \cite{Malik:2002jb}. In the following we will
consider that the transfer rate $Q$ is proportional to the Hubble
rate, as discussed above. Thus, we have
%%%%%%%%%%%%%%%%%%%%%%%%%%%%%%%%%%%%%%%%%%%%%%%%%%%%%%%%%%%%%%%%%%%%%%%%%%%%%
\begin{eqnarray}
Q &=&  - \xi (\dot{\rho}_{c}+\dot{\rho}_{x}) = - \xi \dot{\rho}_{t},\label{interaction}
\end{eqnarray}
%%%%%%%%%%%%%%%%%%%%%%%%%%%%%%%%%%%%%%%%%%%%%%%%%%%%%%%%%%%%%%%%%%%%%%%%%%%%%
where $\xi$ being the coupling parameter of the interaction characterizing the strength and direction of energy transfer between the dark sectors. We note that the negative sign before the coupling parameter in (\ref{interaction}) does not relate anything with the physics of dark matter and dark energy interactions. The typical choice of the interaction (\ref{interaction}) is actually motivared from the phenomenological ground together with the fact that  the background
energy conservation equations are easily solved. In this sense, 
other interaction rates in the literature have been studied such as, $Q \propto \rho_c$ \cite{Valiviita:2009nu}, $Q \propto \rho_x$ \cite{Clemson:2011an}, $Q \propto (\rho_c +\rho_x)$ \cite{Quartin:2008px}, $Q \propto (\rho_c \rho_x)$ \cite{Chimento:2009hj}, $Q \propto (\rho_x\rho_c)(\rho_c +\rho_x)^{-1}$ \cite{Li:2013bya}, $Q \propto \rho_x^2/\rho_c $ \cite{Yang:2017zjs}, 
$Q \propto \dot{\rho}_x$ \cite{Yang:2017ccc}, 
as some particular cases
(also see \cite{Bolotin:2013jpa} for some other interaction models). 
We also mention that the evolution of an inhomogeneous mixture of nonrelativistic
pressureless CDM, coupled to DE in which 
the interaction term proportional to the DE density was studied in Ref. \cite{Izquierdo:2017pnp}. Here, from the spherically symmetric Lema\^{i}tre-Tolman-Bondi metric, the authors found that the interaction $Q$ can be written as  $Q\propto \dot{\rho_x}$ as used in \cite{Yang:2017ccc}. In this sense, from Eqn. (\ref{interaction}) the 
presence of $\dot{\rho}_t$ offers some consequences that differs it from the usual and well known interaction models. Looking at (\ref{interaction}), one can understand that for positive coupling parameter ($\xi >0$), the sign of the interaction rate could be positive, i.e., $Q > 0$ (energy flows from CDM to DE)  if $\dot {\rho}_t <0$, that means the total energy density of the dark fluids should decrease with the evolution of the universe while the interaction rate could be negative, i.e., $Q <0$ (energy flows from DE to CDM) if $\dot{\rho}_t >0$ which means that the total energy density of the dark fluids 
increases with the evolution of the universe. Similarly, for $\xi <0$, one also encounters with the following two possibilities. The interaction rate in this case is positive (i.e., $Q> 0$) for $\dot{\rho}_t >0$ and it is negative (i.e., $Q <0$) if $\dot {\rho}_t <0$. Thus, one can see that the flow of energy between the dark sectors is not only governed by the sign of the coupling parameter, rather it also depends on the evolution of the total dark fluid. This might be considered to be an interesting property of the present interaction since in most of the usual interaction models, the direction of energy flow is actually determined from the coupling parameter only.
The interaction (\ref{interaction})
was explicitly studied in  \cite{G:2014mea}  where a particular case, namely, the interaction between the cosmological constant with matter was considered. However, a careful survey of literature will prove the existence of this interaction in \cite{Chimento:2009hj} much earlier of \cite{G:2014mea}. 

Subsequently, using the same interaction, the background
evolution of the universe was investigated in a generalized way 
where the dark energy equation
of state was considered to be either constant or variable \cite{Sharov:2017iue}.
However, no such perturbation analysis was performed for this interaction and this analysis is an essential issue related to the structure formation of the universe. 
We also observe another interesting feature in this interaction 
and we believe this is worth for further investigations. 
In order to understand this feature,
 we can  rewrite  eq.(\ref{interaction}) in a different way which can be found using the conservation equations (\ref{conservation1}) and (\ref{conservation2})
 where precisely the rate corresponds  to the Hubble parameter such that:

\begin{eqnarray}\label{eq-int}
Q &=& 3\, \xi H\, \Bigl[\rho_{c}+ (1+w_x )\rho_x \Bigr].
\end{eqnarray}
One can now notice that the interaction (\ref{eq-int}) includes the dark energy equation of state $w_x$ aside from the coupling parameter. This differs from the well known interactions where only the energy densities are considered. 
The incorporation of the energy transfer  $Q\propto\dot{\rho}_t$ and hence  
the inclusion of the equation of state could result in a non-interacting scenario (equivalently, $Q =0$) even if the coupling parameter is nonzero. In other words, for $w_{x} = - 1- \rho_c/\rho_x = -1 - r < -1$, where $r= \rho_c/\rho_x >0$, is the coincidence parameter, the non-interacting physics is still realized even for $\xi \neq 0$. We call it the ``\textit{zero coupling condition}''. This kind of interaction is rare in the literature which retrieves the non-interaction cosmology although there exists some non-zero coupling strength. We further notice that the dark energy equation of state in this case belongs to the phantom regime.
We admit that the physics of such zero coupling condition is very strange at least at the present stage, and it surely deserves further investigations. 
We note that a more general interaction scenario recovering the 
above interaction (\ref{interaction}) (or (\ref{eq-int})) was introduced first in Ref. \cite{ Chimento:2009hj} and recently in Ref. \cite{Pan:2017ent} where the authors discussed the analytical solutions for dark matter and dark energy. Certainly, a general interaction recovering different interaction rates as special cases, includes a large number of coupling parameters. The stability of such general interaction model is surely interesting, however, in this work we focus only on the stability of the simplest interaction model that offers an analytic structure.

\section{Exact Solutions}
\label{sec-solutions}

The differential equation (\ref{ode1}) is the main source to understand the evolution of the dark sector, provided this is exactly solved. For constant equation of state in dark energy, the differential equation (\ref{ode1}) is simplified into
\begin{align}
\rho^{\prime\prime}_t+ 3 \Bigl[2+ w_{x}\Bigr]\rho^\prime_t+ 9 \Bigl[1+w_{x} \Bigr]\rho_t  = -\,3 \bar{Q} w_{x} = 3 \xi w_x \rho_t^{\prime},\label{ode1}
\end{align}
and with the use of the interaction (\ref{interaction}), the auxiliary equation becomes $m^2 + 3 \left(  2+w_{x}- \xi w_{x} \right) m + 9 (1+w_x) = 0$. Now, under the condition of $\Delta > 0$ where $\Delta$ is the discriminant of the above auxiliary equation, the exact solution of the above differential equation is,

\begin{eqnarray}
\rho_t&=& \rho_1 a^{m_1}+ \rho_2 a^{m_2},\label{solution-constant-EoS}
\end{eqnarray}
%%%%%%%%%%%%%%%%%%%%%%%%%%%%%%%%%%%%%%%%%%%%%%%%%%%%%%%%%%%%%%%%%%%%%%%%%%%%%%%%%%%%
where $\rho_1$, $\rho_2$ are the constants of integration. The integration constants must be positive, otherwise, if one of them is negative then at some finite scale factor, $\rho_t \equiv 0 \Rightarrow 3 H^2 \approx \rho_b +\rho_r$, which means that the evolution of the universe is governed by the baryons and radition, this is purely unphysical from the observational data we have. Thus, we shall strictly assume that $\rho_1 >0$ and $\rho_2 >0$. The roots of the auxiliary equation, $\big(m_1, m_2\big)$ are given by
\begin{eqnarray}
m_1= \frac{3}{2}\left[-\left(2+w_{x}- \xi w_{x} \right)+ \sqrt{\Delta}\right],\label{m1}\\
m_2= \frac{3}{2}\left[-\left(2+w_{x}- \xi  w_{x} \right)- \sqrt{\Delta}\right],\label{m2}
\end{eqnarray}
where $\Delta = (1 - \xi )^2 w_{x}^2- 4 \xi w_{x}$. In
particular for the case in which $\mid\xi\mid\ll 1$ and as $\mid
w_x\mid\sim\,\mathcal{O}(1) $, we have $\Delta \sim w_x^2>0$. In
this sense,
 for $\Delta >0$,
%%%%%%%%%%%%%%%%%%%%%%%%%%%%%%%%%%%%%%%%%%%%%%%%%%%%%%%%%%%%%%
the exact evolution equations for dark energy and cold dark matter become
\begin{eqnarray}
\rho_{x} = -\left(\frac{1}{3 w_{x}}\right)\Bigg[\rho_1(m_1+ 3) (1+z)^{-m_1} \nonumber\\+ \rho_2(m_2+ 3) (1+z)^{-m_2}\Bigg],\label{DE-constantA}
\end{eqnarray}
%%%%%%%%%%%%%%%%%%%%%%%%%%%%%%%%%%%%%%%%%%%%%%%%%%%%%%%%%%%%%%%%%%%%%
%and
%%%%%%%%%%%%%%%%%%%%%%%%%%%%%%%%%%%%%%%%%%%%%%%%%%%%%%%%%%%%%%%%%%%%
\begin{eqnarray}
\rho_{c}=\left(\frac{1}{3 w_{x}}\right)\Bigg[\rho_1(m_1+ 3+
3w_{x}) (1+z)^{-m_1} \nonumber\\+ \rho_2(m_2+ 3+ 3w_{x})
(1+z)^{-m_2}\Bigg],\label{DM-constantB}
\end{eqnarray}
%%%%%%%%%%%%%%%%%%%%%%%%%%%%%%%%%%%%%%%%%%%%%%%%%%%%%%%%%%%%%%%%%%%%%%%%%%%%%%%%%%%%%%%%%%%%%%%%%%%
where $1+z = a_0a^{-1}= a^{-1}$ (since we have set $a_0 = 1$). Using the present day values of the cosmological parameters, the evolution equations for dark energy and dark matter can respectively be recast as
\begin{widetext}
\begin{eqnarray}
\rho_{x} = -\left(\frac{1}{3 w_{x}}\right)\Bigg[ \left(\frac{(m_2+3+3 w_{x})\rho_{x,0}+(m_2+3)\rho_{c,0}}{m_2-m_1} \right)(m_1+ 3) (1+z)^{-m_1} \nonumber\\+ \left( \frac{(m_1+3+3 w_{x})\rho_{x,0}+(m_1+3)\rho_{c,0}}{m_1-m_2}\right) (m_2+ 3) (1+z)^{-m_2}\Bigg]
\end{eqnarray}
%%%%%%%%%%%%%%%%%%%%%%%%%%%%%%%%%%%%%%%%%
\begin{eqnarray}
\rho_{c}=\left(\frac{1}{3 w_{x}}\right)\Bigg[\left(\frac{(m_2+3+3 w_{x})\rho_{x,0}+(m_2+3)\rho_{c,0}}{m_2-m_1} \right)(m_1+ 3+ 3w_{x}) (1+z)^{-m_1} \nonumber\\+ \left( \frac{(m_1+3+3 w_{x})\rho_{x,0}+(m_1+3)\rho_{c,0}}{m_1-m_2}\right) (m_2+ 3+ 3w_{x}) (1+z)^{-m_2}\Bigg].
\end{eqnarray}
\end{widetext}
Furthermore, in terms of the new quantities ($\rho_1$, $\rho_2$, $m_1$, $m_2$), the usual density parameters for dark energy and dark matter at current time are calculated as
\begin{eqnarray}
\Omega_{x,0} = -\left(\frac{1}{3 w_{x}}\right)\Bigl[\Omega_1(m_1+ 3)+ \Omega_2(m_2+ 3)\Bigr],\label{DEDP}
\end{eqnarray}
\begin{align}
\Omega_{c,0} =  \left(\frac{1}{3 w_{x}}\right)\Bigg[\Omega_1(m_1+ 3+ 3 w_{x}) + \Omega_2(m_2+ 3+ 3 w_{x})\Bigg],\label{DMDP}
\end{align}
%%%%%%%%%%%%%%%%%%%%%%%%%%%%%%%%%%%%%%%%%%%%%%%%%%%%%%%%%%%%%%%%%%%%%%%%%%%%%
 with $\Omega_{x,0}+\Omega_{c,0}= \Omega_1+ \Omega_2$, where
$(\Omega_1, \Omega_2) = (\rho_1/\rho_0, \rho_2/\rho_0)$ and
$\rho_0 = 3H_0^2/8\pi G$. We note that solving the above
equations (\ref{DEDP}) and (\ref{DMDP}) one can easily find
$\Omega_1 = \Omega_1 (m_1, m_2, w_{x})$ and $\Omega_2 = \Omega_2
(m_1, m_2, w_{x})$. The explicit forms for $\Omega_1$ and
$\Omega_2$ are, $\Omega_1= \frac{(m_2+3+3 w_{x})\Omega_{x,0}+(m_2+3)\Omega_{c,0}}{m_2-m_1},\, \Omega_2= \frac{(m_1+3+3 w_{x})\Omega_{x,0}+(m_1+3)\Omega_{c,0}}{m_1-m_2}$.
%%%%%%%%%%%%%%%%%%%%%%%%%%%%%%%%%%%%%%%%%%%%%%%%%%%%%%%%%%%%%%%%%%%%%%%%%%%%%%%%
In particular, we consider the case when dark energy is the
cosmological constant, i.e., the case when $w_x = -1$. For 
convenience, we label IDE as the interacting dark energy 
scenario where DE is not the cosmological constant (i.e., $w_x \neq -1$) while by
IVS we mean the interacting vacuum scenario, that means when 
the DE is represented by the cosmological constant itself.

\section{Dynamics at Large scales: Cosmological Perturbations}
\label{sec-perturbations}

The study of cosmological perturbations unveils the hidden nature of the model. The large scale stability thus has been a very important issue to check the viability of any cosmological model. Indeed, for coupled dark energy one needs to check the same. Precisely, we are interested on the structure formation when the background has a coupling between the dark matter and dark energy governed by the interaction rate specified in equation (\ref{interaction}). Thus, we
consider the perturbed FLRW metric with scalar mode $k$ as  \cite{Mukhanov, Ma:1995ey, Malik:2008im}

\begin{eqnarray}\label{perturbed-metric}
ds^{2}=a^{2}(\tau )\Bigg[-(1+2\phi )d\tau ^{2}+2\partial _{i}Bd\tau dx^{i}\nonumber\\+
\Bigl((1-2\psi )\delta _{ij} +2\partial _{i}\partial _{j}E\Bigr)dx^{i}dx^{j}
\Bigg],
\end{eqnarray}
where $\phi $, $B$, $\psi $, $E$, are
the gauge-dependent scalar perturbation quantities and $\tau $ is the conformal time. Thus, using the metric (\ref{perturbed-metric}), one can find the perturbed equations \cite{Majerotto:2009np, Valiviita:2008iv, Clemson:2011an},
\begin{equation*}
\nabla _{\nu }T_{A}^{\mu \nu }=Q_{A}^{\mu },~~~~\sum\limits_{\mathrm{A}}{%
Q_{A}^{\mu }}=0,
\end{equation*}%
where the symbol $A$ represents the fluid (dark matter or dark energy) and $Q_{A}^{\mu }=(Q_{A}+\delta Q_{A})u^{\mu }+F_A^{\mu}$, where the quantities $Q_A$ is the energy transfer rate and
$F_A^{\mu} = a^{-1} (0, \partial^{i} f_A)$ is the momentum density transfer
relative to the four-velocity $u^{\mu }$, for more discussions in this direction,
we refer to some earlier works
\cite{Majerotto:2009np, Valiviita:2008iv, Clemson:2011an}.
We consider that in the rest frame of dark matter, the momentum
transfer potential is zero \cite{Valiviita:2008iv,Clemson:2011an, Koyama:2009gd}.
Thus, the momentum
transfer potential becomes $k^{2}f_{A}=Q_{A}(\theta -\theta _{c})$. The pressure
perturbation is defined by  \cite{Kodama:1985bj,Hu:1998kj,Valiviita:2008iv}
\begin{equation}
\delta p_{A}=c_{sA}^{2}\delta \rho_{A}+(c_{sA}^{2}-c_{aA}^{2})\rho _{A}^{\prime }(v_{A}+B),
\end{equation}
where $c_{aA}^2$ is the square of the physical sound speed of the fluid `$A$' in the rest frame and it is defined as $c_{aA}^{2}=p_{A}^{%
\prime }/\rho _{A}^{\prime }=w_{x}+w_{x}^{\prime }/(\rho _{A}^{\prime }/\rho
_{A})$.
Now, introducing the density perturbation by $\delta _{A}=\delta \rho _{A}/\rho
_{A}$ and considering no contribution from the anisotropic
stress, i.e., $\pi _{A}=0$,
the density perturbation and the velocity perturbation
equations for the dark matter and dark energy fluids \cite{Majerotto:2009np, Valiviita:2008iv, Clemson:2011an}
\begin{widetext}
\begin{eqnarray}
\delta_A^{\prime} + 3 \mathcal{H} \left(c_{sA}^2 - w_A \right) \delta_A + 9 \mathcal{H}^2 \left(1+w_A \right) \left(c_{sA}^2- c_{aA}^2 \right)\frac{\theta_A}{k^2} + \left(1+w_A \right) \theta_A -3 \left(1+w_A \right) \psi^{\prime} + \left(1+w_A \right) k^2 \left(B- E^{\prime} \right)\nonumber\\ = \frac{a}{\rho_A} \left(\delta Q_A - Q_A \delta _A \right) + \frac{a Q_A}{\rho_A} \left[\phi + 3 \mathcal{H} \left(c_{sA}^2- c_{aA}^2 \right)\frac{\theta_A}{k^2}\right],\\
\theta_A^{\prime} + \mathcal{H} \left(1-3 c_{sA}^2  \right)\theta_A - \frac{c_{sA}^2}{1+w_A} k^2 \delta_A -k^2 \phi = \frac{a}{(1+w_A)\rho_A} \Bigl[ \left(Q_A \theta -k^2 f_A \right) - \left(1+ c_{sA}^2 \right) Q_A \theta_A \Bigr],
\end{eqnarray}
\end{widetext}
where the new quantities $c_{sA}^2$, $c_{aA}^2$, are the adiabatic and physical sound velocity for the fluid $A$, respectively, and $\theta = \theta_{\mu}^{\mu}$ is the volume expansion scalar. Let us note that to avoid from any kind of instabilities, $c_{sA}^2 \geq 0$ has been imposed. We also note that here
$c_{sc}^2 =0$ since we assume cold dark matter (i.e., $w_c$ = 0). In the synchronous gauge, (i.e., $\phi =B=0$, $\psi =\eta $, and $k^{2}E=-h/2-3\eta $), the density and the velocity perturbations for the dark fluids follow
\begin{widetext}
\begin{eqnarray}
\delta _{x}^{\prime } &=&-(1+w_{x})\left( \theta _{x}+\frac{h^{\prime }}{2}%
\right) -3\mathcal{H}(c_{s,x}^{2}-w_{x})\left[ \delta _{x}+3\mathcal{H}%
(1+w_{x})\frac{\theta _{x}}{k^{2}}\right] -3\mathcal{H}w_{x}^{\prime }\frac{%
\theta _{x}}{k^{2}}  \notag \\
&+&\frac{aQ}{\rho _{x}}\left[ -\delta _{x}+\frac{\delta Q}{Q}+3\mathcal{H}%
(c_{s,x}^{2}-w_{x})\frac{\theta _{x}}{k^{2}}\right] , \\
\theta _{x}^{\prime } &=&-\mathcal{H}(1-3c_{s,x}^{2})\theta _{x}+\frac{%
c_{s,x}^{2}}{(1+w_{x})}k^{2}\delta _{x}+\frac{aQ}{\rho _{x}}\left[ \frac{%
\theta _{c}-(1+c_{s,x}^{2})\theta _{x}}{1+w_{x}}\right] , \label{theta-x}\\
\delta _{c}^{\prime } &=&-\left( \theta _{c}+\frac{h^{\prime }}{2}\right) +%
\frac{aQ}{\rho _{c}}\left( \delta _{c}-\frac{\delta Q}{Q}\right) , \label{eqn:delta-c}\\
\theta _{c}^{\prime } &=&-\mathcal{H}\theta _{c},
\end{eqnarray}%
\end{widetext}
where the term $\delta Q/Q$ includes the perturbation term for the Hubble
expansion rate $\delta H$. Now inserting the interaction rate (\ref{interaction}) into the above equations, one can write down the explicit perturbation equations as

\begin{widetext}
\begin{eqnarray}
\delta'_x
&=&-(1+w_x)\left(\theta_x+\frac{h'}{2}\right)
-3\mathcal{H}(c^2_{sx}-w_x)\left[\delta_x+3\mathcal{H}(1+w_x)\frac{\theta_x}{k^2}\right] \nonumber \\
&+&3\mathcal{H}\xi\left[\frac{\rho_c}{\rho_x}+(1+w_x)\right]\left[\frac{\rho_c(\delta_c-\delta_x)}{\rho_c+(1+w_x)\rho_x}+\frac{\theta+h'/2}{3\mathcal{H}}+3\mathcal{H}(c^2_{sx}-w_x)\frac{\theta_x}{k^2}\right], \\
\theta'_x
&=&-\mathcal{H}(1-3c^2_{sx})\theta_x+\frac{c^2_{sx}}{(1+w_x)}k^2\delta_x
+3\mathcal{H}\xi\left[\frac{\rho_c}{\rho_x(1+w_x)}+1\right]\left[\theta_c-(1+c^2_{sx})\theta_x\right], \\
\delta'_c
&=&-\left(\theta_c+\frac{h'}{2}\right)
+3\mathcal{H}\xi\left[1+(1+w_x)\frac{\rho_x}{\rho_c}\right]\left[\frac{(1+w_x)\rho_x(\delta_c-\delta_x)}{\rho_c+(1+w_x)\rho_x}-\frac{\theta+h'/2}{3\mathcal{H}}\right], \label{delta_c}\\
\theta'_c
&=&-\mathcal{H}\theta_c~.
\label{eq:perturbation}
\end{eqnarray}

\end{widetext}

Let us now focus on the growth-rate of matter perturbations for the prescribed
interaction in this work.  Here, we neglect the clustering of dark energy
with the assumption of $c_{sx}^2 = 1$. However, depending on the strength of the interaction, the dark energy perturbations could be an
important issue, but on the sub-Hubble scale, such perturbation
is not important provided that the sound speed of dark energy perturbations is
assumed to be positive \cite{Koyama:2009gd}. The
evolution equation for $\delta_c$ can be written as

\begin{widetext}
\begin{eqnarray}
&&\delta ''_c+\left\{1-3\xi\left[1+(1+w_x)\frac{\rho_x}{\rho_c}\right]\right\}\mathcal{H}\delta '_c
=4\pi Ga^2\rho_b\delta_b + \nonumber \\
&&4\pi Ga^2\rho_c\delta_c \left\{1+2\xi\frac{\rho_{t}}{\rho_c}\left[1+(1+w_x)\frac{\rho_x}{\rho_c}\right]
\left[ \frac{\mathcal{H}'}{\mathcal{H}^2}+1-3w_x+3\xi\left(1+\frac{\rho_x}{\rho_c}\right) +3\xi(1+w_x)\frac{\rho_x}{\rho_c}\left(1+\frac{\rho_x}{\rho_c}\right) \right] \right\},
\label{eq:deltacprime2}
\end{eqnarray}
\end{widetext}
where $\mathcal{H} = a H$, is the conformal Hubble parameter and $H$ can be found from (\ref{friedmann1}). It is evident that putting $\xi =0$ into (\ref{eq:deltacprime2}), one gets back the evolution equation for $\delta_c$ for the non-interacting cosmologies, i.e. $ \delta_m ''+\mathcal{H} \delta_m ' = 4 \pi G \rho_m \delta_m$ (Note that, $\rho_m = \rho_c +\rho_b$). Furthermore, one can also measure the deviations in the expansion history through
\begin{eqnarray}
\frac{\mathcal{H}_{eff}}{\mathcal{H}}=1-3\xi\left[1+(1+w_x)\frac{\rho_x}{\rho_c}\right],
\label{eq:Heff}
\end{eqnarray}
and also in the gravitational constant $G$ as

\begin{align}
\frac{G_{eff}}{G}=1+2\xi \left( \frac{\rho_{t}}{\rho_c} \right) \left[1+(1+w_x)\frac{\rho_x}{\rho_c}\right]
\Bigg[ \frac{\mathcal{H}'}{\mathcal{H}^2}+1-3w_x \nonumber\\+3\xi\left(1+\frac{\rho_x}{\rho_c}\right) +3\xi(1+w_x)\frac{\rho_x}{\rho_c}\left(1+\frac{\rho_x}{\rho_c}\right) \Bigg].
\label{eq:Geff}
\end{align}

\begin{figure*}
\includegraphics[width=0.35\textwidth]{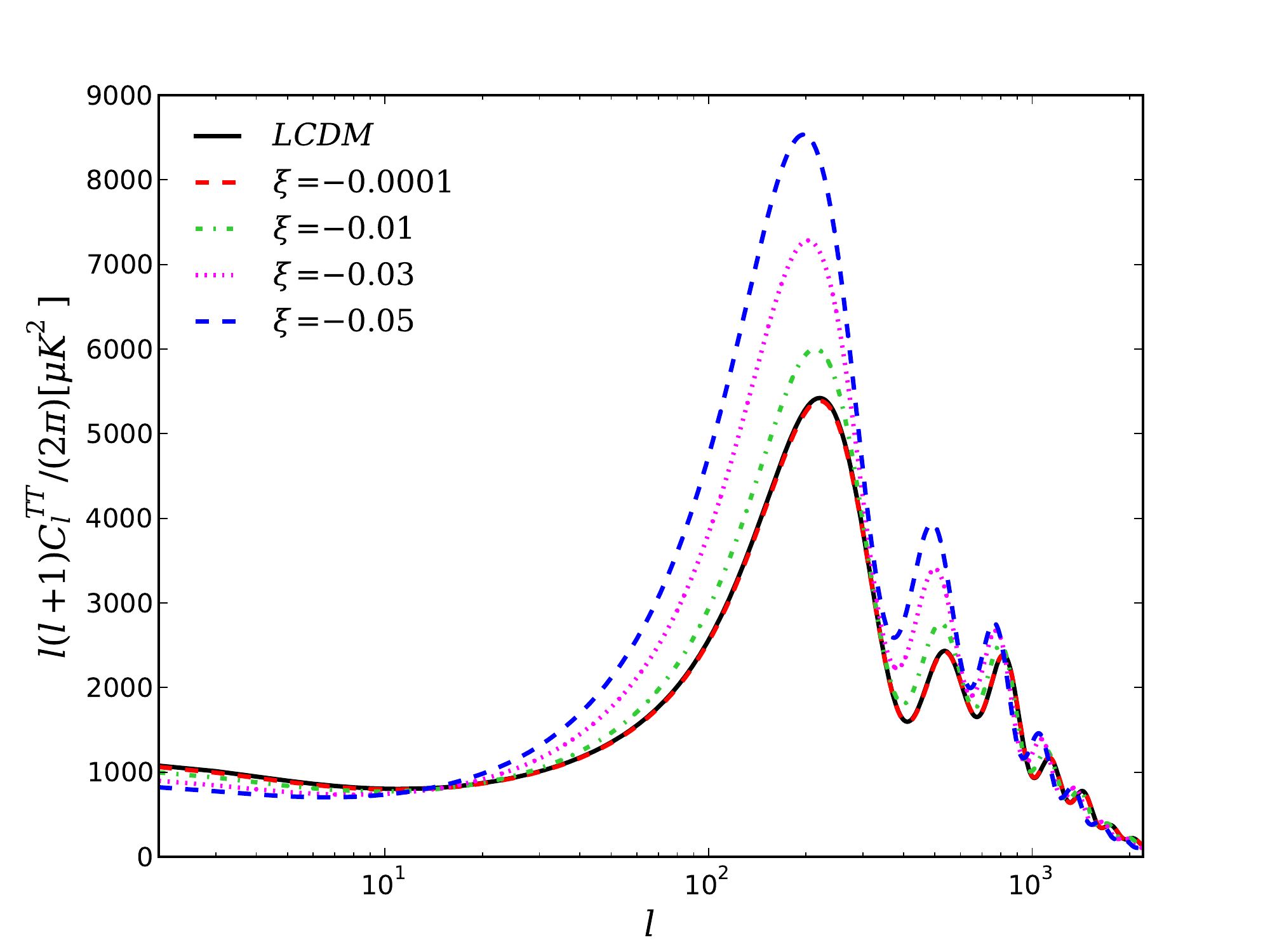}
\includegraphics[width=0.35\textwidth]{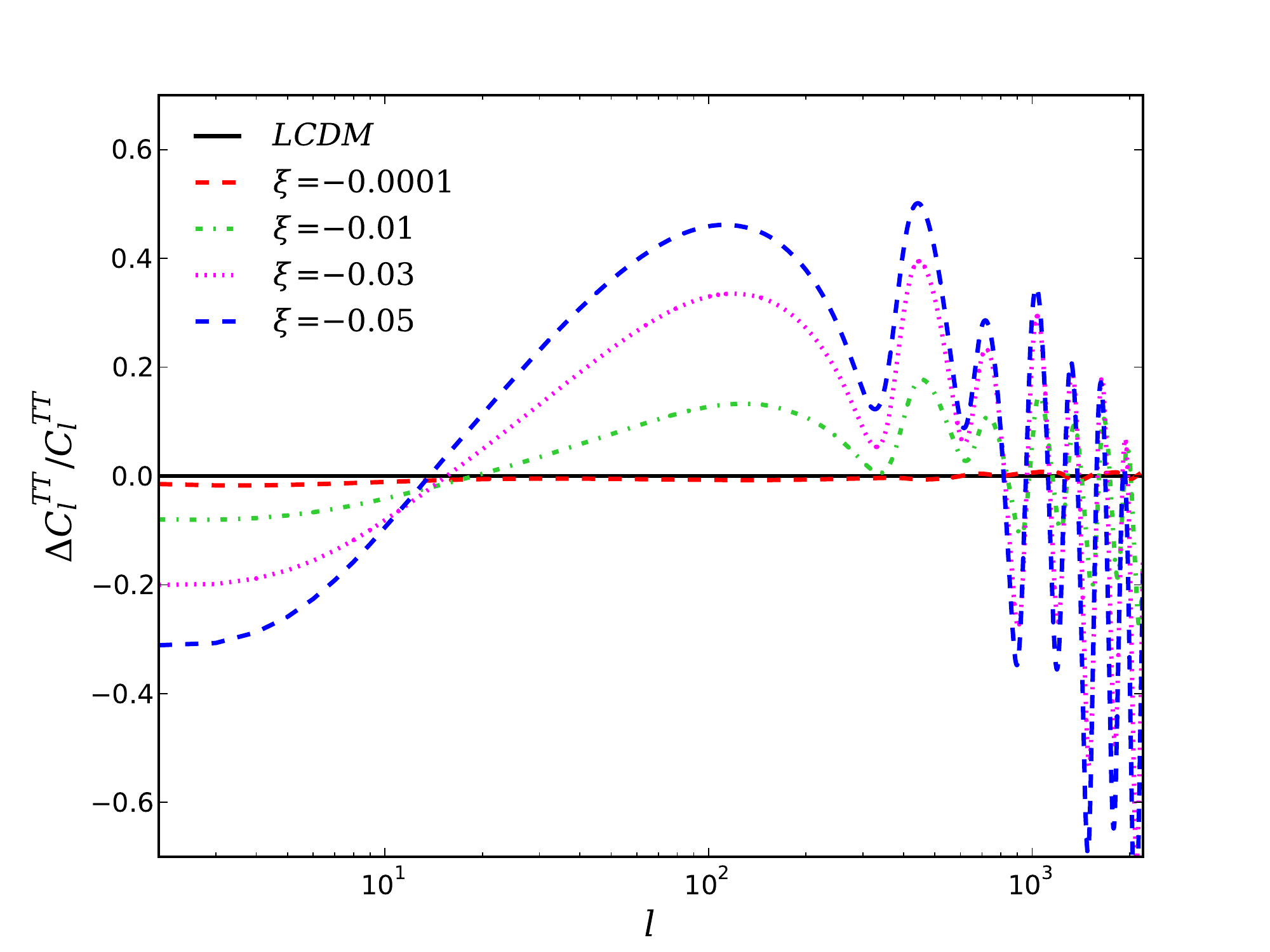}
\caption{Color Online $-$ The behaviour of the IDE scenario in the large scales has been presented for different measures of the coupling parameter $\xi$. \texttt{Left Panel:} Here we display the evolutions of the CMB TT spectra for different values of the coupling parameter representing its strength. We see that with the increase in the magnitude of the coupling parameter, the interaction scenario effectively deviates from the usual non-interacting $\Lambda$CDM cosmology. We note that the curves presenting $\xi =-0.0001$ and $\Lambda$CDM are almost indistinguishable from one another. 
\texttt{Right Panel:} Here, the relative deviation in the CMB TT spectra in compared to the non-interacting $\Lambda$CDM model has been shown. This confirms the observation as found in the left panel of this figure. In this figure, we observe that a very small difference between the curves presenting $\xi =-0.0001$ and $\Lambda$CDM exists but that is very hard to detect. }
\label{fig:CMB-ide}
\end{figure*}
\begin{figure*}
\includegraphics[width=0.35\textwidth]{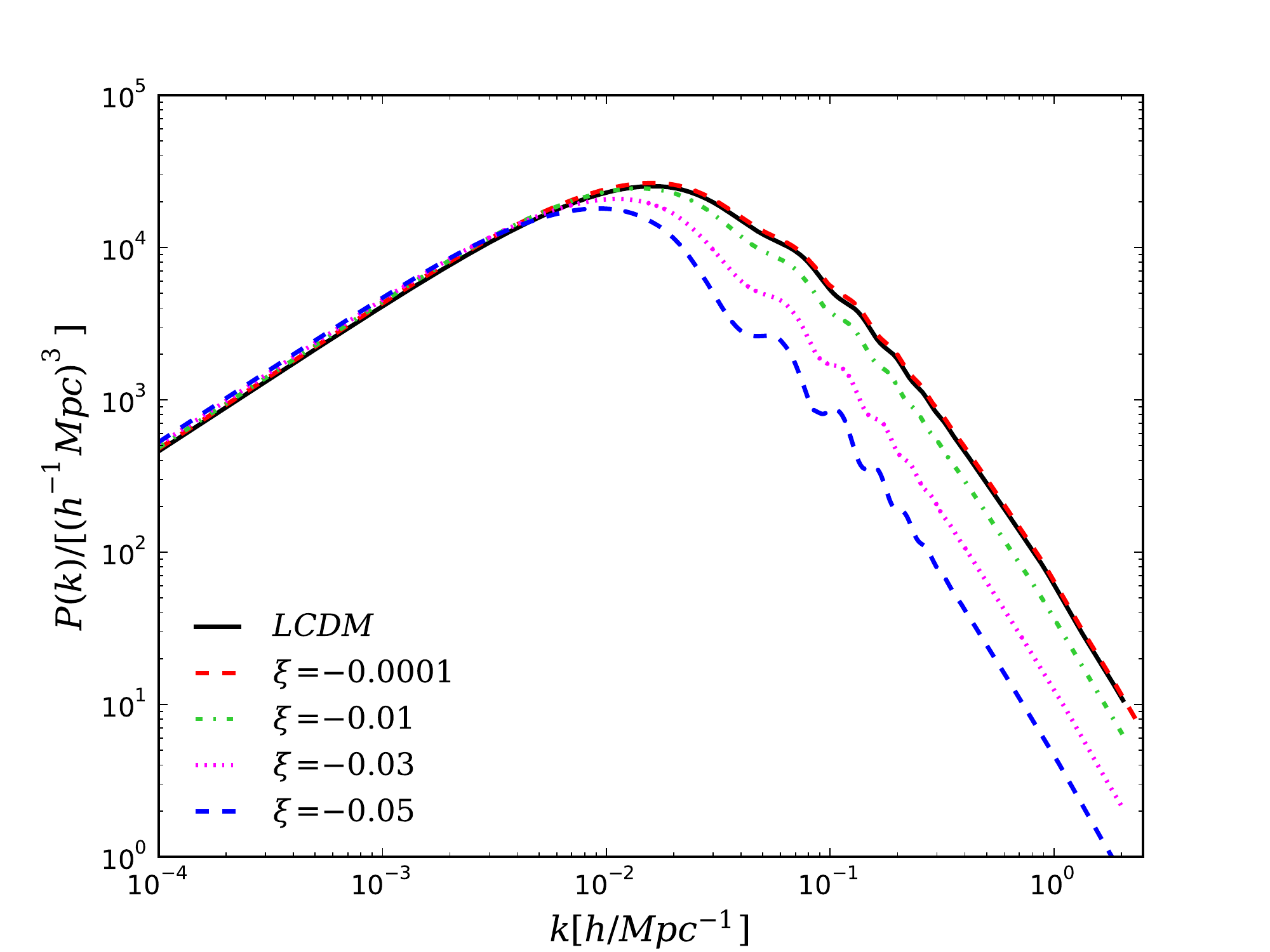}
\includegraphics[width=0.35\textwidth]{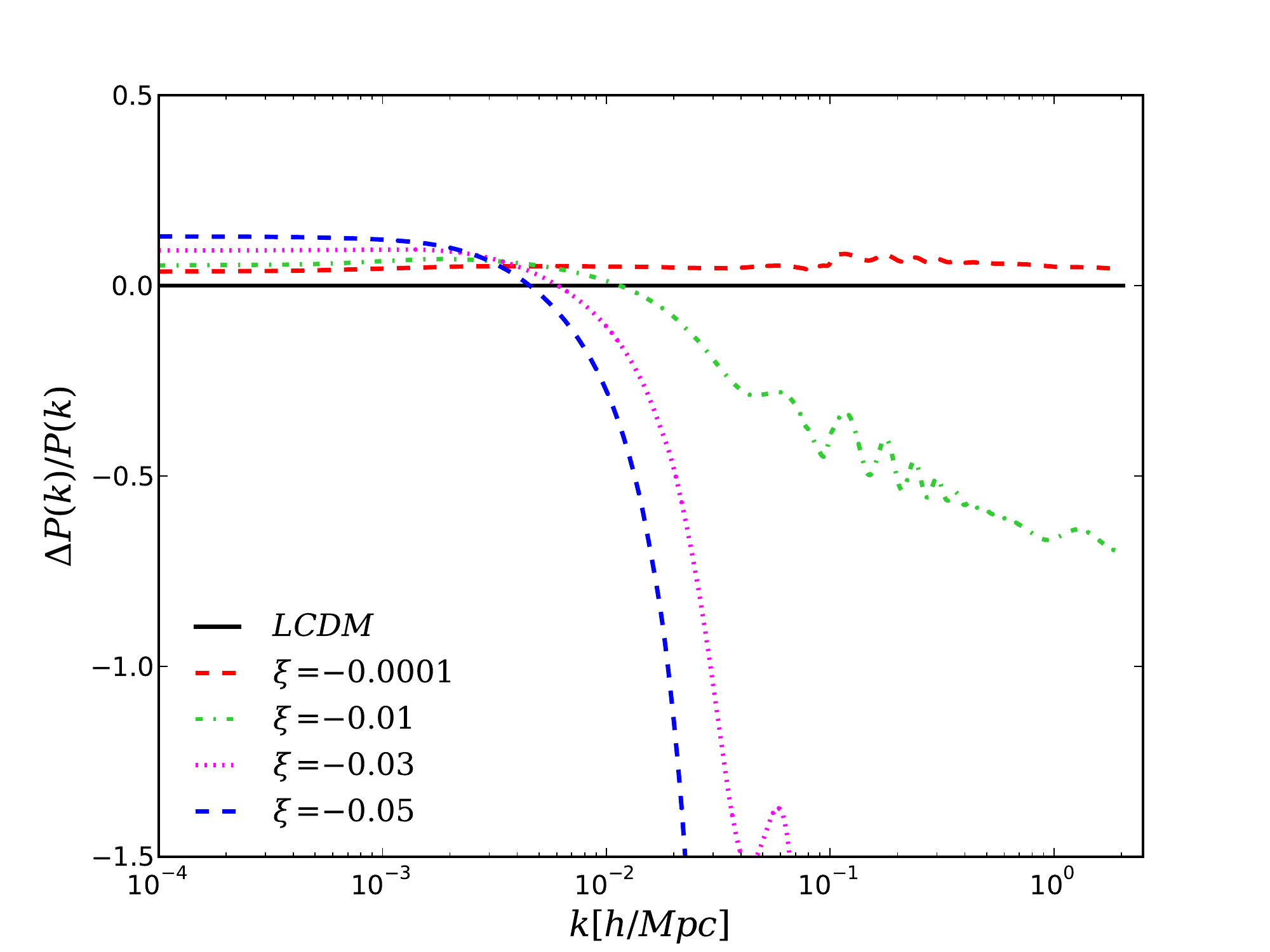}
\caption{Color Online $-$ The behaviour of the IDE scenario in the large scales has been presented for different measures of the coupling parameter. \texttt{Left Panel:} We show the evolutions of the matter power spectra for different coupling strengths of the interaction model. We find that with the increase of the coupling strength, the interaction scenario has a deviation from the usual non-interacting scenario (i.e., $\xi =0$) $\Lambda$CDM. Let us note that the curves presenting $\xi =-0.0001$ and $\Lambda$CDM are almost indistinguishable from one another. \texttt{Right Panel:} The relative deviation in the matter power spectra in compared to the  non-interacting $\Lambda$CDM model has been shown and we find similar observation as realized from its left panel. In this figure, we observe that a very small difference between the curves presenting $\xi =-0.0001$ and $\Lambda$CDM exists, and it is clearly visible. }
\label{fig:Mpower-ide}
\end{figure*}
\begin{figure*}
\includegraphics[width=0.35\textwidth]{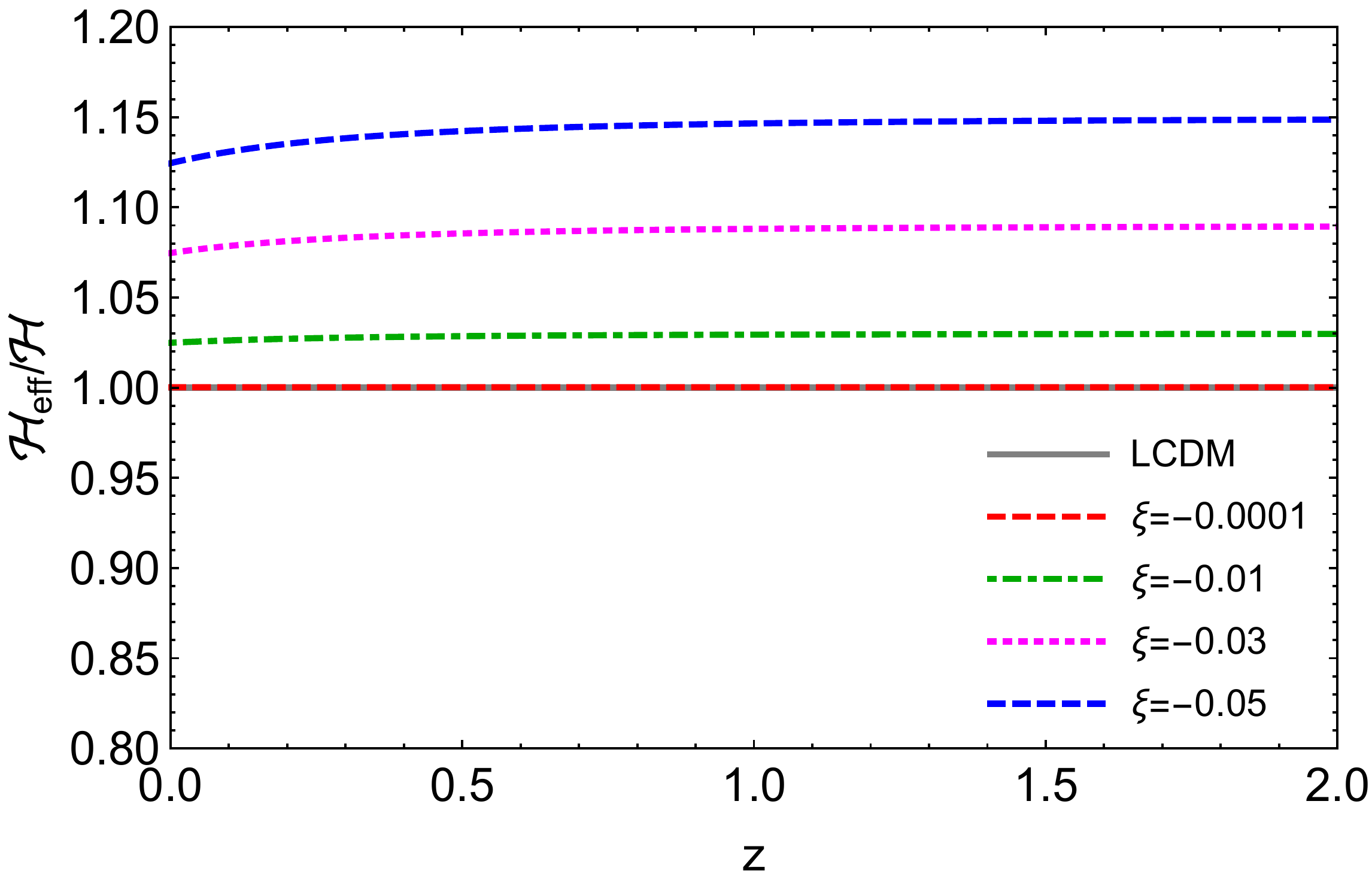}
\includegraphics[width=0.35\textwidth]{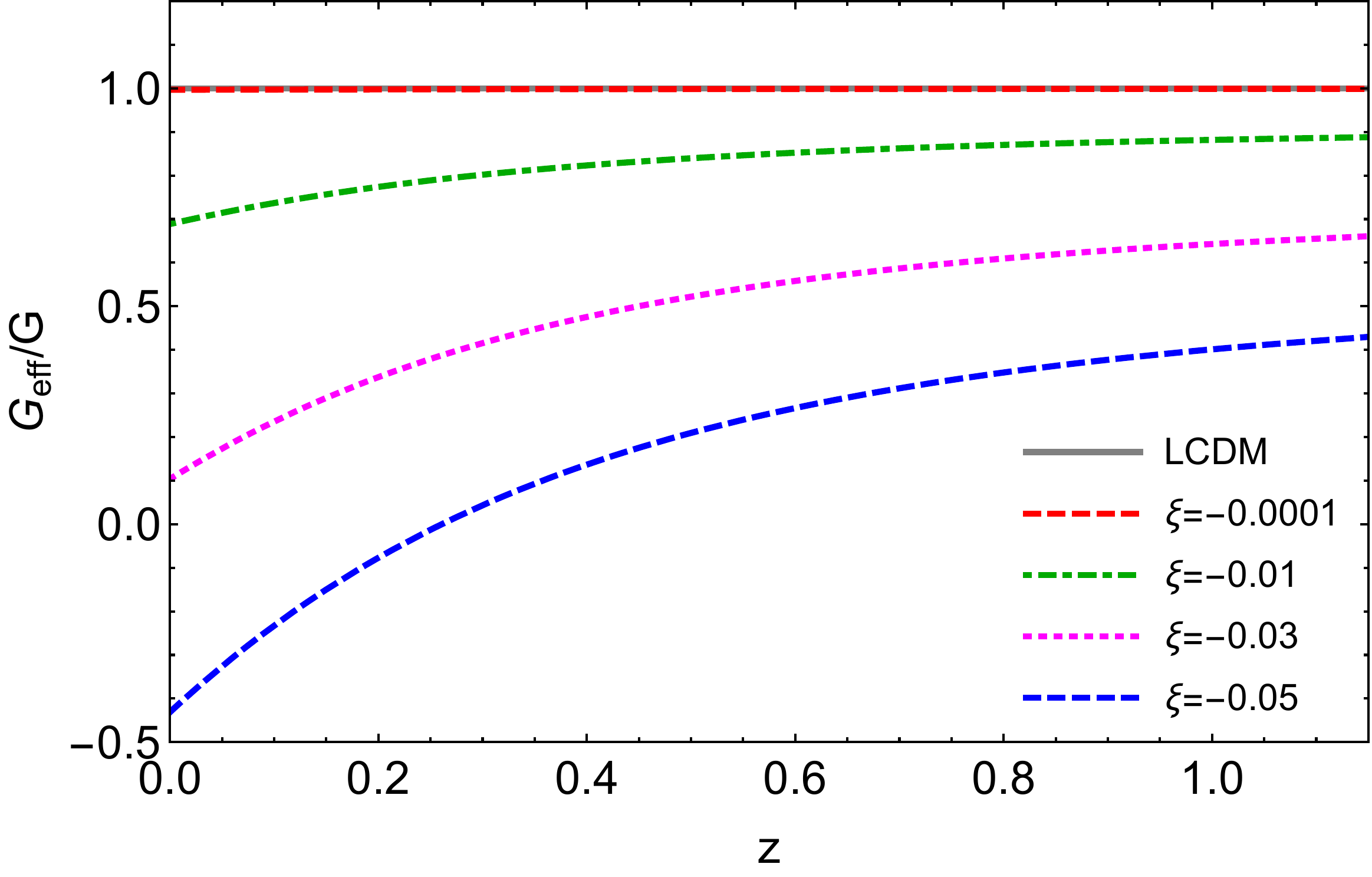}
\caption{Color Online $-$ \texttt{Left Panel:} The dynamical evolution of the quantity $\mathcal{H}_{\text{eff}}/\mathcal{H}$ has been depicted in presence of different coupling parameters of the interaction rate (\ref{interaction}). The curves from upper to lower respectively stand for $\Lambda$CDM ($\xi =0$) model, $\xi = -0.0001, -0.01, -0.03, -0.05$. We notice that the curves presenting non-interacting $\Lambda$CDM and $\xi= -0.0001$  are practically indistinguishable from one another. \texttt{Right Panel:} The evolution of the quantity $G_{\text{eff}}/G$ has been shown for different coupling parameters of the interaction rate (\ref{interaction}). The curves from lower to upper levels respectively stand for $\Lambda$CDM ($\xi =0$) model, $\xi = -0.0001, -0.01, -0.03, -0.05$. Similar to the left panel, here we also notice that the curves presenting $\Lambda$CDM and $\xi= -0.0001$ are practically indistinguishable from each other. From both the panels, we arrive at a common conclusion which states that, as $\xi$ increases (considering its magnitude), the model starts deviating from the non-interacting $\Lambda$CDM cosmology and the coupling parameter $\xi =-0.05$ can be safely excluded from the consideration. }
\label{fig-Heff-Geff}
\end{figure*}
\begin{figure*}
\includegraphics[width=0.35\textwidth]{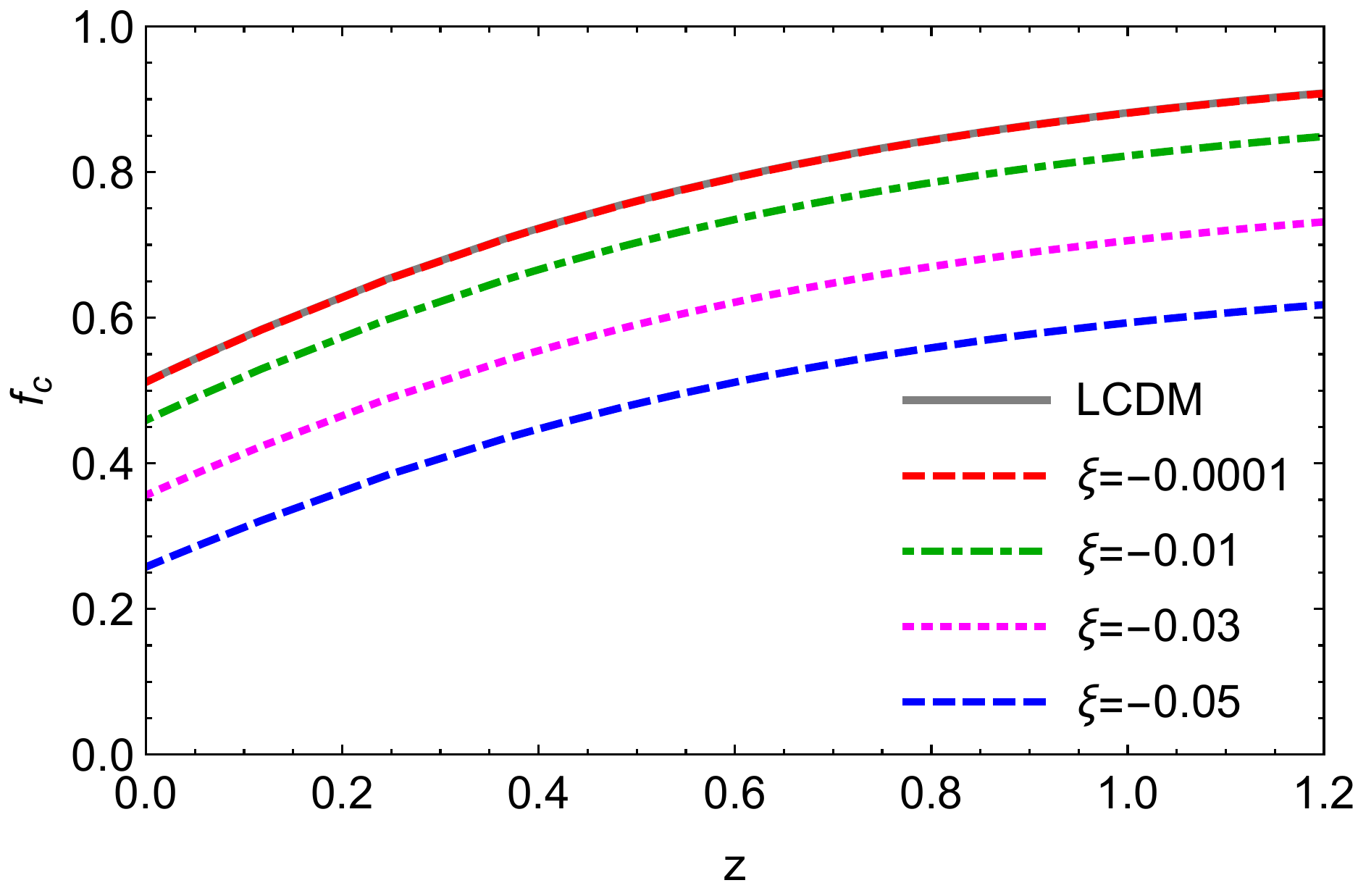}
\caption{Color Online $-$  The evolution of growth rate for the cold dark matter in presence of the interaction rate (\ref{interaction}) has been shown for different values of the coupling strength. The curves from upper to lower respectively stand for the non-interacting $\Lambda$CDM model (where $\xi=0$) and with other coupling parameters $\xi= -0.0001, -0.01, -0.03, -0.05$. Here too, the curves for $\Lambda$CDM and $\xi= -0.0001$ are indistinguishable from one another. 
From the figure we observe that  as long as the strength or magnitude of the coupling parameter increases, the growth rate for the cold dark matter sector significantly deviates from
$\xi = 0$ (no-interaction, $\Lambda$CDM). The physical scenario indicates that with the increase of the coupling strength, the growth-rate for the cold dark matter decreases with the evolution of the universe. }
\label{fig-fc}
\end{figure*}
\begin{figure*}
\includegraphics[width=0.35\textwidth]{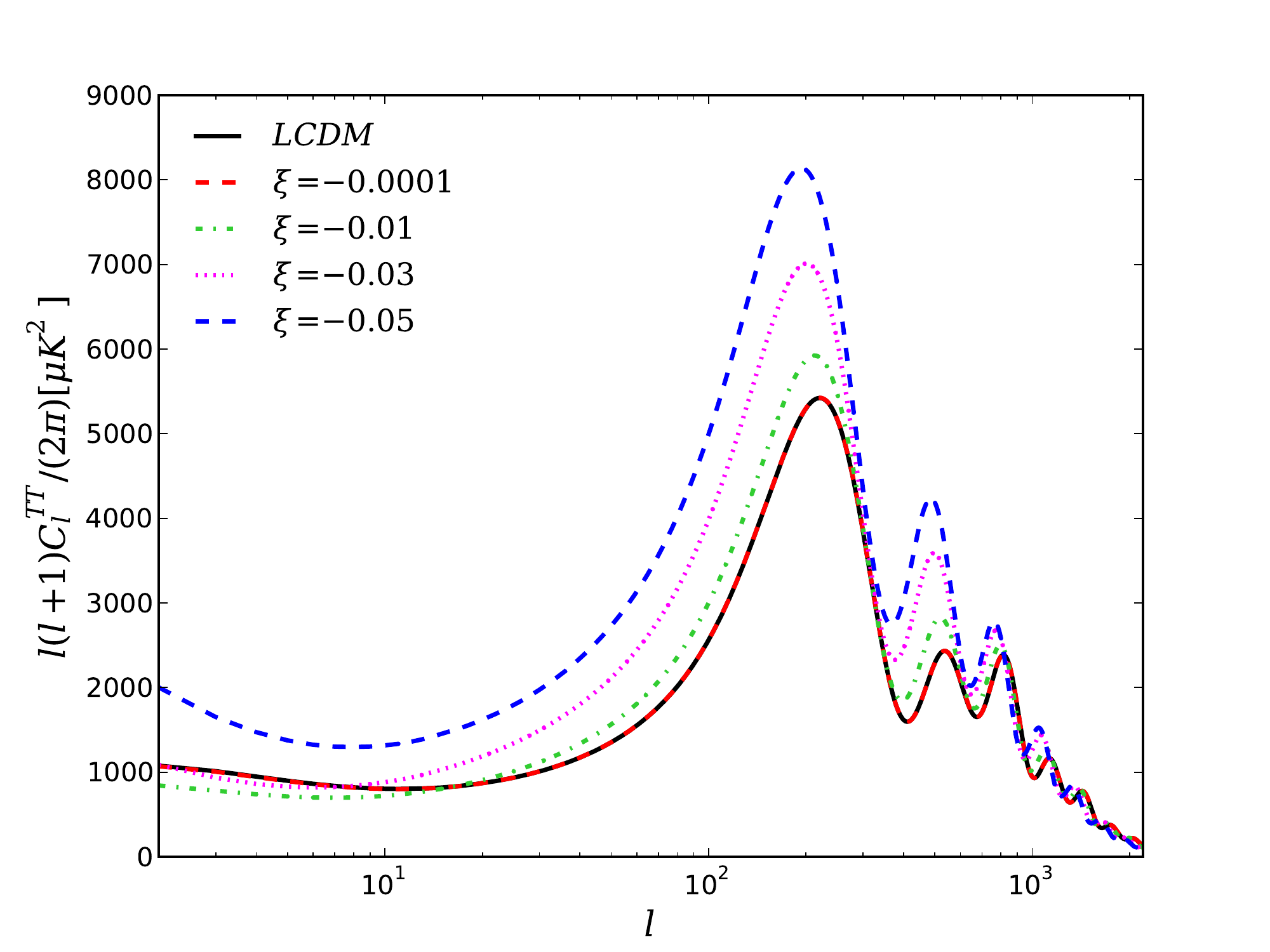}
\includegraphics[width=0.35\textwidth]{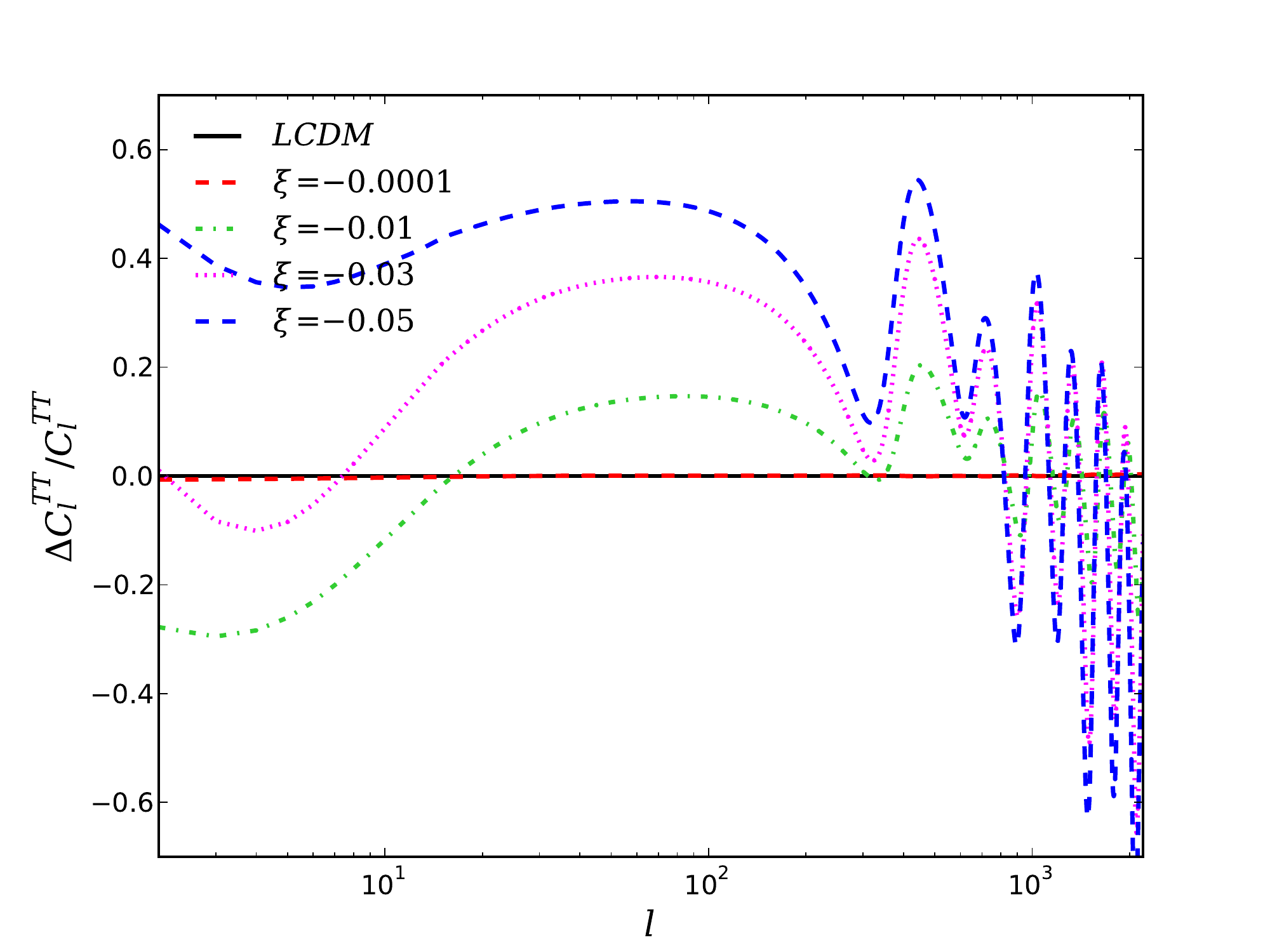}
\caption{Color Online $-$ The behaviour of the interacting vacuum scenario (IVS) in the large scales  has been presented for different measures of the coupling parameter $\xi$. \texttt{Left Panel:} In this plot, we show the evolutions of the CMB TT spectra for different coupling strengths of the interaction model. One can clearly see that as the magnitude or strength of the coupling parameter increases, te deviation of the interaction model becomes prominent from the non-interacting $\Lambda$CDM cosmology.
We note that the curves presenting $\xi =-0.0001$ and $\Lambda$CDM cannot be differentiated from one another.  \texttt{Right Panel:} The relative deviation in the CMB TT spectra in compared to the non-interacting $\Lambda$CDM model has been shown here. From this plot, one can easily conclude that the increament in $\xi$ results in significant deviation from the corresponding non-interacting scenario. Here, we observe that the curves presenting $\xi =-0.0001$ and $\Lambda$CDM overlap with each other.}
\label{fig:CMB-ivs}
\end{figure*}
\begin{figure*}
\includegraphics[width=0.35\textwidth]{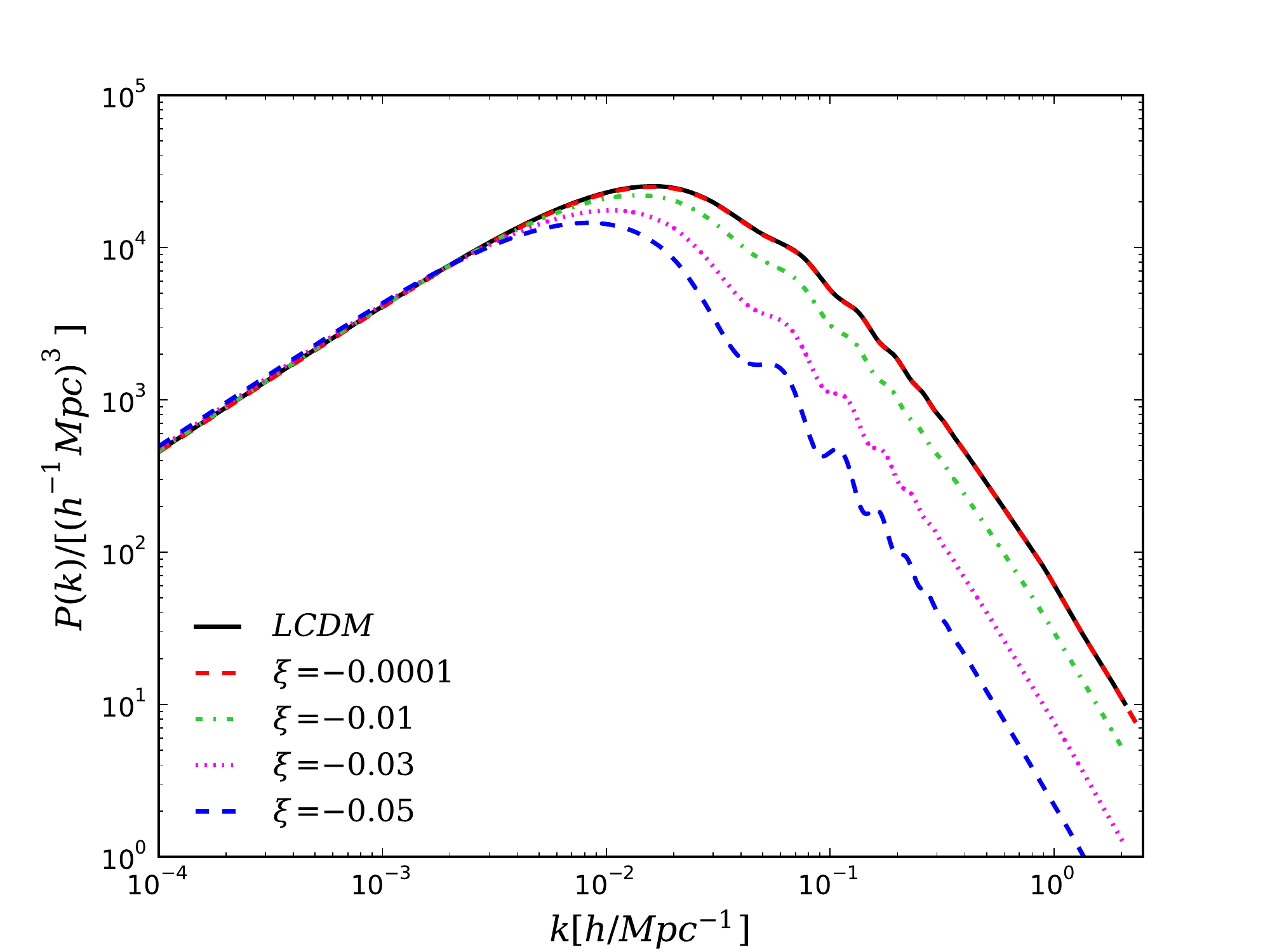}
\includegraphics[width=0.35\textwidth]{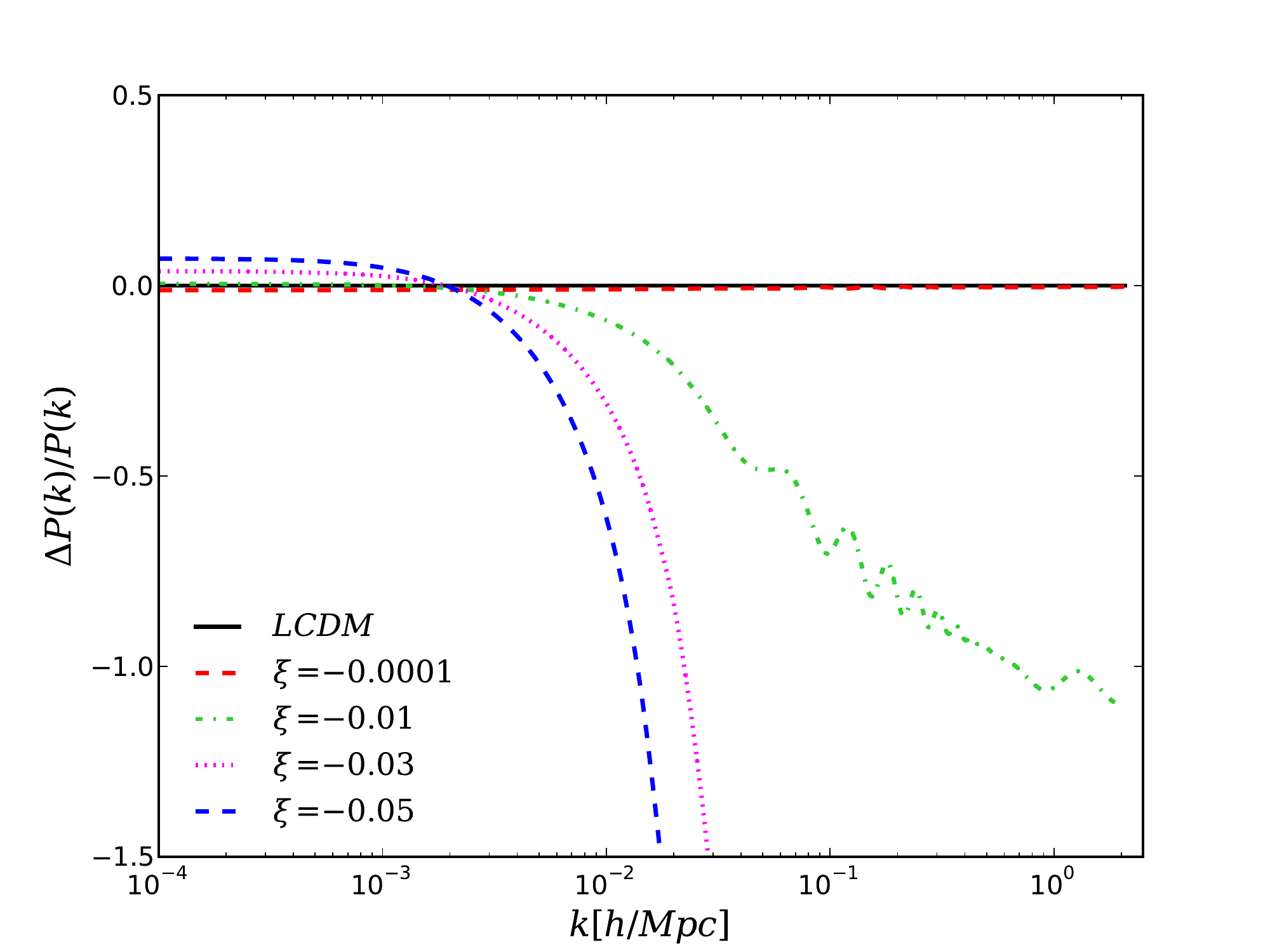}
\caption{Color Online $-$ The behaviour of the interacting vacuum scenario (IVS) in the large scales has been displayed for different measures of the coupling parameter $\xi$. \texttt{Left Panel:} We show the evolutions of the matter power spectra for different coupling strengths of the interaction model which shows that with the increase of the coupling parameter, the interaction model results in significant deviation from the non-interacting $\Lambda$CDM cosmology. We notice that the curves presenting $\xi =-0.0001$ and $\Lambda$CDM cannot be differentiated from one another.
\texttt{Right Panel:} The relative deviation in the matter power spectra in compared to the non-interacting $\Lambda$CDM model has been shown with similar conclusions as observed from the left panel of this figure. From this plot we see that the curves presenting $\xi =-0.0001$ and $\Lambda$CDM cannot be distinguished from one another although a very minimal difference between them is present. }
\label{fig:Mpower-ivs}
\end{figure*}
\begin{figure*}
\includegraphics[width=0.35\textwidth]{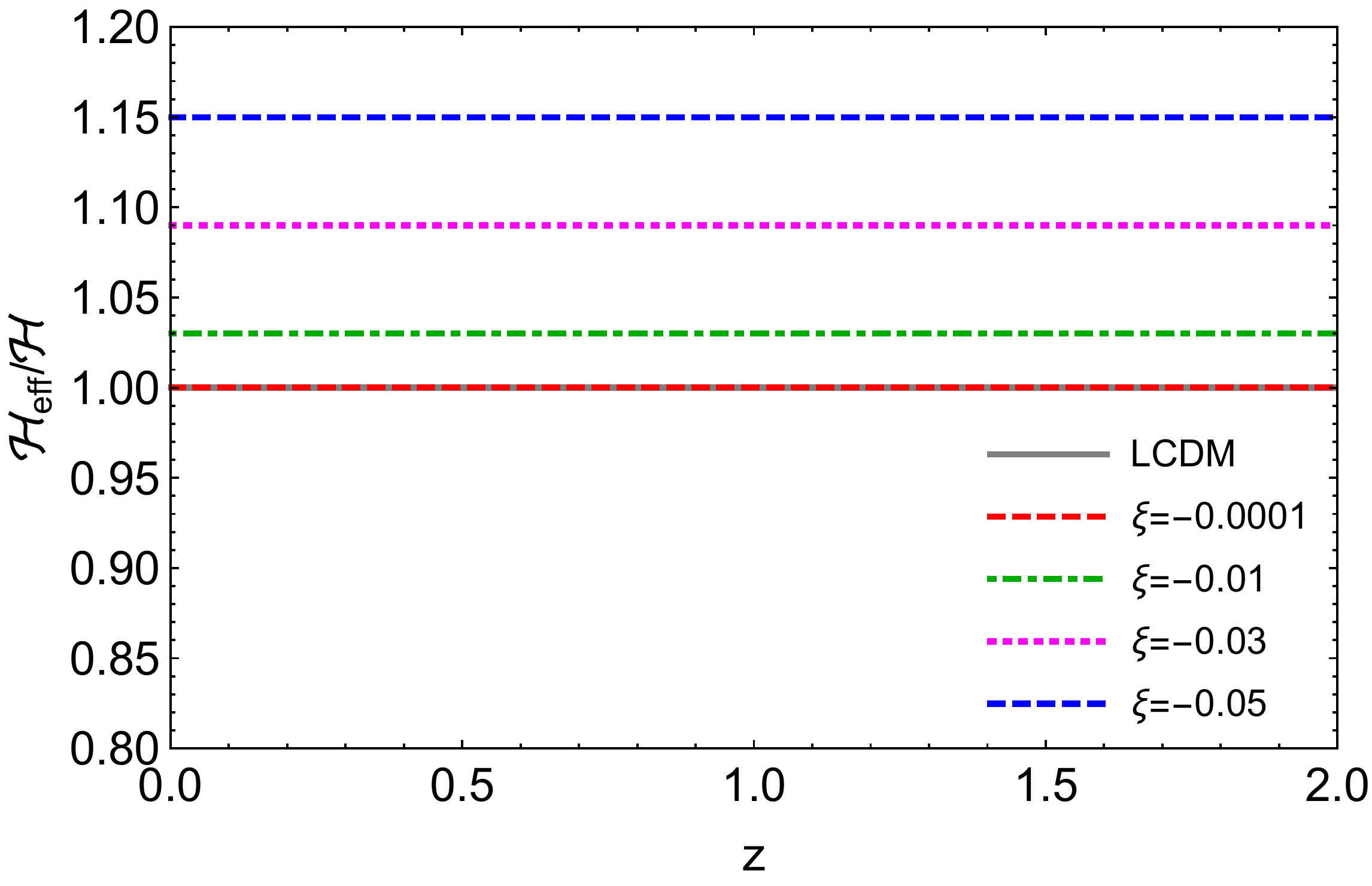}
\includegraphics[width=0.35\textwidth]{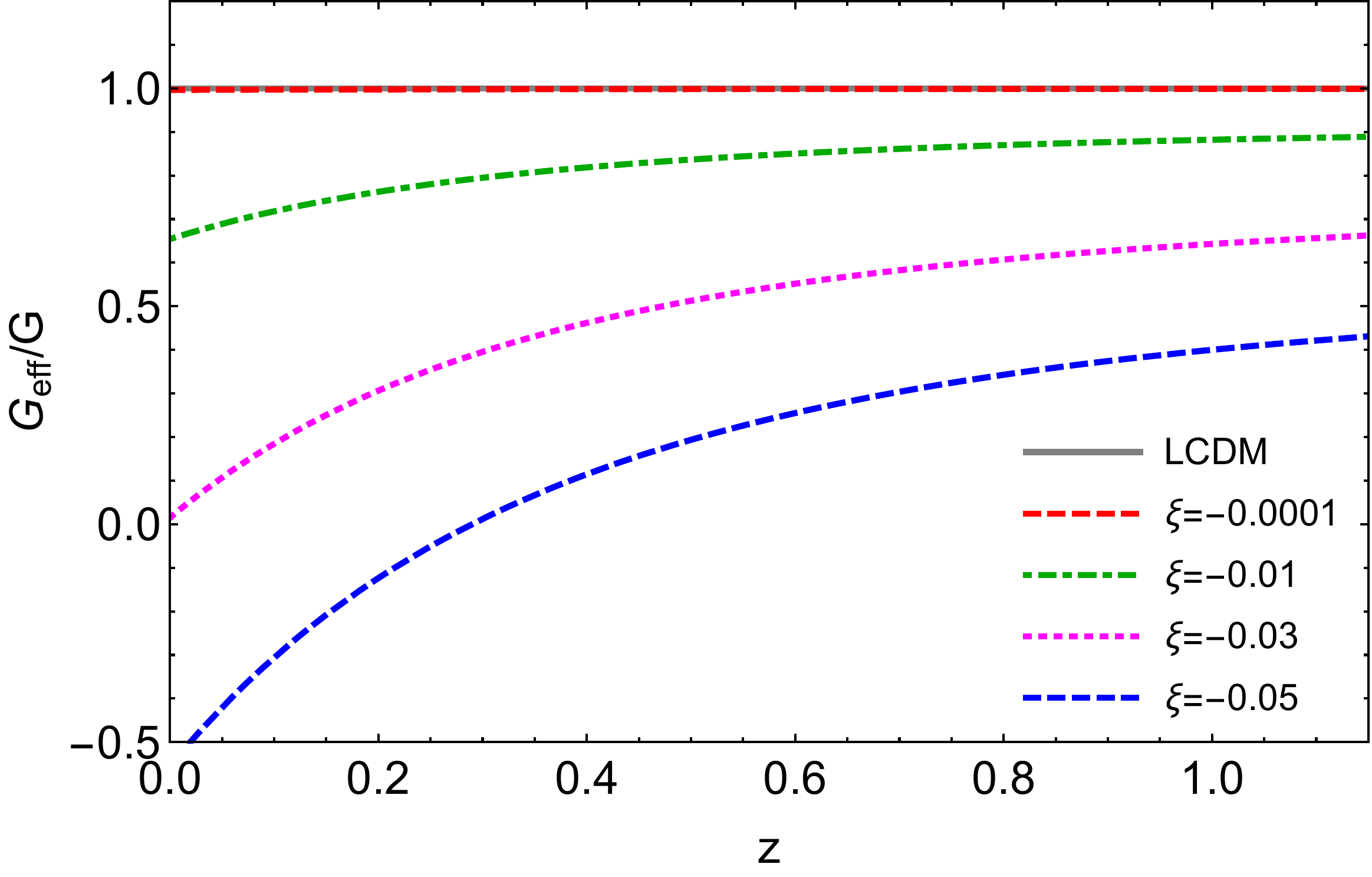}
\caption{Color Online $-$ \texttt{Left Panel:} The dynamical evolution of the modified expansion history $\mathcal{H}_{\text{eff}}$ has  been depicted in presence of different couplings of the interaction rate (\ref{interaction}) for the interacting vacuum scenario. The curves from upper to lower respectively stand for the non-interacting $\Lambda$CDM model ($\xi=0$) and for other coupling parameters, $\xi= -0.0001, -0.01, -0.03, -0.05$. \texttt{Right Panel:} The evolution of the quantity $G_{\text{eff}}/G$ has been shown for different coupling parameters for the interacting vacuum scenario. The curves from lower to upper levels respectively stand for the non-interacting $\Lambda$CDM model ($\xi=0$) and for other coupling parameters, $\xi= -0.0001, -0.01, -0.03, -0.05$. From both the panels, we arrive at a common conclusion which states that, as $\xi$ increases, the model starts deviating from the non-interacting $\Lambda$CDM cosmology. In both left and right panels, one can see that the curves presenting $\xi =-0.0001$ and $\Lambda$CDM cannot be differentiated from one another}
\label{fig-Heff-Geff-ivs}
\end{figure*}
\begin{figure*}
\includegraphics[width=0.35\textwidth]{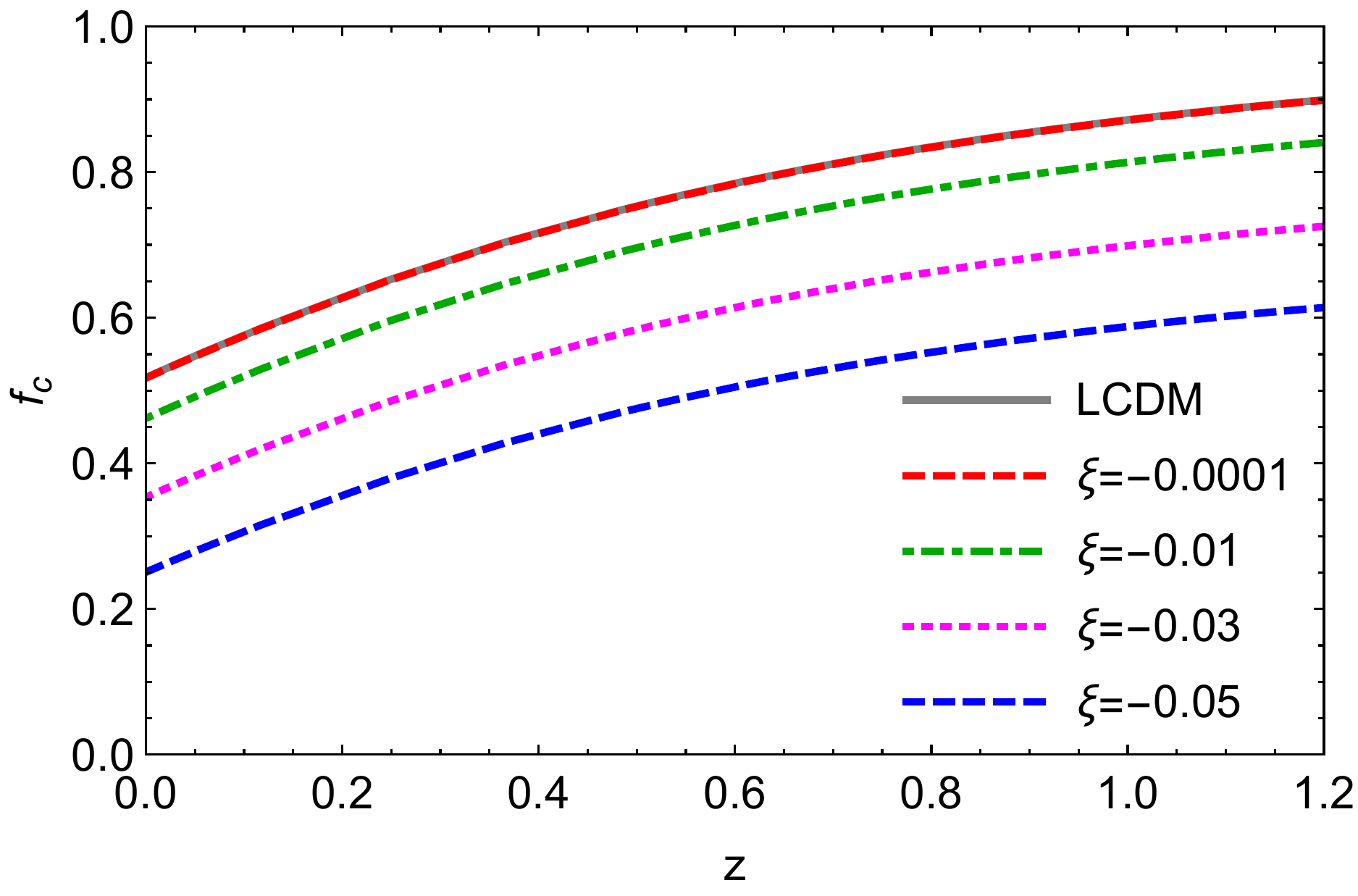}
\caption{Color Online $-$ For the interacting vacuum scenario we display the evolution of growth rate for the cold dark matter for different coupling strengths. The curves from upper to lower respectively stand for non-interacting $\Lambda$CDM model ($\xi=0$) and for other coupling parameters, $\xi= -0.0001, -0.01, -0.03, -0.05$. However, we observe that if the coupling strength increases, then the growth-rate for the cold dark matter decreases with the evolution of the universe. Similar to previous observations, here too we see that the curves presenting $\xi =-0.0001$ and $\Lambda$CDM cannot be distinguished from one another. }
\label{fig-fc-ivs}
\end{figure*}

One can see that $\xi =0$ in both (\ref{eq:Heff}) and (\ref{eq:Geff}) recovers the standard evolutions of the corresponding quantities where no interaction is present. Further, we consider the growth rate of cold  dark matter defined by $f_c \equiv \frac{d}{d\ln a}(\ln \delta_c) $. One may notice that the presence of interaction into the dark sector automatically modifies the Euler equation, that means, the dark matter may not follow the geodesics \cite{Koyama:2009gd}. Thus, in presence of interaction, the above quantities give a qualitative nature of the interaction rate and its behaviour in compared to the non-interacting cosmologies quantified by $\xi =0$. 

Let us first focus on the dynamics of the IDE  model in the large scale of the universe.  The behaviour of this model has been displayed through the evolution of the CMB TT spectra and the matter power spectra. In the left panel of Fig. \ref{fig:CMB-ide}, we show the behaviour of IDE model through the CMB TT spectra which shows that as long as the coupling strength of the interaction increases, the model is significantly deviates from the non-interacting $\Lambda$CDM model. The deviation is also clear  
from the relative deviation $\Delta C_{l}^{TT}/C_{l}^{TT}$ (here $\Delta C_{l}^{TT}$ measures the deviation of the model from $\Lambda$CDM), one can see that the nonzero deviation from $\Lambda$CDM is prominent for all the coupling parameters considered in the analysis which is evident in the low multipoles $l$. On the other hand, for large coupling strength, the model significantly deviates from $\Lambda$CDM which is clear from both the left and right panels of Fig. \ref{fig:CMB-ide}. Similarly, for different coupling strengths of the interaction rate,  we have shown the evolution of the matter power spectra in the left panel of Fig. \ref{fig:Mpower-ide}. Again we see that for a large coupling strength, the model significantly deviates from the $\Lambda$CDM cosmology and this deviation is prominent for large $k$,
while for very small coupling, the interaction model is very close to $\Lambda$CDM. However, the deviation from the $\Lambda$CDM, even for a very small but nonzero coupling strength, still exists which is clear from the relative deviation shown in the right panel of Fig. \ref{fig:Mpower-ide}. The analyses from both CMB and matter power spectra as well as from the corresponding relative deviations, one may argue that the coupling strength $\xi =-0.05$ is very high and can be excluded from the picture.

Furthermore, we depict the modified expansion history $\mathcal{H}_{\text{eff}}$ (eqn. (\ref{eq:Heff})) and the effective gravitational constant (eqn. (\ref{eq:Geff})) respectively in the left and right panels of Fig. \ref{fig-Heff-Geff}. Both the plots in Fig. \ref{fig-Heff-Geff}
show that indeed for large coupling strength, the modified expansion history and the effective gravitational constant offer significant changes from that of the non-interacting  $\Lambda$CDM cosmology. Finally, from the growth rate of cold dark matter, $f_c$, displayed in Fig. \ref{fig-fc} we observe similar trend, that means, here too, large coupling strength implies the deviation of the model from non-interacting $\Lambda$CDM cosmology. We conclude that for large coupling strength, the growth rate of cold dark matter significantly decreases. 

The dynamics of this interaction scenario in the large scale of the universe has also been investigated. In the left panel of Fig. \ref{fig:CMB-ivs} we have shown the variation in the CMB TT spectra for different strengths of the coupling parameter, $\xi$ and compared them with the non-interacting $\Lambda$CDM scenario. We see that as $\xi$ increases its strength, a significant changes in the CMB TT spectra is observed with respect to the non-interacting scenario while for lower coupling strengths, the deviation from the non-interacting $\Lambda$CDM model becomes low. However, since the observational data predict a very small coupling parameter allowing its zero value in the 68.3\% confidence level, thus, it is expected that a small deviation from the $\Lambda$CDM model should be present. In order to measure such small deviation, we measure the relative deviation of the interacting model with different coupling strengths with respect to the $\Lambda$CDM model, and this is shown in the right panel of Fig. \ref{fig:CMB-ivs}.  This plot (right panel of Fig. \ref{fig:CMB-ivs}) practically tells that $\xi \neq 0$, however small it is, the deviation from $\Lambda$CDM should exist, although it is also true that such deviation is very very small which is not much significant. A similar pattern is found when the large scale dynamics is described in terms of the matter power spectra shown in Fig. \ref{fig:Mpower-ivs}.  The left panel of the Fig. \ref{fig:Mpower-ivs}
shows a qualitative changes in the matter power spectra for different coupling strengths while the right panel tells us how much the model with different coupling strengths are far from $\Lambda$CDM. Overall, from the analyses for IVS one can realize that the coupling strength $\xi =-0.05$ presents a significant deviation from the $\Lambda$CDM cosmology and which according to the present observational data is not reliable, and hence this strong mangitude of the coupling parameter should be avoived. 

\begin{table}
\begin{center}
\begin{tabular}{c|c|c}
Parameter                    & Prior (IDE) & Prior (IVS) \\
\hline
$\Omega_{b} h^2$         & $[0.005,0.1]$ & $[0.005,0.1]$\\
$\tau$                       & $[0.01,0.8]$  & $[0.01,0.8]$\\
$n_s$                        & $[0.5, 1.5]$ & $[0.5, 1.5]$\\
$\log[10^{10}A_{s}]$         & $[2.4,4]$ & $[2.4,4]$\\
$100\theta_{MC}$             & $[0.5,10]$ & $[0.5,10]$ \\
$w_x$                        & $(-3, -1)$ & $-$ \\
$\xi$                        & $[-1, 0]$  & $[-1, 0]$

\end{tabular}
\end{center}
\caption{The table summarizes the flat priors on the cosmological parameters for the interacting scenario with $w_x < -1$ (IDE) and interacting vacuum scenario (IVS). }
\label{priors-I}
\end{table}

\section{Data and Results}
\label{sec-results}

In this section we first describe the
astronomical data with the statistical technique
to constrain the present
interacting scenarios and the results of the analyses.
We include the following sets of astronomical data.

\begin{itemize}

\item {\it Cosmic microwave background (CMB) observations:} We use CMB data from the Planck's 2015 observations \cite{Adam:2015rua, Aghanim:2015xee}. Precisely, we use the
likelihoods $C^{TT}_l$, $C^{EE}_l$, $C^{TE}_l$ in addition to low$-l$ polarization data (i.e. Planck TT, TE, EE+ low TEB).

\item {\it Baryon acoustic oscillations (BAO) data:} For BAO data, the estimated ratio $r_s/D_V$ as a `standard ruler' has been used in which $r_s$ is the co-moving sound horizon at the baryon drag epoch and $D_V$ is the effective distance given by $D_V(z)=\left[(1+z)^2D_A(a)^2\frac{z}{H(z)}\right]^{1/3}$. Here  $D_A$ is the angular diameter distance. In this analysis we use four data points from different astronomical surveys, namely, from the 6dF Galaxy Survey (6dFGS) redshift measurement at  $z_{\emph{\emph{eff}}}=0.106$
\cite{Beutler:2011hx}, the Main Galaxy Sample of Data Release 7 of Sloan
Digital Sky Survey (SDSS-MGS)measurement at $z_{\emph{\emph{eff}}}=0.15$
\cite{Ross:2014qpa}, and the CMASS and LOWZ samples from the latest Data
Release 12 (DR12) of the Baryon Oscillation Spectroscopic Survey (BOSS) measurements at
$z_{\mathrm{eff}}=0.57$ \cite{Gil-Marin:2015nqa} and $z_{\mathrm{eff}%
}=0.32$ \cite{Gil-Marin:2015nqa}.

\item {\it Redshift space distortion (RSD) data:} We use the RSD data from two observational surveys namely CMASS sample \cite{Gil-Marin:2016wya} and the LOWZ sample \cite{Gil-Marin:2016wya}. The effective redshifts for the CMASS and LOWZ samples are respectvely at  $z_{\mathrm{eff}}=0.57$ and
and $z_{\mathrm{eff}}=0.32$. We note that when these two RSD data points are 
considered in the analysis, then DR12 of BOSS from BAO will not be considered.  

\item {\it $H_0$ from Hubble Space Telescope (HST):} The local Hubble constant measured from the HST by Riess et al. \cite{Riess:2016jrr} that yields $H_0= 73.02 \pm 1.79$  km/s/Mpc with 2.4\% precision.

\item {\it Joint light curve analysis (JLA):} This is the Supernovae Type Ia sample that contains 740 data points spread in the redshift interval $z \in [0.01, 1.30]$ \cite{Betoule:2014frx}. This low redshifts sample is the first indication for an accelerating universe.

\item {\it Hubble parameter measurements from cosmic chronometers (CC):}
We choose the cosmic chronometers to measure the Hubble parameter values at different redshifts. The cosmic chronometers are basically the galaxies that are most old and huge massive and the technique that we apply to measure the Hubble parameter values, is the differential age evolution of the galaxies. For a detailed description we refer to \cite{Moresco:2016mzx} and the references cited therein for more information about their implementation.  In this work we consider thirty Hubble parameter values spread in $z \in (0, 2)$ and they are found in \cite{Moresco:2016mzx}.

\item {\it Weak lensing (WL):}  The weak gravitational lensing
data from the Canada$-$France$-$Hawaii Telescope Lensing Survey (CFHTLenS) \cite{Heymans:2013fya,Asgari:2016xuw}.

\end{itemize}

In order to extract the observational constraints of the interacting scenarios, we
use the publicly available Monte Carlo Markov Chain (MCMC) package \texttt{cosmomc} \cite{Lewis:2002ah, Lewis:1999bs} equipped with a convergence diagnostic followed by the Gelman and Rubin statistics \cite{Gelman-Rubin}. The parameters space for interacting dark energy and interacting vacuum scenarios respectively are

\begin{align}
\mathcal{P}_1 \equiv\Bigl\{\Omega_bh^2, \Omega_{c}h^2, 100 \theta_{MC}, \tau, w_x, \xi, n_s, log[10^{10}A_S]\Bigr\},
\label{eq:parameter_space1}
\end{align}
and
\begin{align}
\mathcal{P}_2 \equiv\Bigl\{\Omega_bh^2, \Omega_{c}h^2, 100 \theta_{MC}, \tau, \xi, n_s, log[10^{10}A_S]\Bigr\},
\label{eq:parameter_space2}
\end{align}
where in both (\ref{eq:parameter_space1}) and (\ref{eq:parameter_space2}), $\Omega_bh^2$, $\Omega_{c}h^2$, are respectively the
baryons density and the cold dark matter density;
$100 \theta_{MC}$, $\tau$, $n_s$, $A_S$, are respectively  the  ratio of sound horizon to the angular diameter distance, optical depth, scalar spectral index, and the amplitude of the initial power spectrum. The parameter $\xi$ is the coupling strength while $\mathcal{P}_1$ has one extra parameter $w_x$. Thus, we see that the interacting dark energy has eight free parameters and the interacting vacuum scenario has seven free parameters. During the MCMC analysis, we generally fix some priors on the model parameters.
In Table \ref{priors-I} we show the priors fixed on various cosmological parameters while constraining both the interacting models, namely, IDE and IVS. The priors mainly on $w_x$ and $\xi$ play an essential role in the analysis because the early time instabilities associated with the model, if any, significantly depend on the parameters space of $(w_x, \xi)$.  Now, if we closely look at the 
model (\ref{eq-int}), one can see that the interaction model (\ref{eq-int}) actually incorporates two separate interaction rates, namely, $Q \propto \rho_c$ and $Q \propto \rho_x$, hence, the stability of the entire model (\ref{eq-int}) depends on the region where both of them do not lead any early time instabilities. However, one can note that for some specific regions of the parameters space of $w_x$ and $\xi$, early time instabilities can be avoided 
\cite{Gavela:2009cy} while the entire region for $\xi$ allowing both positive and negative values may not be always suitable to avoid such instability. This actually depends on the interaction model.  
Thus, motivated by this fact, we divided several regions of the model parameters $w_x$ and $\xi$ to test the stability of the IDE scenario, for instance,
``$w_x$ free and $\xi$ free''; ``$w_x>-1$ and $\xi$ free''; ``$w_x>-1$ and $\xi \geq 0$''; ``$w_x>-1$, $\xi \leq 0$''; ``$w_x<-1$ and $\xi$ free''; finally with ``$w_x<-1$ and $\xi \leq 0$''.  We found that only for the region ``$w_x<-1$, $\xi \leq 0$'', the model does not lead to any early time instabilities while for the other regions the model meets with early time instabilities. Quite interestingly, this allowed region (i.e., $w_x < -1$) has an additional feature. It has been found that in presence of a non-gravitational interaction in the dark sectors, when the dark energy equation of state is allowed to cross the cosmological constant boundary, that means for $w_x< -1$, the tension on $H_0$ can be alleviated \cite{Kumar:2017dnp, DiValentino:2017iww}.  In this context we would like to add that some previous studies have found that for the non-interacting cosmologies with constant dark energy equation-of-state ($w_x$), the region $w_x> -1$ is also allowed and even preferred by some observational data \cite{Aubourg:2014yra,Hee:2016nho,Zhao:2017cud}. 
Now, following the similar trend, for the interacting vacuum scenario, we performed similar analyses with different priors on $\xi$, namely, $\xi \geq 0$, $\xi \leq 0$ and $\xi$ to be free. We found that for $\xi \leq 0$, early time instabilities do not appear. 

\begingroup
\squeezetable
\begin{center}
\begin{table*}
\begin{tabular}{cccccccccccc}
\hline\hline
Parameters & CMB & $\begin{array}[c]{c}
\text{CMB+BAO}\\+\text{HST} \end{array}$ & $\begin{array}[c]{c}
\text{CMB+BAO}\\+\text{RSD} \end{array}$ & $\begin{array}[c]{c}
\text{CMB+BAO}\\+\text{RSD+HST} \end{array}$ & $\begin{array}[c]{c}
\text{CMB+BAO}\\+ \mbox{RSD+HST}\\+\text{JLA+CC} \end{array}$ & $\begin{array}[c]{c}
\text{CMB+BAO}\\+ \mbox{RSD+HST}\\+\text{JLA+CC}\\+\text{WL} \end{array}$\\ \hline

$\Omega_c h^2$  & $    0.1260_{-    0.0059}^{+    0.0035}$ &$ 0.1204_{-    0.0015}^{+    0.0017}$ & $    0.1205_{-    0.0013}^{+    0.0014}$ & $    0.1201_{-    0.0014}^{+    0.0013}$ & $    0.1197_{-    0.0013}^{+    0.0012}$ & $    0.1191_{-    0.0011}^{+    0.0011}$\\

$\Omega_b h^2$ & $    0.0223_{-    0.0002}^{+    0.0002}$ &$ 0.02231_{-    0.0002}^{+    0.0002}$ & $    0.0223_{-    0.0002}^{+    0.0002}$ & $    0.0223_{-    0.0002}^{+    0.0002}$ & $    0.0223_{-    0.0002}^{+    0.0002}$ & $    0.0223_{-    0.0001}^{+    0.0001}$\\

$100\theta_{MC}$ & $ 1.0310_{-    0.0005}^{+    0.0007}$ &$    1.0405_{-    0.0005}^{+    0.0006}$ & $    1.0405_{-    0.0003}^{+    0.0003}$ & $ 1.0405_{-    0.0003}^{+    0.0003}$ & $    1.0406_{-    0.0004}^{+    0.0003}$ & $    1.0406_{-    0.0003}^{+    0.0003}$ \\

$\tau$ & $    0.0711_{-    0.0187}^{+    0.0184}$ & $ 0.0811_{-    0.0204}^{+    0.0214}$ & $    0.0687_{-    0.0163}^{+    0.0167}$ &  $    0.0621_{-    0.0160}^{+    0.0171}$ & $    0.0820_{-    0.0160}^{+    0.0164}$ & $    0.0636_{-    0.0159}^{+    0.0163}$\\

$n_s$ & $    0.9678_{-    0.0056}^{+    0.0057}$ & $0.9739_{-    0.0051}^{+    0.0051}$ & $    0.9728_{-    0.0038}^{+    0.0039}$ & $    0.9730_{-    0.0041}^{+    0.0041}$  & $    0.9746_{-    0.0035}^{+    0.0037}$ & $    0.9751_{-    0.0036}^{+    0.0037}$\\

${\rm{ln}}(10^{10} A_s)$ & $    3.0824_{-    0.0362}^{+    0.0356}$ & $3.1032_{-    0.0386}^{+    0.0418}$ & $    3.0770_{-    0.0316}^{+    0.0349}$ & $    3.0642_{-    0.0309}^{+    0.0338}$ & $    3.1043_{-    0.0314}^{+    0.0332}$ & $    3.0658_{-    0.0308}^{+    0.0318}$\\

$\Omega_{m0}$ & $    0.3523_{-    0.0693}^{+    0.0394}$ &$    0.2865_{-    0.0092}^{+    0.0092}$ & $    0.3105_{-    0.0098}^{+    0.0100}$ & $    0.2990_{-    0.0091}^{+    0.0083}$ & $    0.2942_{-    0.0074}^{+    0.0075}$ & $    0.2994_{-    0.0073}^{+    0.0073}$\\

$\sigma_8$ & $    0.8221_{-    0.0350}^{+    0.0392}$ &$    0.8635_{-    0.0192}^{+    0.0192}$ & $    0.8279_{-    0.0136}^{+    0.0137}$ & $    0.8311_{-    0.0143}^{+    0.0146}$ & $    0.8516_{-    0.0160}^{+    0.0162}$ &  $    0.8250_{-    0.0147}^{+    0.0132}$\\

$H_0$ &  $   65.5213_{-    3.9333}^{+    4.5145}$ & $70.7651_{-    1.1482}^{+    1.1132}$ & $   67.9685_{-    1.0243}^{+    0.8324}$ & $   69.1889_{-    0.8904}^{+    0.8698}$ & $   69.6402_{-    0.8523}^{+    0.8265}$ & $   68.8940_{-    0.8176}^{+    0.6849}$\\

$w_x$ & $   -1.1093_{-    0.0509}^{+    0.0828}$ &$   -1.1511_{-    0.0586}^{+    0.0529}$ &  $   -1.0603_{-    0.0201}^{+    0.0427}$ & $   -1.0940_{-    0.0394}^{+    0.0407}$ & $   -1.0960_{-    0.0365}^{+    0.0375}$ & $   -1.0608_{-    0.0238}^{+    0.0289}$\\

$\xi$ & $> -0.004884$ & $>-0.001285$ & $> -0.001384$ & $> -0.001278$ & $> -0.000959$ & $> -0.000935$\\

\hline\hline
\end{tabular}
\caption{The table summarizes the observational constraints on the cosmological parameters of IDE  at 68.3\% confidence-level for different combinations of the observational data. For the coupling parameter, we only report their values at 95.4\% lower confidence-level.}\label{tab:results-I}
\end{table*}
\end{center}
\endgroup
\begin{figure*}
\includegraphics[width=0.7\textwidth]{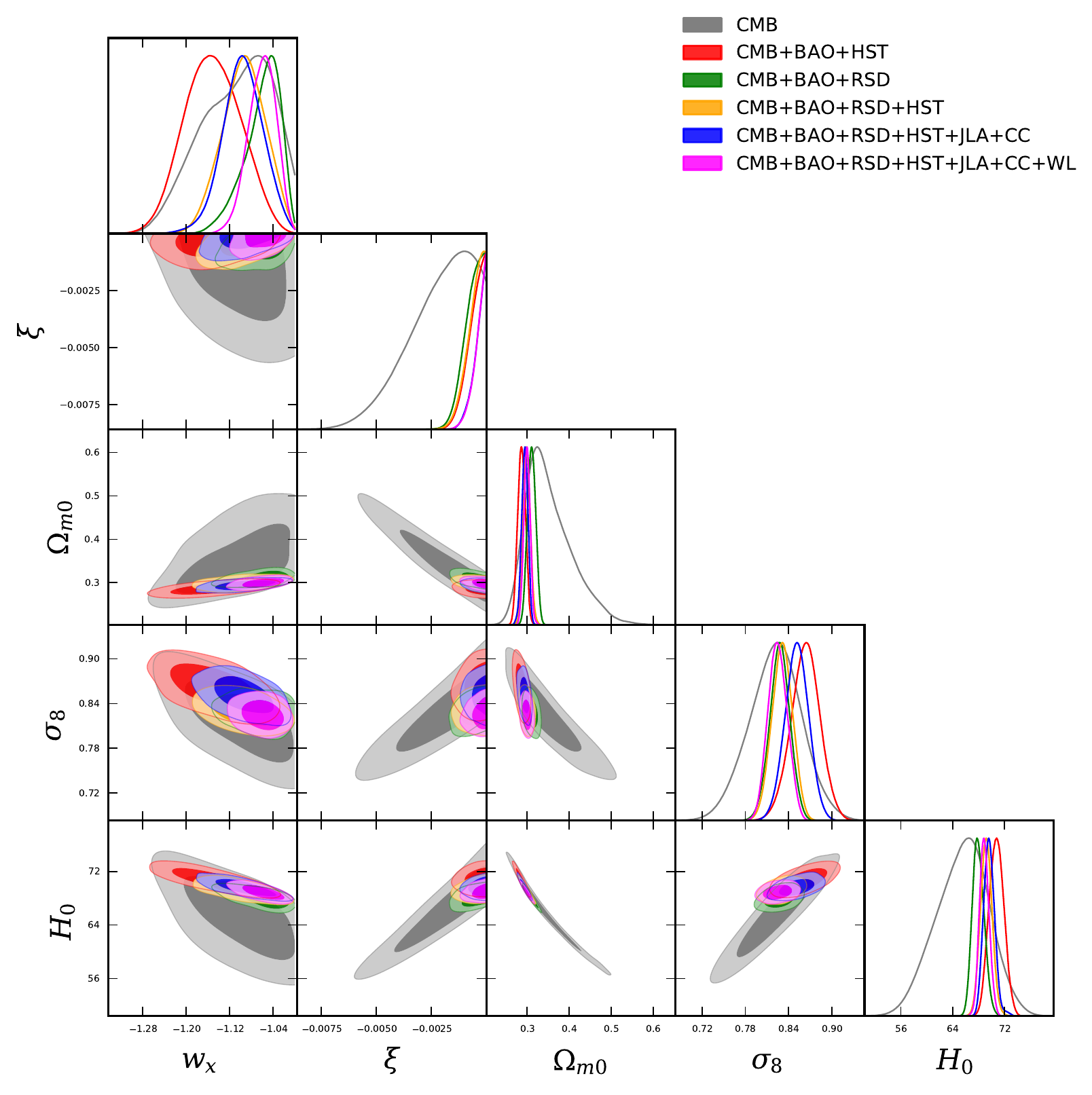}
\caption{Color Online $-$ Contour plots for different combinations of the cosmological parameters in the 68.3\% and 95.4\% confidence levels have been displayed for distinct observational combinations. Additionally, we also show the one-dimensional posterior distributions for those parameters at the extreme right corners of each row.  From the two-dimensional contour plots one can notice that the addition of any external data to CMB decreases the error bars of the cosmological parameters in a significant way. }
\label{fig-contour1}
\end{figure*}
\begin{figure*}
\includegraphics[width=0.35\textwidth]{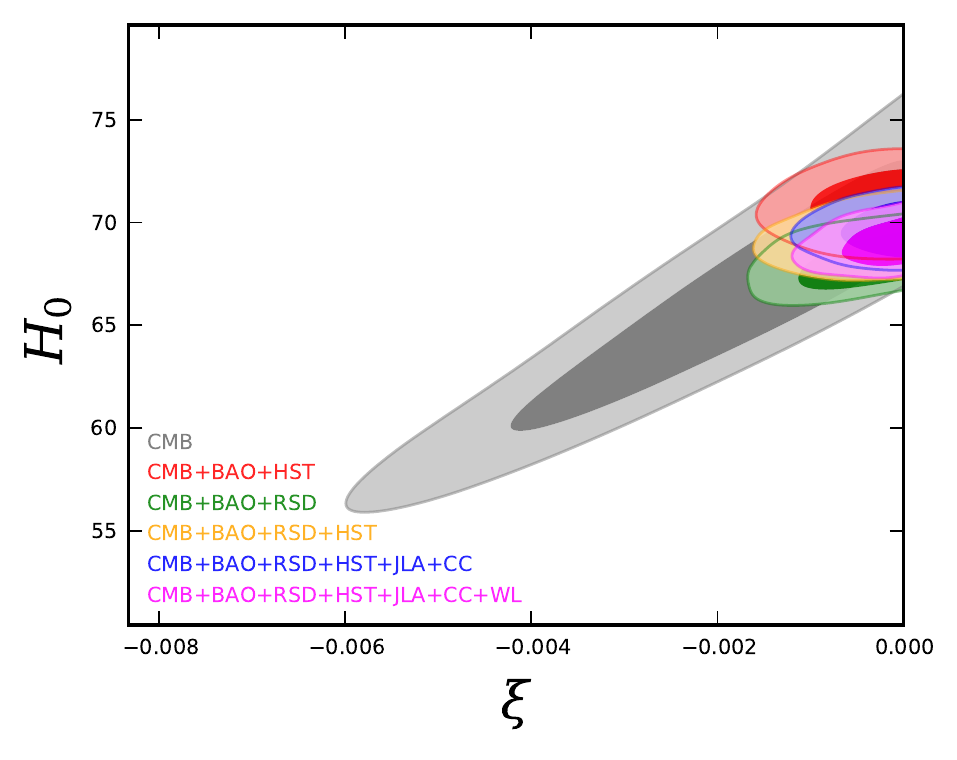}
\includegraphics[width=0.35\textwidth]{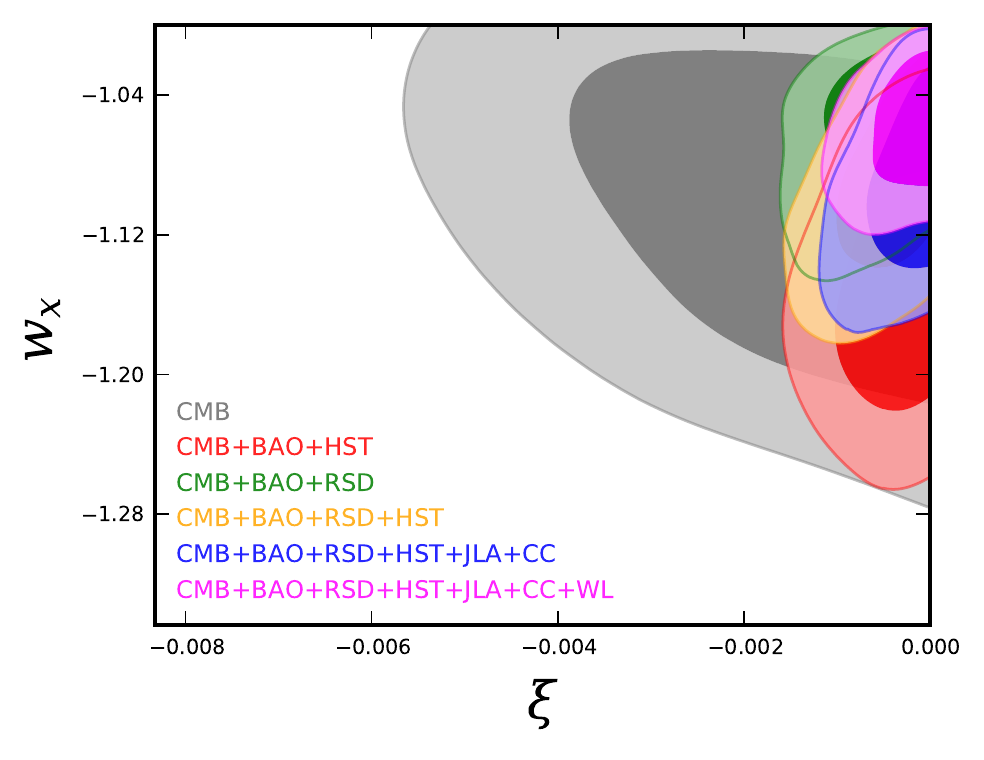}
\includegraphics[width=0.35\textwidth]{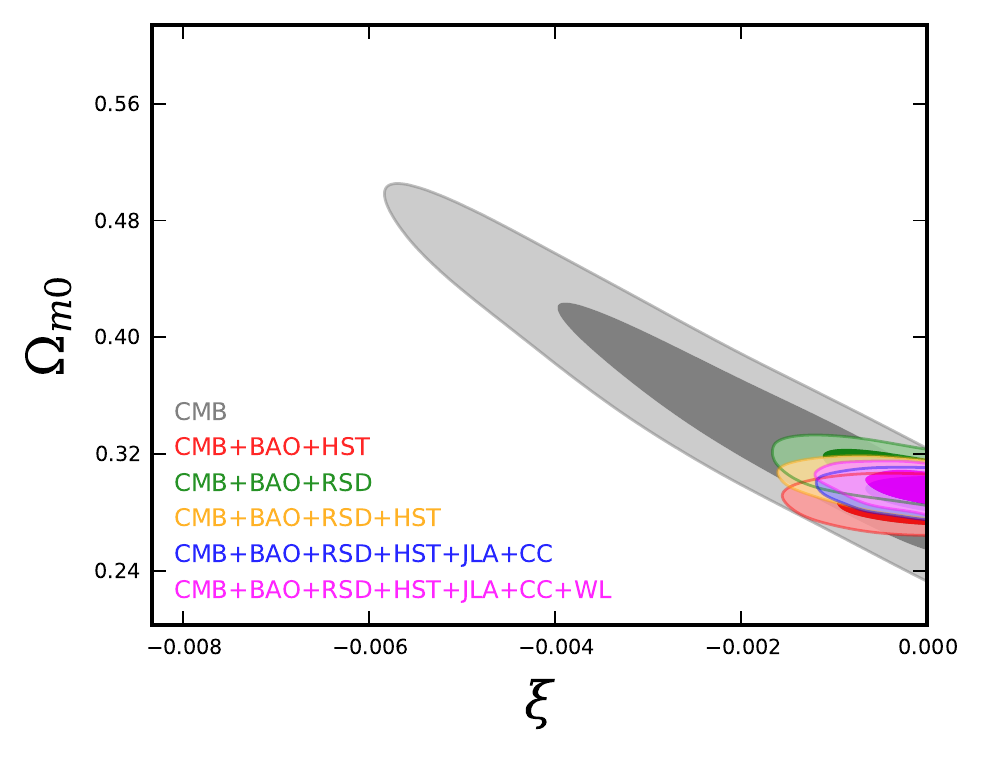}
\caption{Color Online $-$ The dependence of the coupling strength on some important cosmological parameters has been shown in the ($\xi, H_0$), $(\xi, w_x)$ and ($\xi, \Omega_{m0}$) planes at 68.3\% and 95.4\% confidence levels using different combinations of the observational data displayed above. We observe that the correlations between the parameters shown in the plots exist. \textit{Upper left panel:} We see that the CMB data allow a nonzero interaction in the dark sector for lower values of the Hubble parameter, however, from the combined analysis no conclusive statement can be made on the dependence of $H_0$ and the coupling strength $\xi$. \textit{Upper right panel:} The plot shows that the allowance of $w_x < -1$ is an indication of  an interaction in the dark sector. \textit{Lower panel:} One can notice that only CMB data indicate that coupling strength has a direct dependence on the density parameter $\Omega_{m0}$ while the combined analysis cannot make any deciding relation between the parameters involved. Thus, in order to clarify such issues we have shown three-dimensional scattered plots in Fig. \ref{fig-scattered-ide} with  detailed discussions. }
\label{fig-xi-H0-Om}
\end{figure*}
\begin{figure*}
\includegraphics[width=0.35\textwidth]{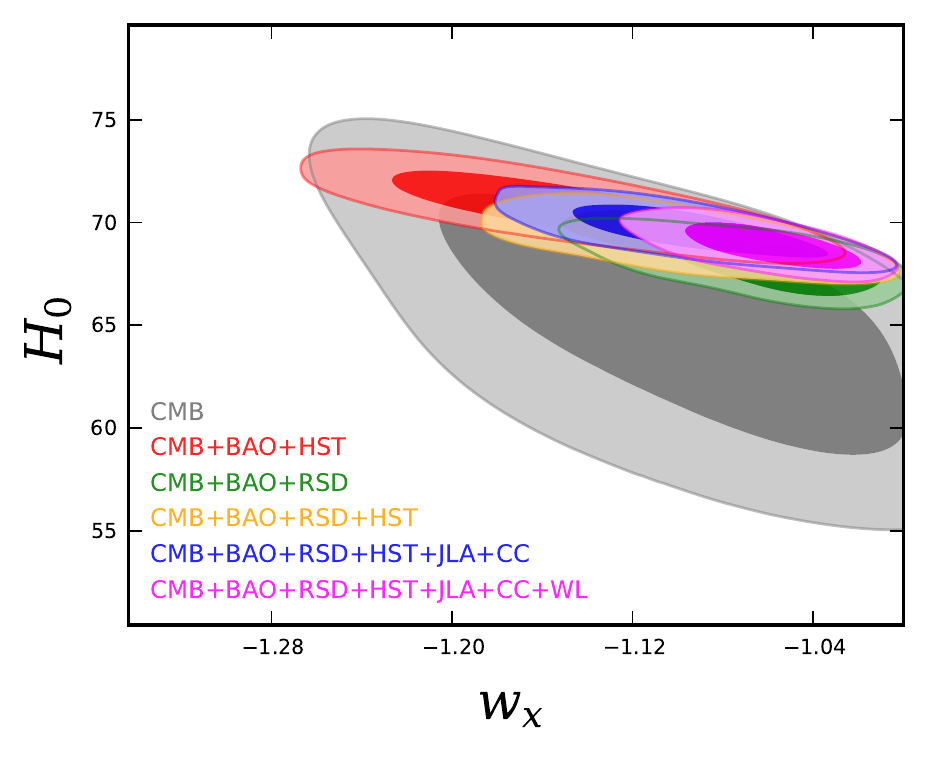}
\includegraphics[width=0.354\textwidth]{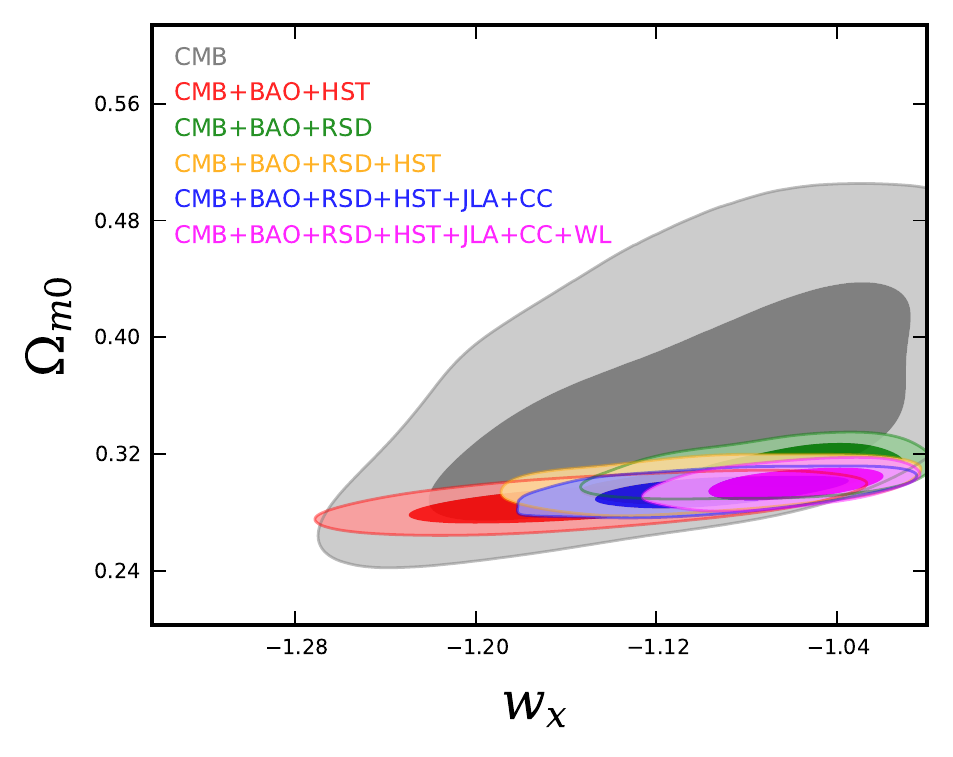}
\caption{Color Online $-$ 68.3\% and 95.4\% confidence-level contour plots in the two dimensional planes ($w_x, H_0$), ($\Omega_{m0}, w_x$) have been displayed for different
combined analyses. \textit{Left panel:} One may notice that for lower values of the Hubble parameter, the dark energy equation of state increases, that means $|w_x|$ decreases. \textit{Right panel:} Here we notice that as $\Omega_{m0}$ decreases, the dark energy equation of state moves toward more phantom region. }
\label{fig-wx-H0-Om}
\end{figure*}
\begin{figure*}
\includegraphics[width=0.35\textwidth]{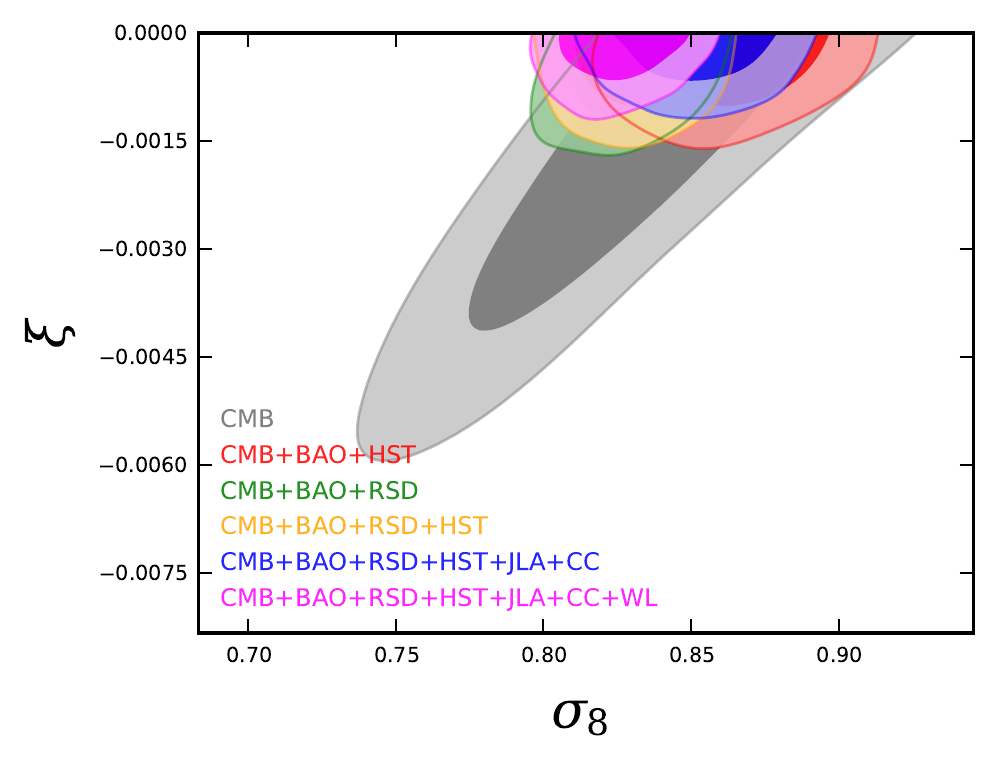}
\includegraphics[width=0.33\textwidth]{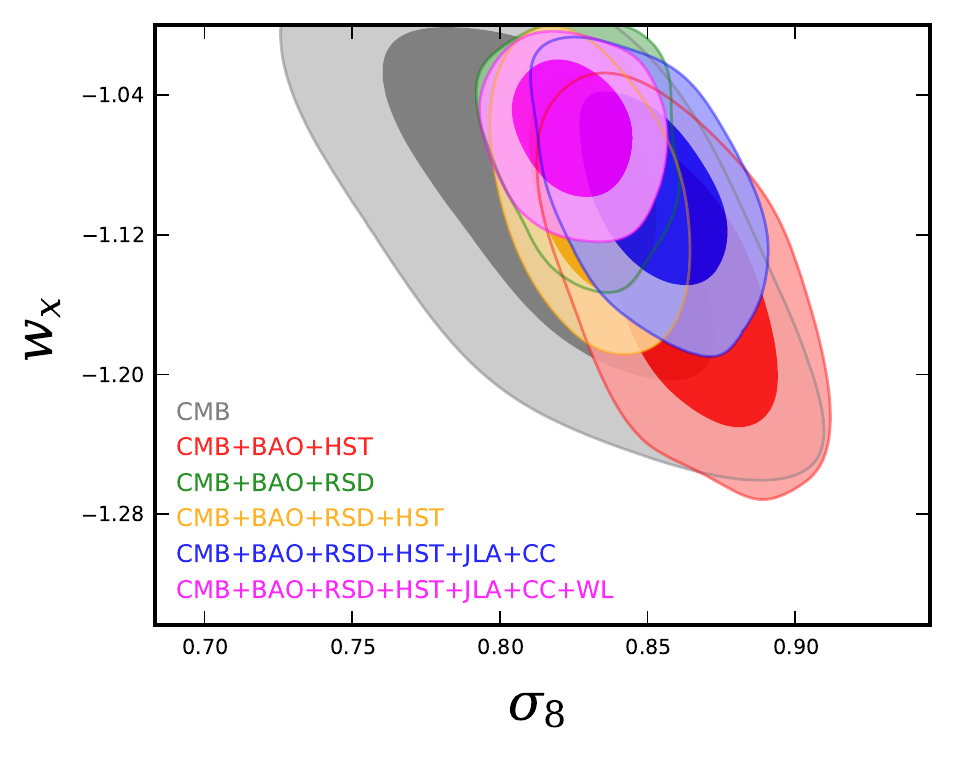}
\includegraphics[width=0.34\textwidth]{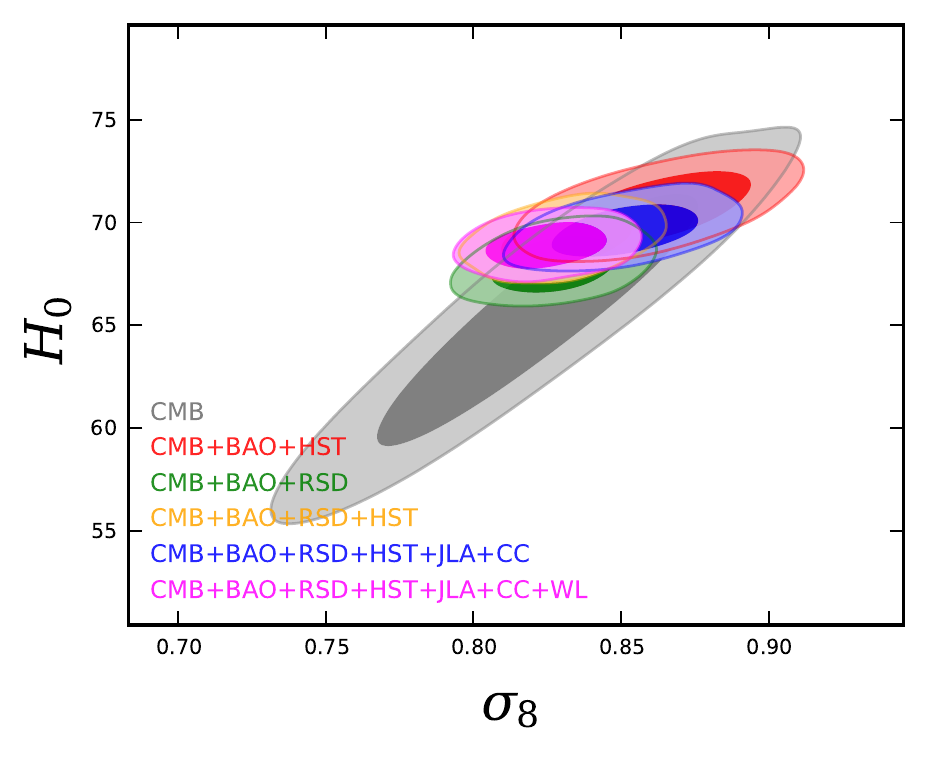}
\caption{Color Online $-$ 68.3\% and 95.4\% confidence-level contour plots in the two dimensional ($\sigma_8, w_x$), $(\sigma_8, H_0)$ and $(\sigma_8, \xi)$ planes have been shown.
\textit{Upper Left Panel:} One may notice that if the strength of the interaction increases, then $\sigma_8$ takes lower values. \textit{Upper Right Panel:} It is clearly seen that as long as the dark energy equation of state moves toward a more phantom region, the parameter $\sigma_8$ takes bigger values. \textit{Lower Panel:} The larger values of the Hubble parameter allows larger values of $\sigma_8$. }
\label{fig-sigma8}
\end{figure*}
\begin{figure*}
\includegraphics[width=0.325\textwidth]{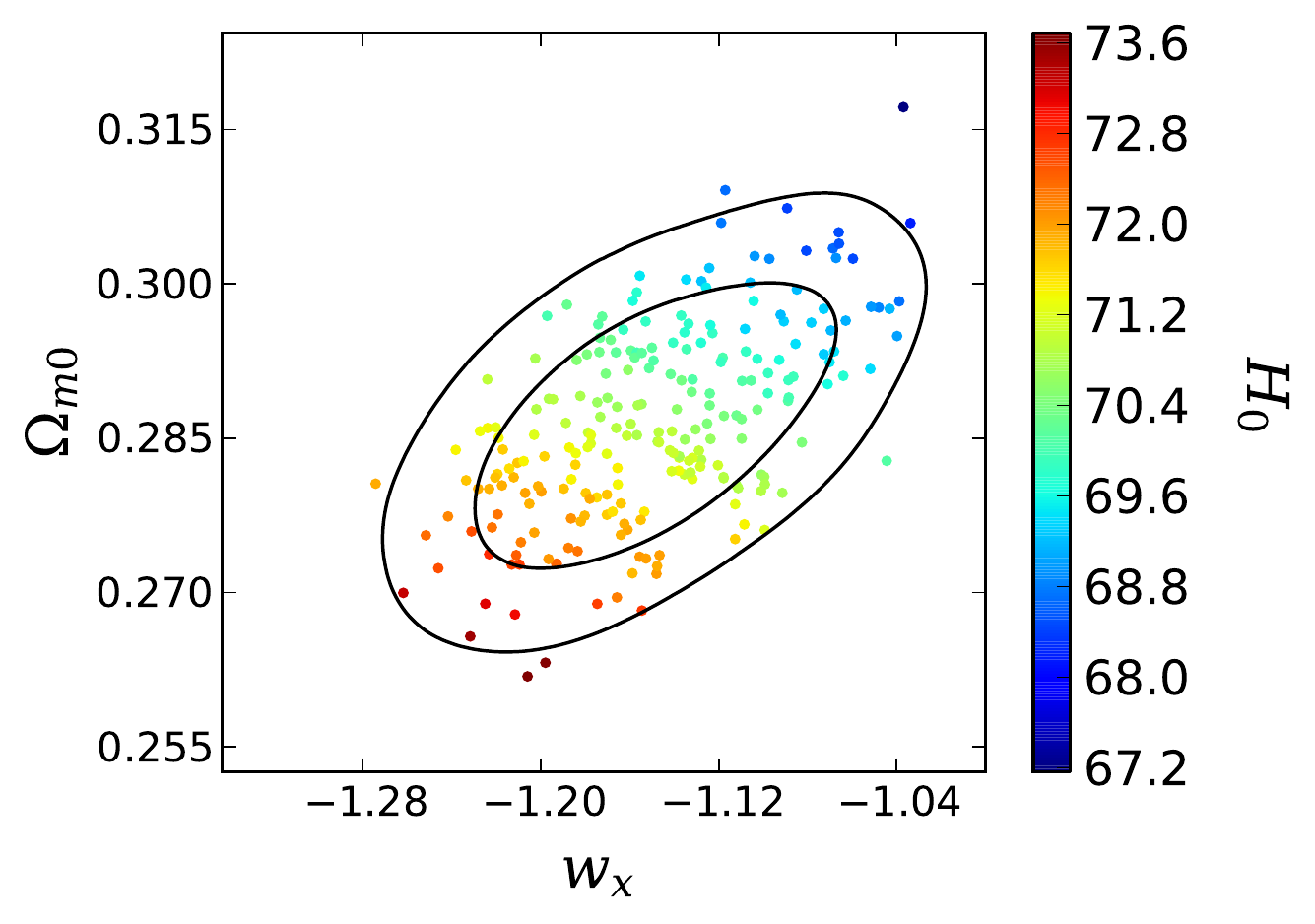}
\includegraphics[width=0.325\textwidth]{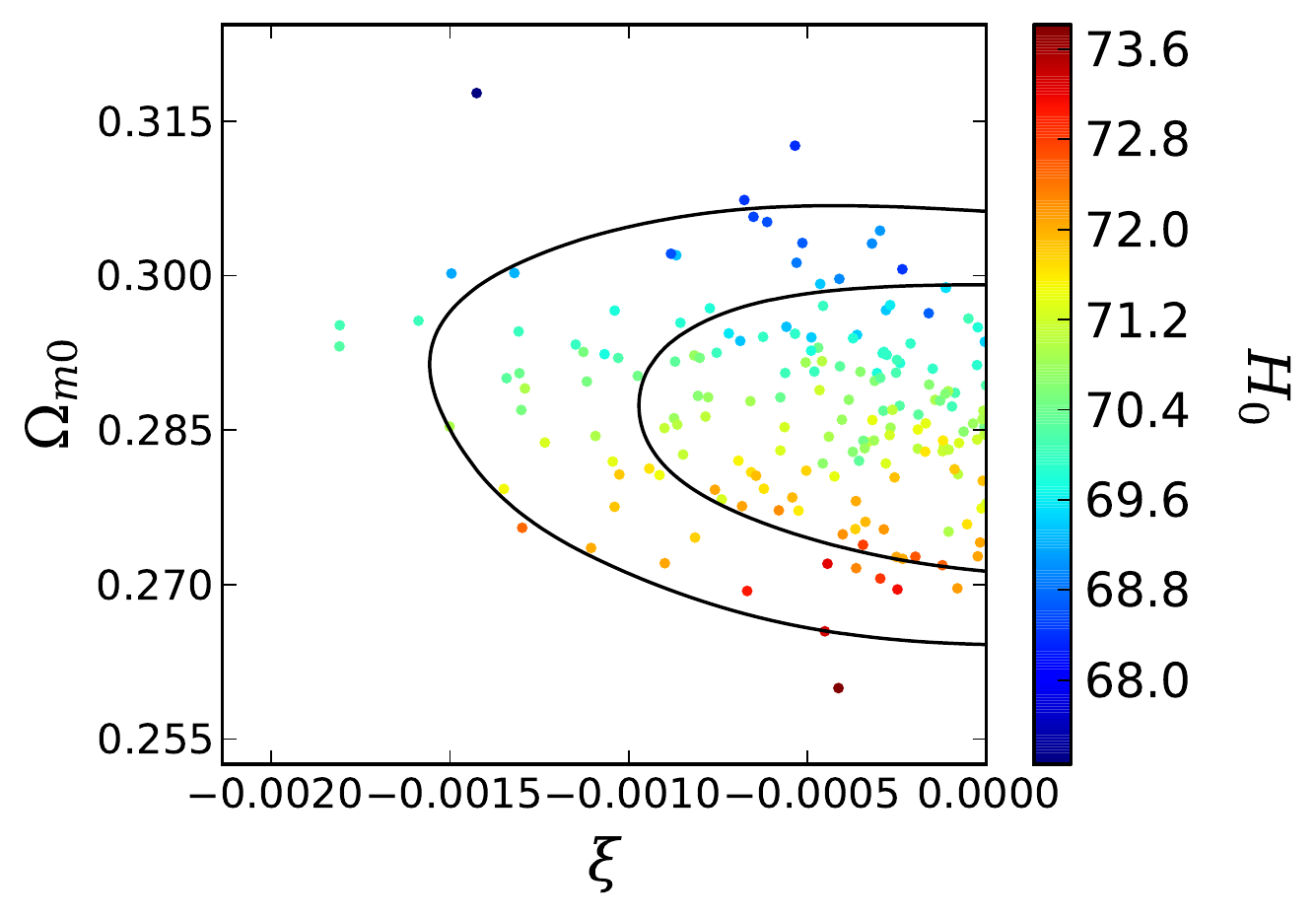}
\includegraphics[width=0.325\textwidth]{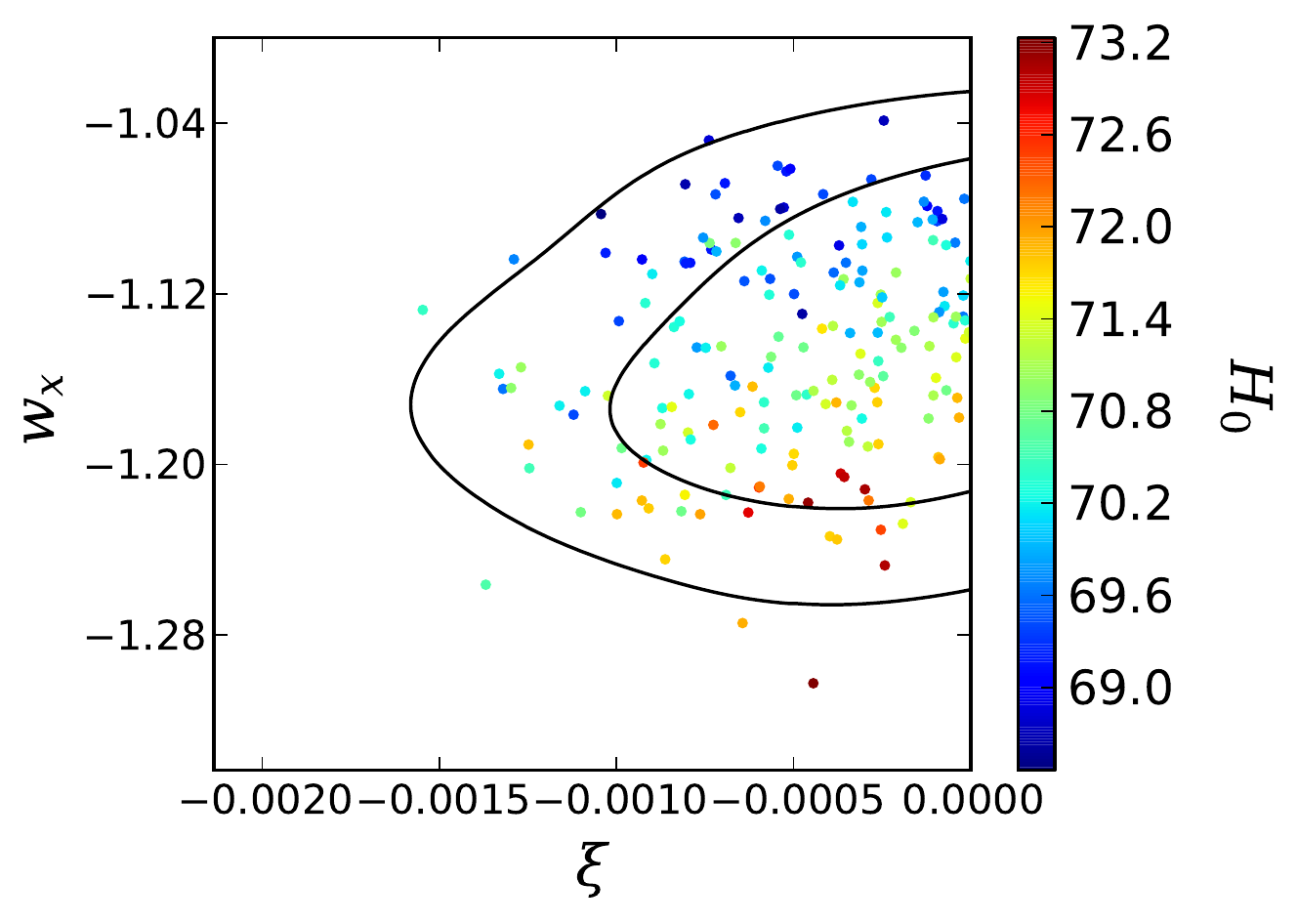}
\includegraphics[width=0.325\textwidth]{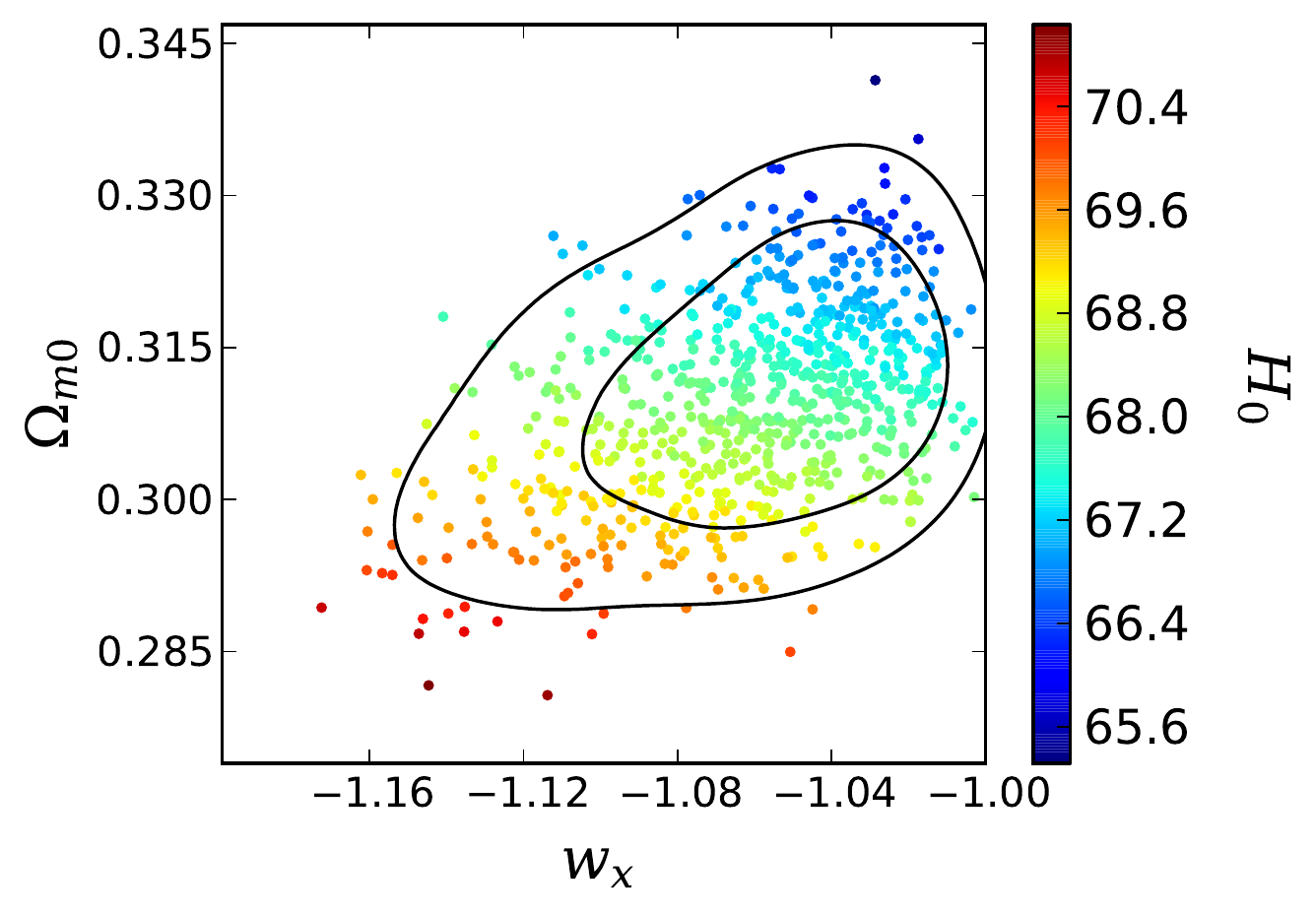}
\includegraphics[width=0.325\textwidth]{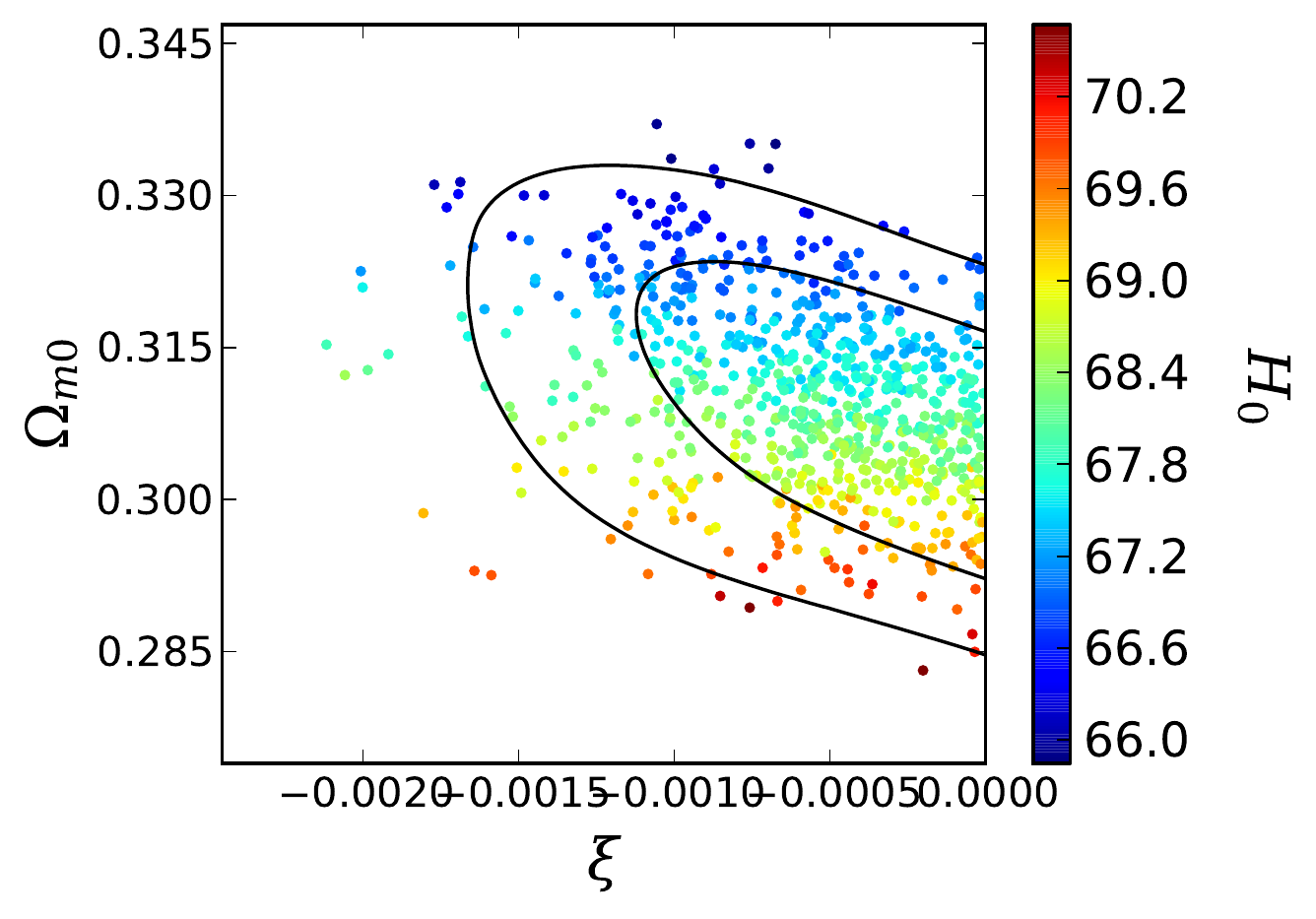}
\includegraphics[width=0.325\textwidth]{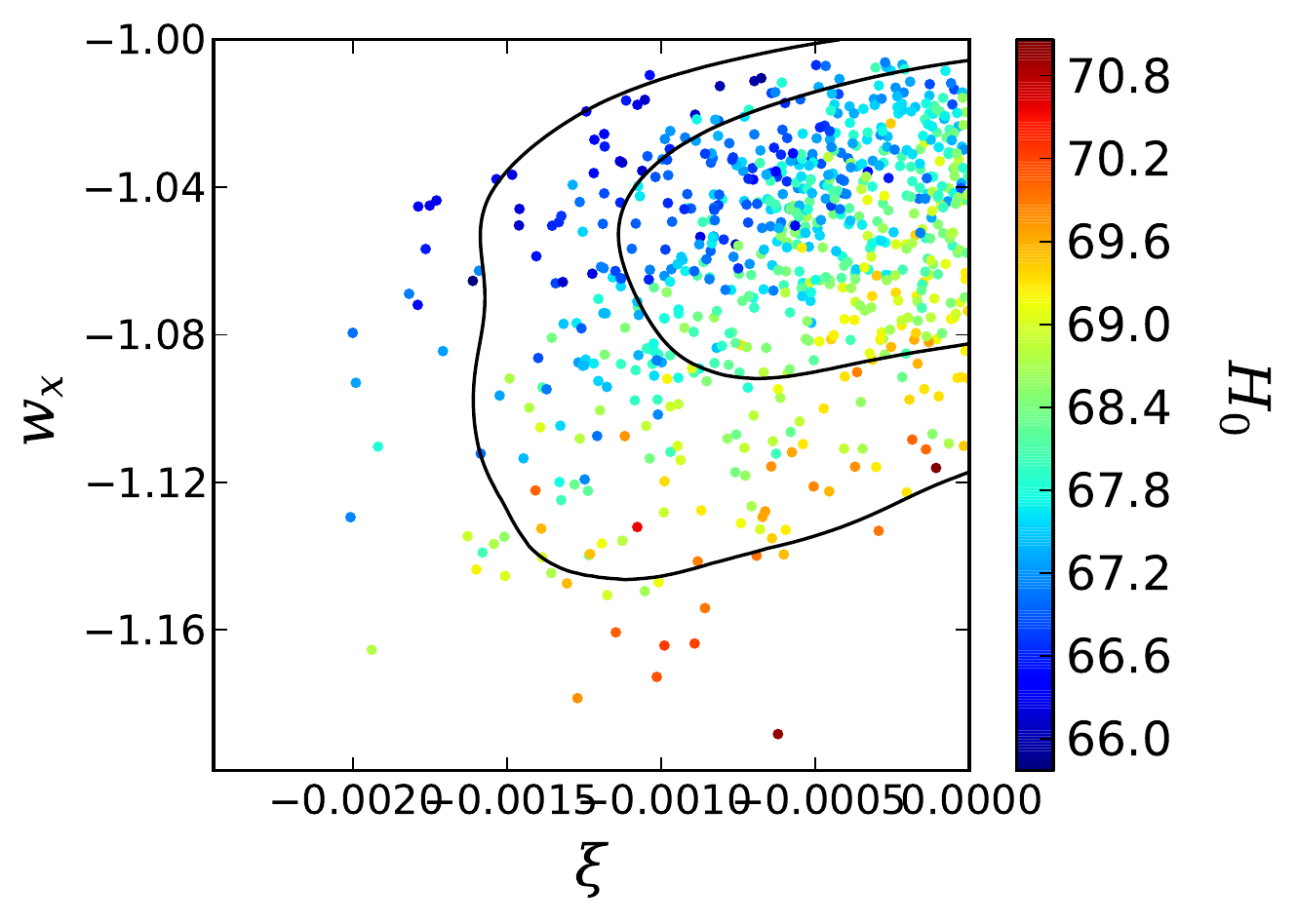}
\includegraphics[width=0.325\textwidth]{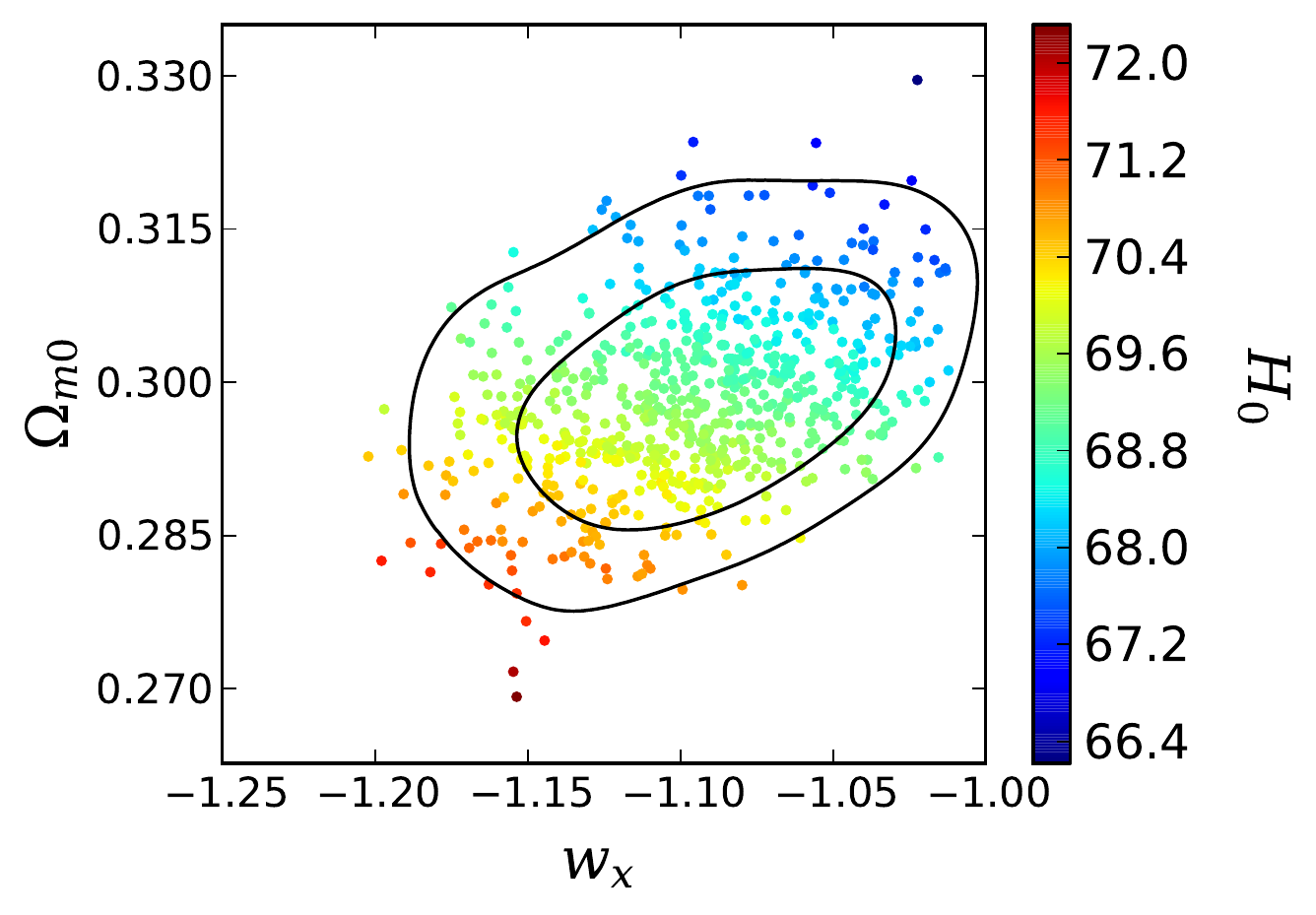}
\includegraphics[width=0.325\textwidth]{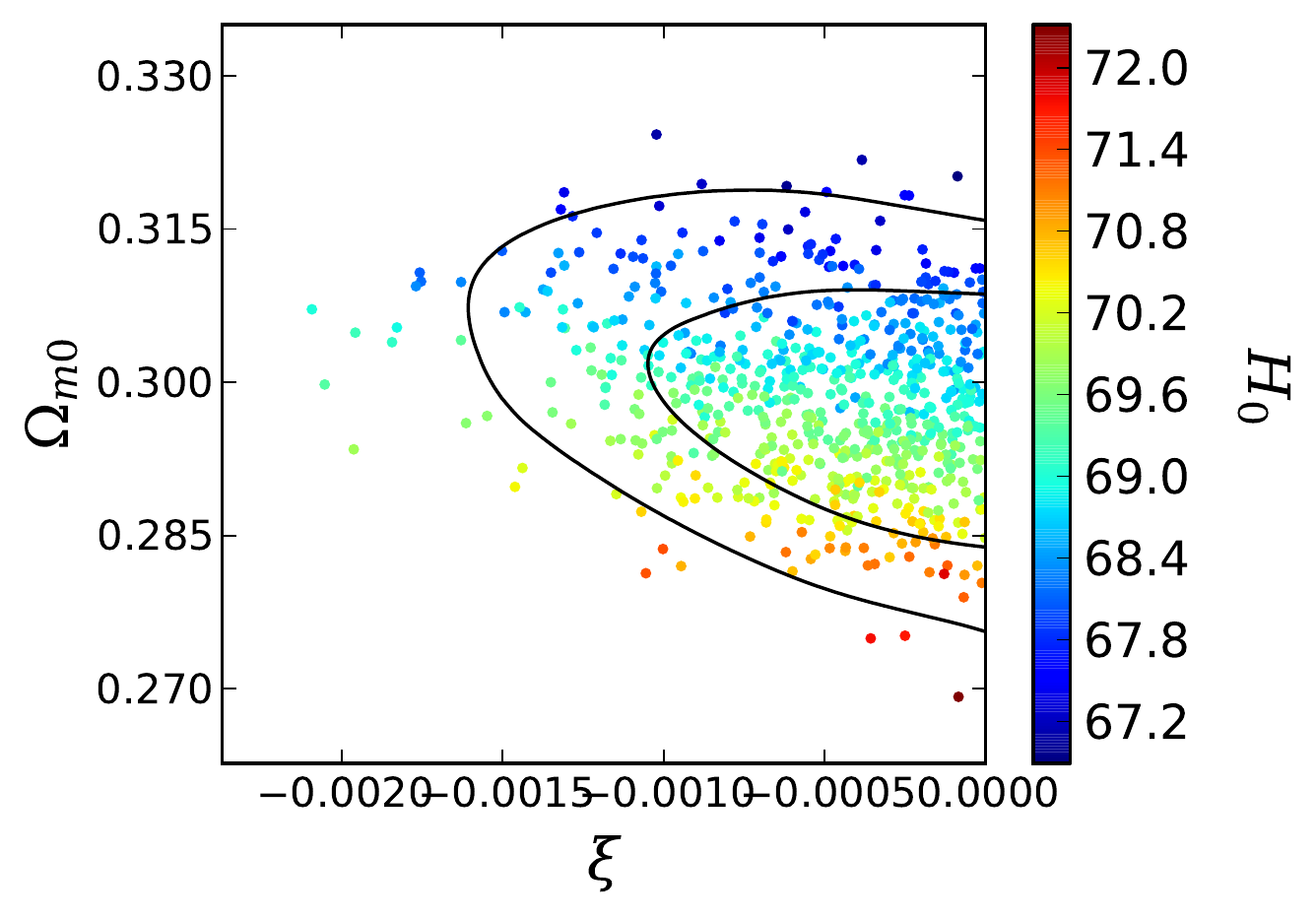}
\includegraphics[width=0.325\textwidth]{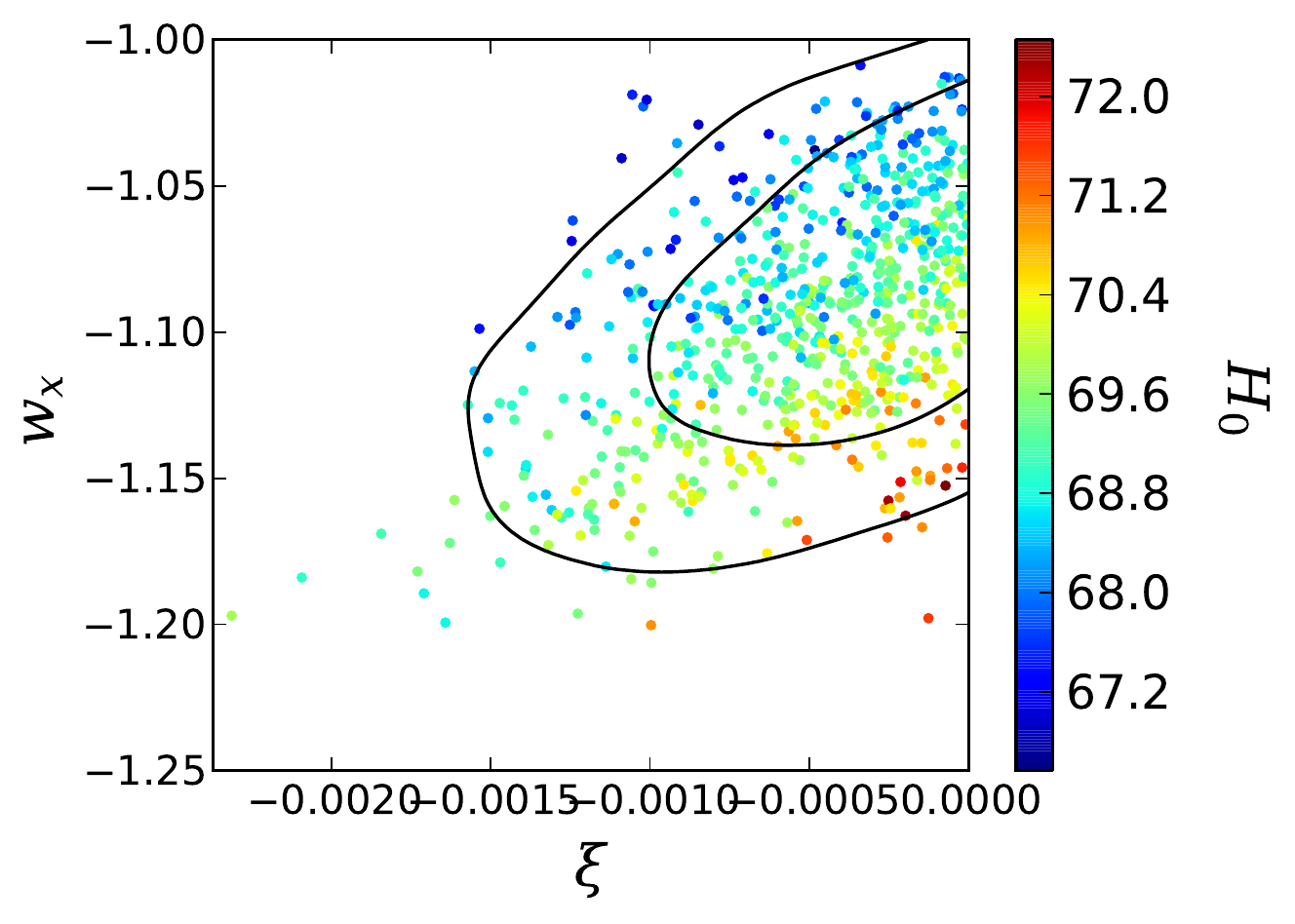}
\includegraphics[width=0.325\textwidth]{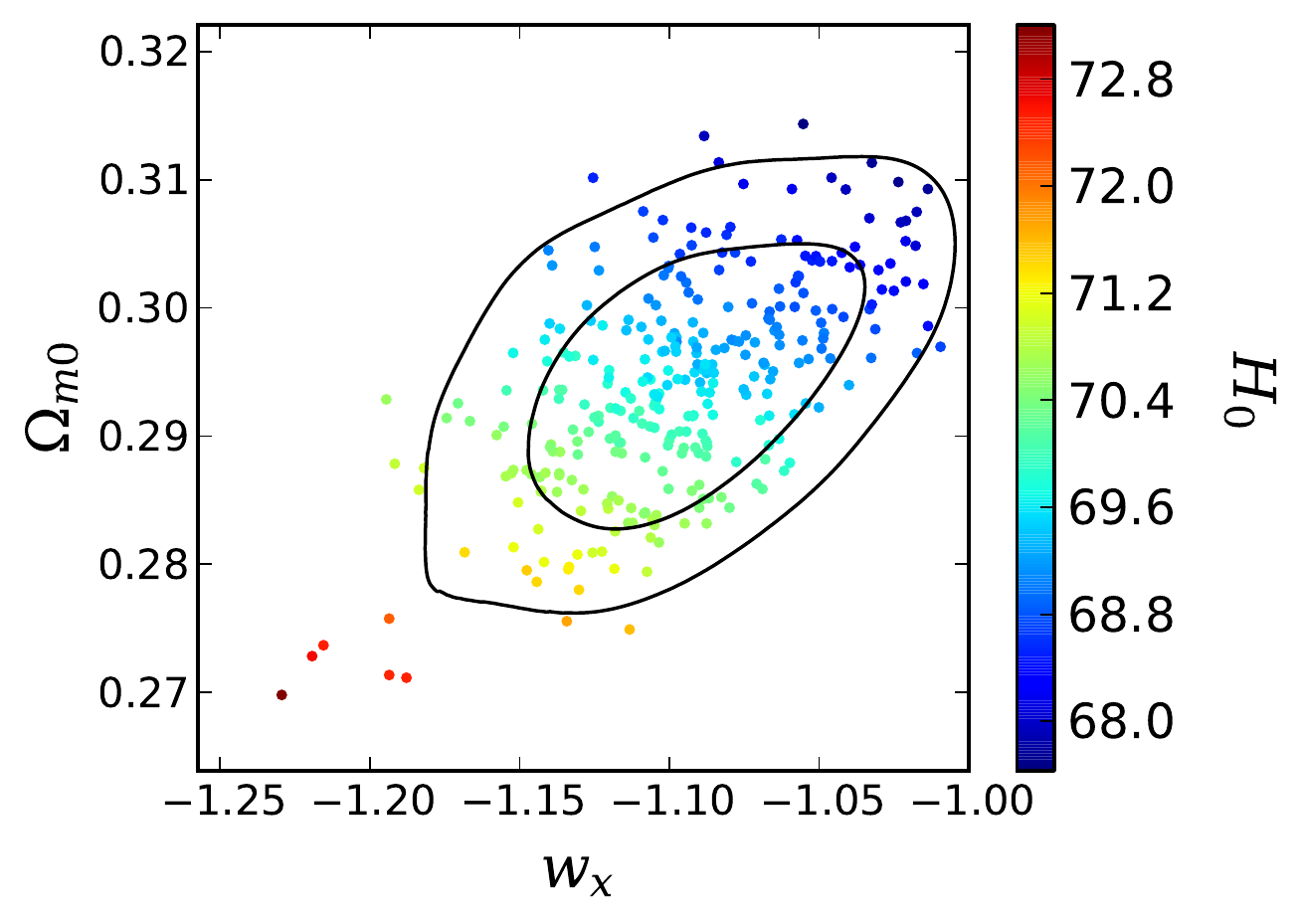}
\includegraphics[width=0.325\textwidth]{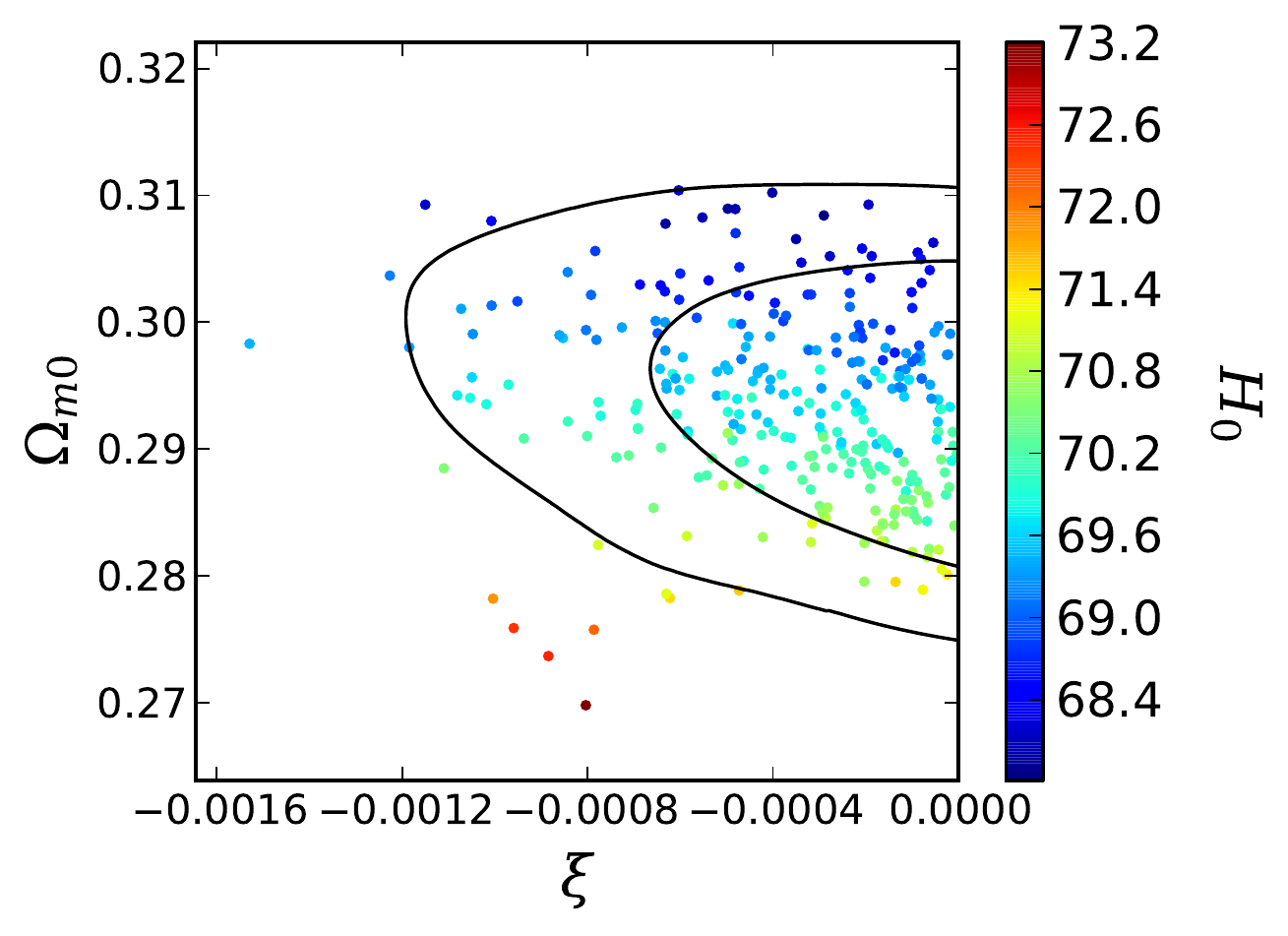}
\includegraphics[width=0.325\textwidth]{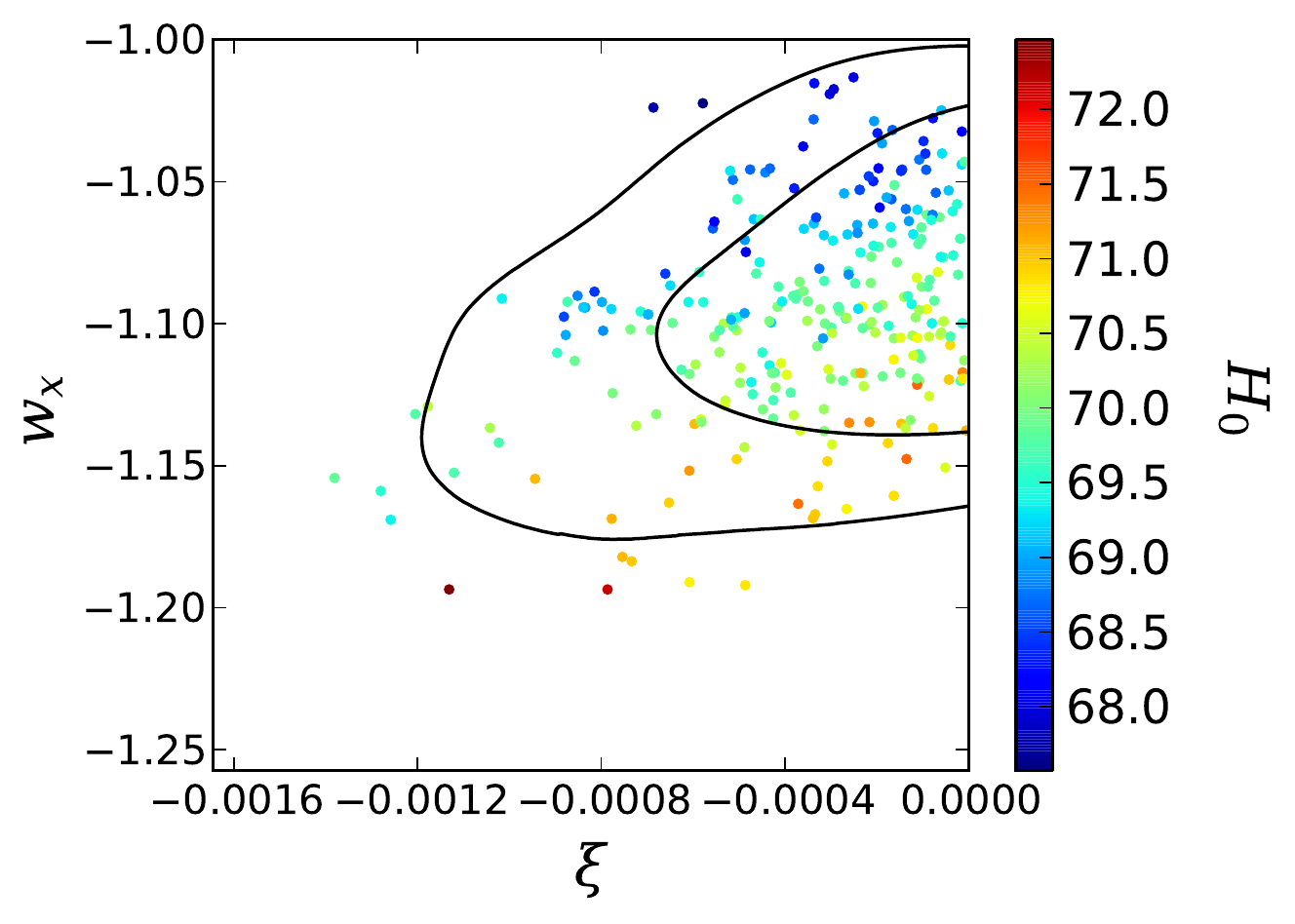}
\includegraphics[width=0.325\textwidth]{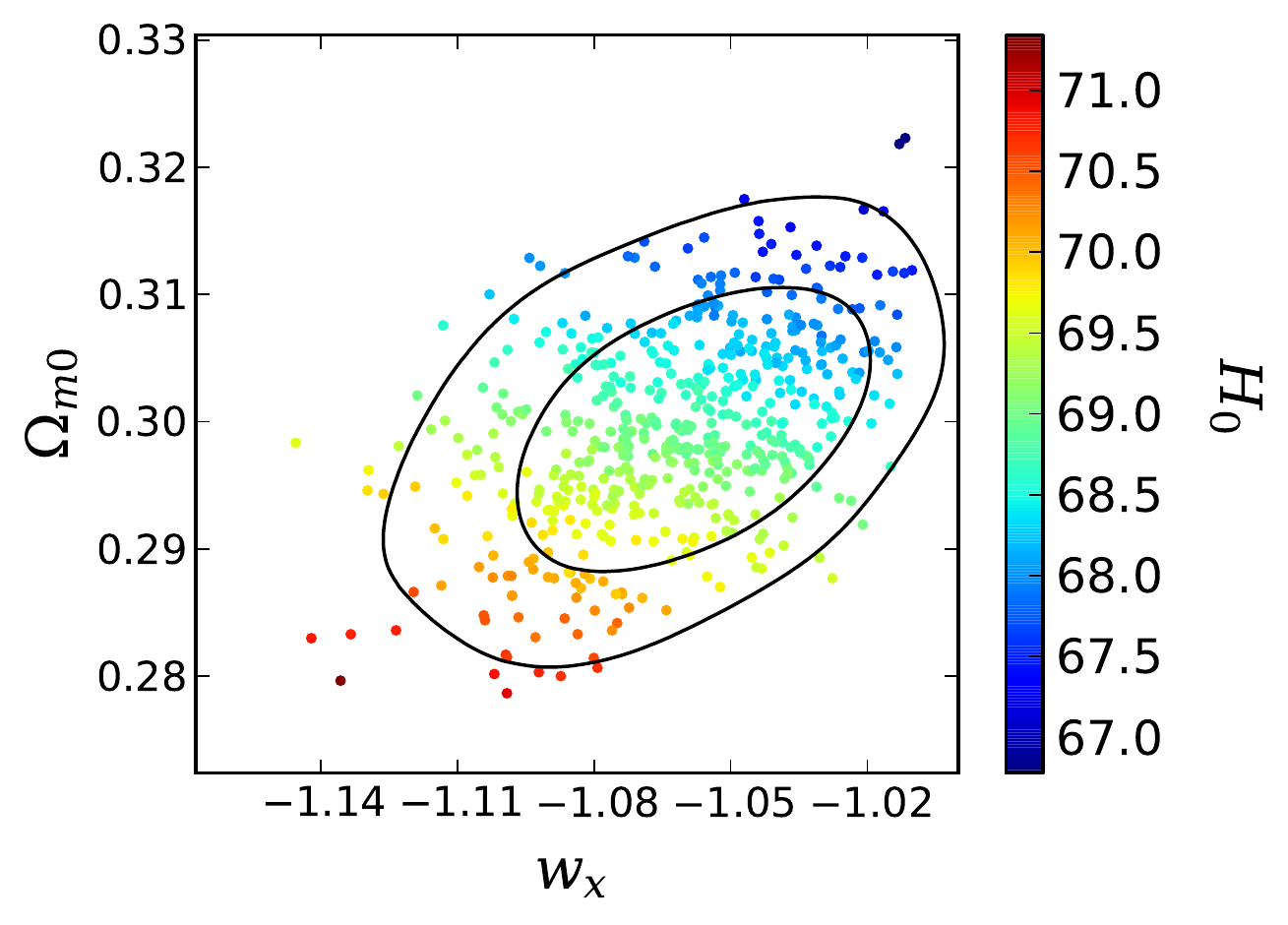}
\includegraphics[width=0.325\textwidth]{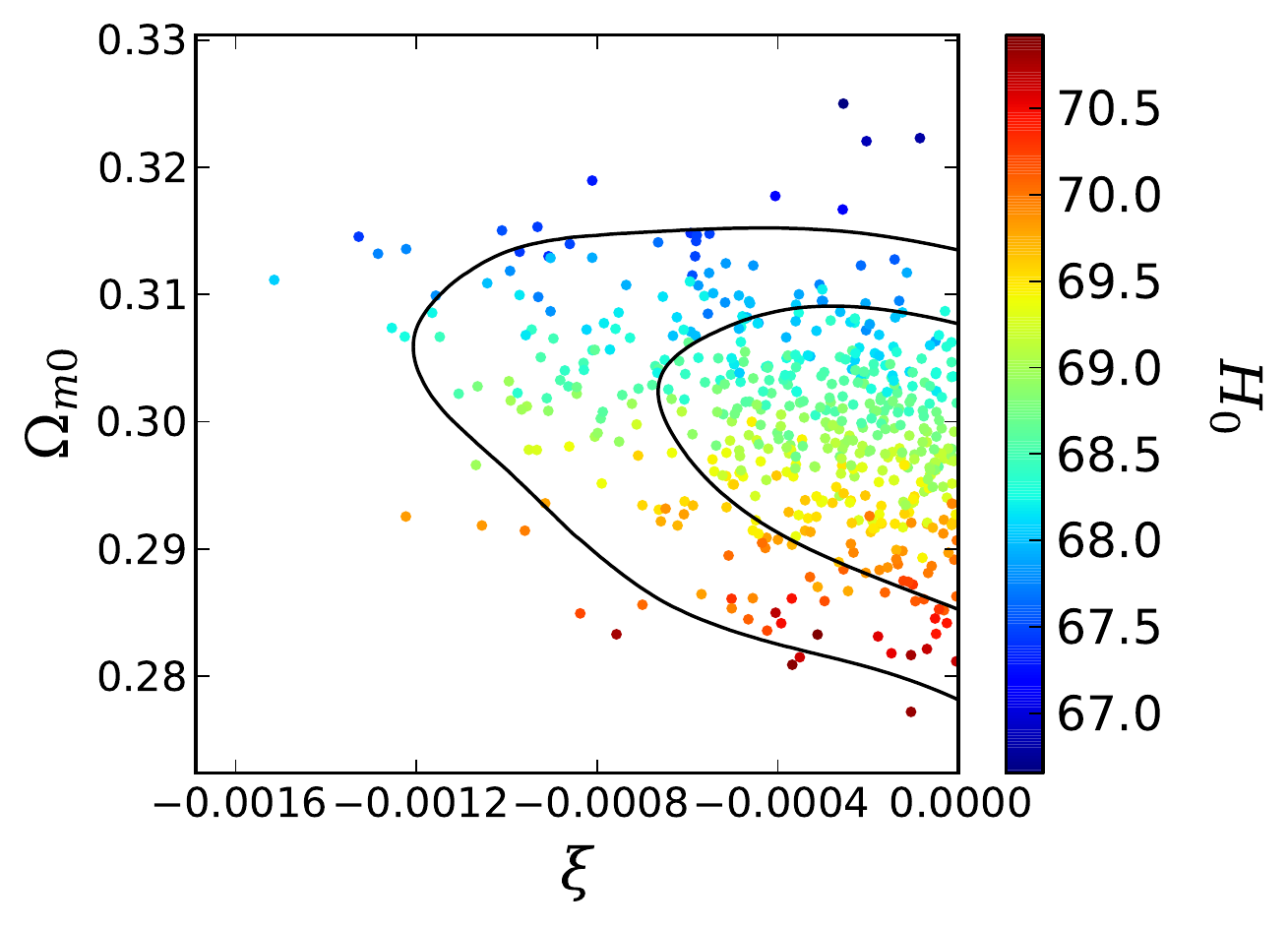}
\includegraphics[width=0.325\textwidth]{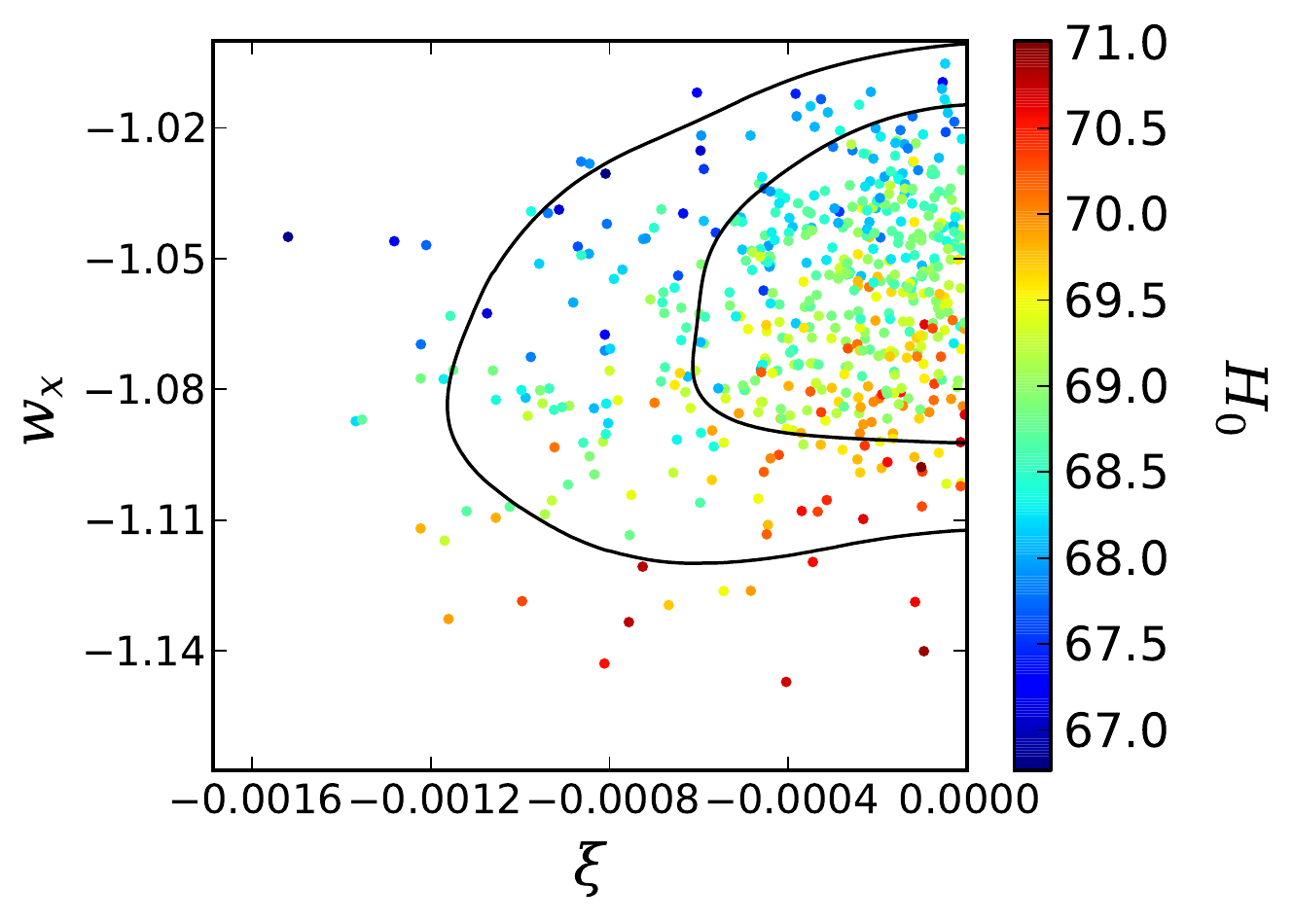}
\caption{Color Online $-$ In each panel we show the three dimensional scattered plots colored by the $H_0$ values of the markov chain monte carlo (mcmc) chains of the corresponding combined analysis. The combined analysis from the top to the bottom panels are respectively (i) CMB+BAO+HST, (ii) CMB+BAO+RSD, (iii) CMB+BAO+RSD+HST, (iv) CMB+BAO+RSD+HST+JLA+CC, and (v) CMB+BAO+RSD+HST+JLA+CC+WL. \texttt{First column:} From the mcmc chains of all combined analyses, one may notice that as the values of $H_0$ decrease (represented by the points in blue) the dark energy equation of state moves toward the cosmological constant boundary. \texttt{Second column:} The mcmc chains of all combined analyses infer that the lower values of $H_0$ prefer a nonzero coupling in the dark sector which is statistically consistent with zero. \texttt{Last column:} With the lower values of $H_0$, the dark energy equation of state moves toward the cosmological constant boundary and a nonzero coupling in the dark sector is favored which is indeed very close to zero. }
\label{fig-scattered-ide}
\end{figure*}

\subsection{IDE: Results}
\label{sec-int-general}

In Table \ref{tab:results-I} we summarize the 68\% confidence-levels constraints on the cosmological parameters for $\xi \leq 0$ and $w_x <-1$  using a variety of astronomical data displayed in the table. In Fig. \ref{fig-contour1}, 68.3\% and 95.4\% confidence-level contour plots for different combinations of the model parameters have been shown including one dimensional posterior distribution for some selected parameters of the interacting scenario as well.  We notice that the combined data set CMB+ext, where ``ext'' is the combination of any two data sets from  BAO, RSD, HST, JLA, CC, WL,  significantly reduces the allowed region in the parameters space.

From the analyses presented in Table \ref{tab:results-I}, one can easily state that the coupling parameter is very low. The coupling parameter, $\xi$ (at 95.4\% lower CL), are constrained to be (see Table \ref{tab:results-I}): 
\begin{itemize}
\item $\xi>-0.004884$ (CMB only), 
\item $\xi>-0.001285$ (CMB+BAO+HST), 
\item \text{$\xi>-0.001384$} (CMB+BAO+RSD), 
\item $\xi>-0.001278$ (CMB+BAO+RSD+HST), 
\item $\xi >-0.000959$ (CMB+BAO+RSD+HST+JLA+CC), 
\item $\xi>-0.000935$, for the last combined analysis (CMB+BAO+RSD+HST+JLA+CC+WL),
\end{itemize}
while we must note that within 68.3\% CL, $\xi= 0$ is allowed, that means, effectively, IDE may recover the non-interacting $w_x$CDM cosmology. 
In Fig. \ref{fig-xi-H0-Om}, we show the dependece of $\xi$ over other cosmological parameters for this model. 
Now, from the constraints on the dark energy equation of state summarized in Table \ref{tab:results-I}, it is quite clear-cut that, $w_x$ assumes values that are
close to ``$-1$''. In Fig. \ref{fig-wx-H0-Om}, we show the dependence of $w_x$ with
other important cosmological parameters for a better understanding. From the left panel of Fig. \ref{fig-wx-H0-Om} we
see that as $H_0$ decreases, $w_x$ approaches toward the cosmological constant limit, while from the right panel of Fig. \ref{fig-wx-H0-Om} we observe that $\Omega_{m0}$ takes large values as $w_x \rightarrow -1$. Further, in Fig. \ref{fig-sigma8}, we explicitly show the two dimensional contour plots in the planes ($\sigma_8$, $w_x$),
$(\sigma_8, H_0)$ and $(\sigma_8, \xi)$ in order to
measure the variations in $\sigma_8$ in presence of the coupling. Our analysis shows that, an increased coupling strength effectively lowers the values of $\sigma_8$, that means the model significantly deviate from the $\Lambda$CDM model. One can also observe that for more phantom state in the dark energy equation of state the values of $\sigma_8$ increase. In addition, we also observe that, in presence of the coupling, higher values of the Hubble parameter also indicate higher values of $\sigma_8$.

Furthermore, we analyzed the mcmc chains for all combined analyses focusing on the behaviour of the coupling strength, dark energy equation of state and the density parameter for the matter sector monitored by the Hubble parameter values. The analysis has been displayed in Fig. \ref{fig-scattered-ide}. Precisely, such analysis provides with the qualitative behaviour of the interacting model in terms of the coupling strength and the dark energy equation of state. The analysis shows that the lower values of the Hubble parameter signal for a non-zero interaction in the dark sector but the dark energy equation of state still lies within a close neighborhood of the cosmological constant boundary ``$-1$''. Also, the density parameter for matter takes bigger values for lower values of the Hubble parameter as well.

Lastly, we compare the $\chi^2_{min}$ values bewtween IDE and $\Lambda$CDM model obtained for different combined analyses (see Table \ref{tab:chi2}). We observe that for some combined analyses, the $\chi^2_{min}$ achieved for IDE is bigger than the $\Lambda$CDM model. One may notice that almost all combined analyses return a greater $\chi_{min}^2$ for IDE in compared to the standard $\Lambda$CDM.

\begingroup
\squeezetable
\begin{center}
\begin{table*}
\begin{tabular}{cccccccccccc}
\hline\hline
Model  & CMB & $\begin{array}[c]{c}
\text{CMB+BAO}\\+\text{HST} \end{array}$ & $\begin{array}[c]{c}
\text{CMB+BAO}\\+\text{RSD} \end{array}$ & $\begin{array}[c]{c}
\text{CMB+BAO}\\+\text{RSD+HST} \end{array}$ & $\begin{array}[c]{c}
\text{CMB+BAO}\\+ \mbox{RSD+HST}\\+\text{JLA+CC} \end{array}$ & $\begin{array}[c]{c}
\text{CMB+BAO}\\+ \mbox{RSD+HST}\\+\text{JLA+CC}\\+\text{WL} \end{array}$\\ \hline

IDE:
$\chi^2_{min~\mbox{(best-fit)}}$ & 12960.778 & 12981.276 & 12975.450 & 12982.168 & 13689.092 & 13723.708\\

IVS: $\chi^2_{min~\mbox{(best-fit)}}$ & 12961.606 & 12980.844 & 12971.080 & 12982.742 & 13693.894 & 13724.124\\

$\Lambda$CDM: $\chi^2_{min~\mbox{(best-fit)}}$ & 12964.062 & 12978.886 & 12974.124 & 12981.336 & 13693.560 & 13722.170\\
\hline\hline
\end{tabular}
\caption{Table displaying the $\chi^2_{min}$ obtained for the best-fit values of the parameters of the two interacting dark energy scenarios and the non-interacting $\Lambda$CDM cosmolofy. }\label{tab:chi2}
\end{table*}
\end{center}
\endgroup
\begingroup
\squeezetable
\begin{center}
\begin{table*}
\begin{tabular}{cccccccccccc}
\hline\hline
Parameters & CMB & $\begin{array}[c]{c}
\text{CMB+BAO}\\+\text{HST} \end{array}$ & $\begin{array}[c]{c}
\text{CMB+BAO}\\+\text{RSD} \end{array}$ & $\begin{array}[c]{c}
\text{CMB+BAO}\\+\text{RSD+HST} \end{array}$ & $\begin{array}[c]{c}
\text{CMB+BAO}\\+ \mbox{RSD+HST}\\+\text{JLA+CC} \end{array}$ & $\begin{array}[c]{c}
\text{CMB+BAO}\\+ \mbox{RSD+HST}\\+\text{JLA+CC}\\+\text{WL} \end{array}$\\ \hline

$\Omega_c h^2$ & $ 0.1225_{-    0.0031}^{+    0.0021}$ & $    0.1178_{-    0.0010}^{+    0.0010}$ & $    0.1193_{-    0.0011}^{+    0.0011}$ & $    0.1183_{-    0.0012}^{+    0.0011}$ & $    0.1182_{-    0.0011}^{+    0.0011}$ & $    0.1178_{-    0.0010}^{+    0.0010}$\\

$\Omega_b h^2$ & $    0.0223_{-    0.0002}^{+    0.0002}$ & $    0.0224_{-    0.0001}^{+    0.0001}$ & $    0.0223_{-    0.0001}^{+    0.0001}$ & $    0.0224_{-    0.0001}^{+    0.0002}$ & $    0.0224_{-    0.0001}^{+    0.0001}$ & $    0.0224_{-    0.0001}^{+    0.0001}$\\

$100\theta_{MC}$ & $ 1.0402_{-    0.0004}^{+    0.0004}$ & $    1.0408_{-    0.0003}^{+    0.0003}$ & $    1.0405_{-    0.0003}^{+    0.0003}$ & $    1.0407_{-    0.0003}^{+    0.0003}$ & $    1.0407_{-    0.0004}^{+    0.0003}$ & $    1.0407_{-    0.0003}^{+    0.0003}$ \\

$\tau$ & $0.0765_{-    0.0178}^{+    0.0192}$ & $    0.0915_{-    0.0155}^{+    0.0185}$ & $    0.0781_{-    0.0157}^{+    0.0138}$ & $    0.0796_{-    0.0163}^{+    0.0165}$ & $    0.0793_{-    0.0162}^{+    0.0156}$ & $    0.0750_{-    0.0160}^{+    0.0169}$\\

$n_s$ & $0.9695_{-    0.0049}^{+    0.0048}$ & $    0.9788_{-    0.0038}^{+    0.0037}$ & $    0.9750_{-    0.0036}^{+    0.0035}$ & $    0.9771_{-    0.0038}^{+    0.0039}$ & $    0.9773_{-    0.0041}^{+    0.0039}$ & $    0.9783_{-    0.0038}^{+    0.0035}$\\

${\rm{ln}}(10^{10} A_s)$ & $ 3.0966_{-    0.0339}^{+    0.0373}$ & $    3.1214_{-    0.0311}^{+    0.0373}$ & $    3.0960_{-    0.0283}^{+    0.0275}$ & $    3.0979_{-    0.0319}^{+    0.0311}$ & $    3.0965_{-    0.0350}^{+    0.0316}$ & $    3.0863_{-    0.0316}^{+    0.0327}$\\

$\Omega_{m0}$ & $0.3425_{-    0.0271}^{+    0.0159}$ & $    0.3064_{-    0.0064}^{+    0.0063}$ & $    0.3167_{-    0.0071}^{+    0.0074}$ & $    0.3096_{-    0.0077}^{+    0.0068}$ & $    0.30911013_{-    0.0077}^{+    0.0065}$ & $    0.3064_{-    0.0065}^{+    0.0061}$\\

$\sigma_8$ & $ 0.8118_{-    0.0170}^{+    0.0212}$ & $    0.8262_{-    0.0127}^{+    0.0156}$ & $    0.8169_{-    0.0120}^{+    0.0120}$ & $    0.8167_{-    0.0125}^{+    0.0124}$ & $    0.8161_{-    0.0143}^{+    0.0134}$ & $    0.8108_{-    0.0127}^{+    0.0128}$\\

$H_0$ & $   65.2375_{-    1.1629}^{+    1.8234}$ & $   67.8090_{-    0.4817}^{+    0.5004}$ & $   67.0308_{-    0.5437}^{+    0.5368}$ & $   67.5675_{-    0.5259}^{+    0.5413}$ & $   67.6067_{-    0.5032}^{+    0.5796}$ & $   67.7953_{-    0.4735}^{+    0.4983}$\\

$\xi$ & $\xi > -0.001953$ & $\xi > -0.000490$ & $\xi > -0.000726$ & $\xi > -0.000549$ & $\xi > -0.000563$ & $\xi > -0.000557$\\

\hline\hline
\end{tabular}
\caption{The table summarizes the observational constraints of the cosmological parameters for the interacting vacuum scenario (IVS) at 68.3\% confidence-level for different combinations of the observational data. For the coupling parameter $\xi$, we report only their values at 95.4\% lower confidence-level. }\label{tab:results-II}
\end{table*}
\end{center}
\endgroup

\subsection{IVS: Results}
\label{sec-vaccum-scenario}

\begin{figure*}
\includegraphics[width=0.7\textwidth]{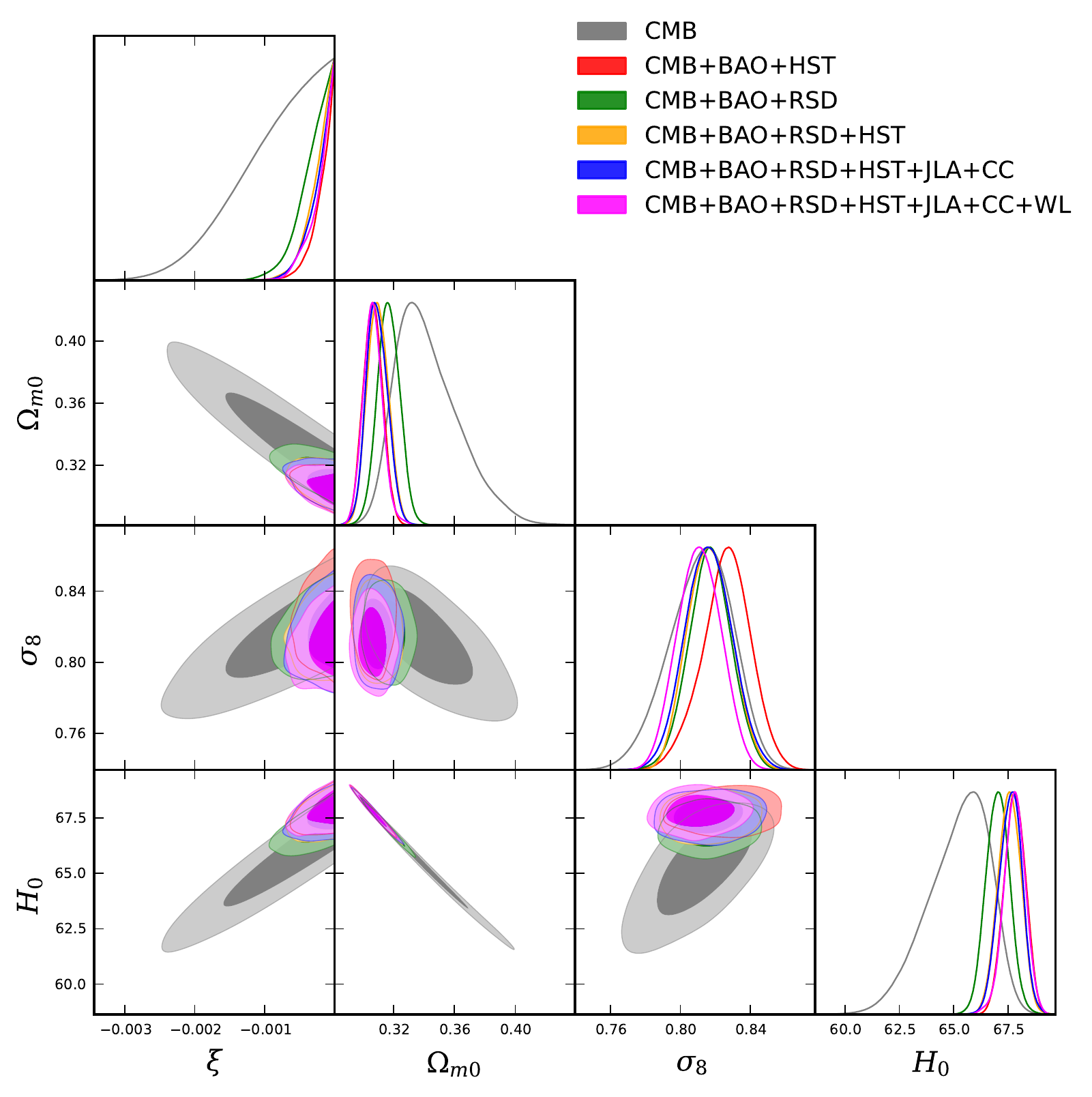}
\caption{Color Online $-$ Contour plots for different combinations of the cosmological parameters in the 68.3\% and 95.4\% confidence levels for the interacting vacuum scenario (IVS) have been displayed for distinct observational combinations. Additionally, we also show the one-dimensional posterior distributions for those parameters at the extreme right corners of each row.  From the two-dimensional contour plots one can see that the addition of any external data to CMB decreases the error bars of the cosmological parameters.}
\label{fig-contour-vacuum1}
\end{figure*}

As a particular case we consider the simplest possibility when dark energy is the cosmological constant. Now,  we have
also constrained this interacting scenario using the same combined analyses as employed in section \ref{sec-int-general}. The results have been summarized in Table \ref{tab:results-II} and Fig. \ref{fig-contour-vacuum1} shows the two dimensional contour plots at 68.3\% and 95.4\% confidence-levels for different combinations of the free model parameters using the six different combined analyses. Additionally, in the extreme right corner of each row of the Fig. \ref{fig-contour-vacuum1} we further show the one-dimensional posterior distributions for some selected model parameters of this interacting scenario. From Fig. \ref{fig-contour-vacuum1} one can see that the addition of any other external data to CMB significantly decreases the allowed region in the parameters space and hence the parameters are well constrained when any external data set is added to CMB. 
    
From the analysis we notice that the coupling strength of the interaction is very very small and it is very close to zero. In particular, in 95.4\% lower confidence-level,  we find that, 
\begin{itemize}
\item $\xi  >  -0.001953$ (for CMB alone), 
\item $\xi > -0.000490$ (CMB+BAO+HST), 
\item $\xi > -0.000726$ (CMB+BAO+RSD), 
\item $\xi > -0.000549$ (CMB+BAO+RSD+HST), 
\item $\xi > -0.000563$ (CMB+BAO+RSD+HST+JLA+CC), 

and  finally,

\item $\xi > -0.000557$, for the last combined analysis (CMB+BAO+RSD+HST+JLA+CC+WL). 
\end{itemize}

Additionally, we must mention that within 68.3\% confidence level, the non-interacting scenario (i.e., $\xi =0$)  is recovered (excluding the CMB analysis).  Thus, one 
can see that this interaction scenario is effectively very close to the non-interacting 
$\Lambda$CDM scenario.

\begin{figure*}
\includegraphics[width=0.34\textwidth]{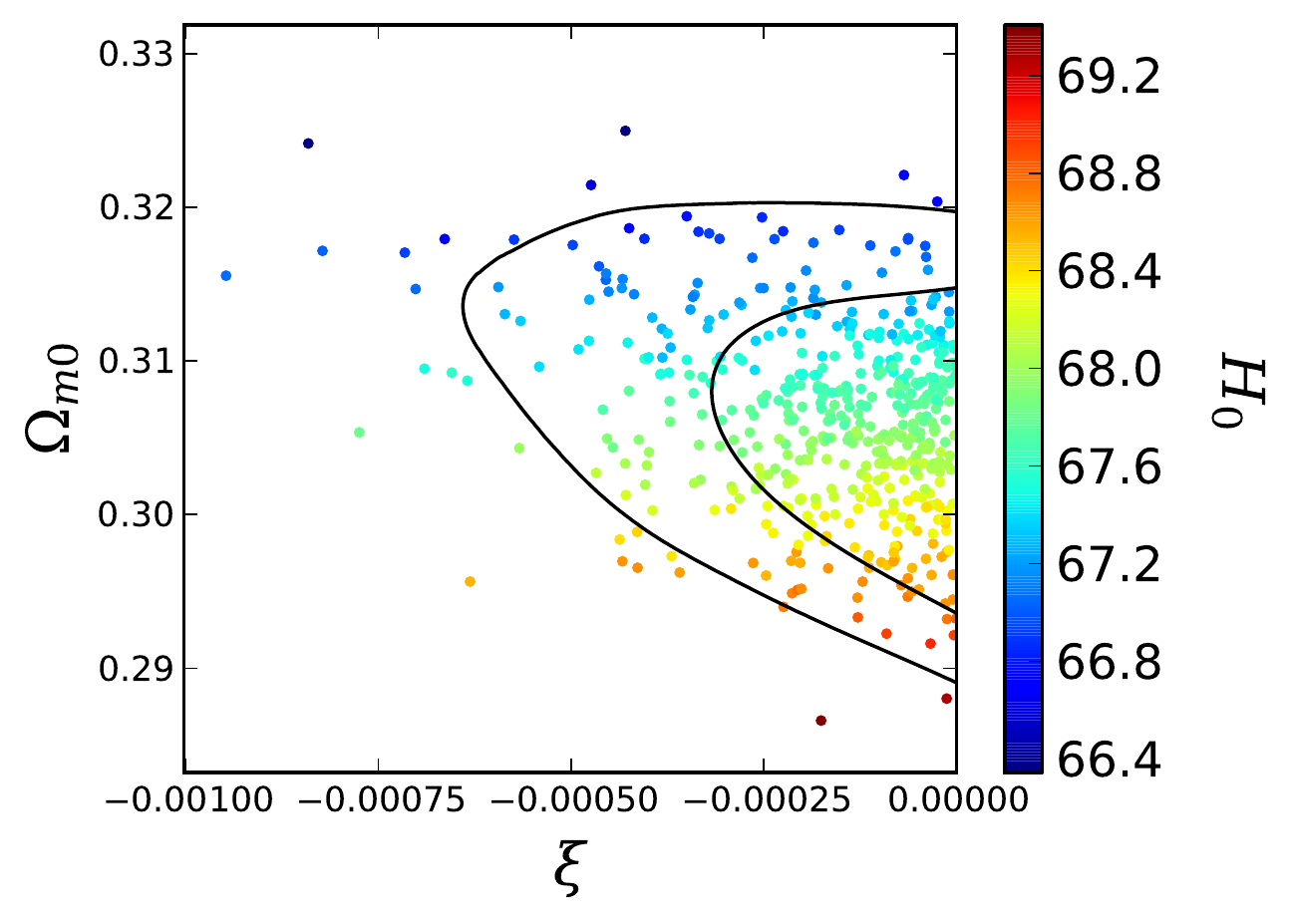}
\includegraphics[width=0.34\textwidth]{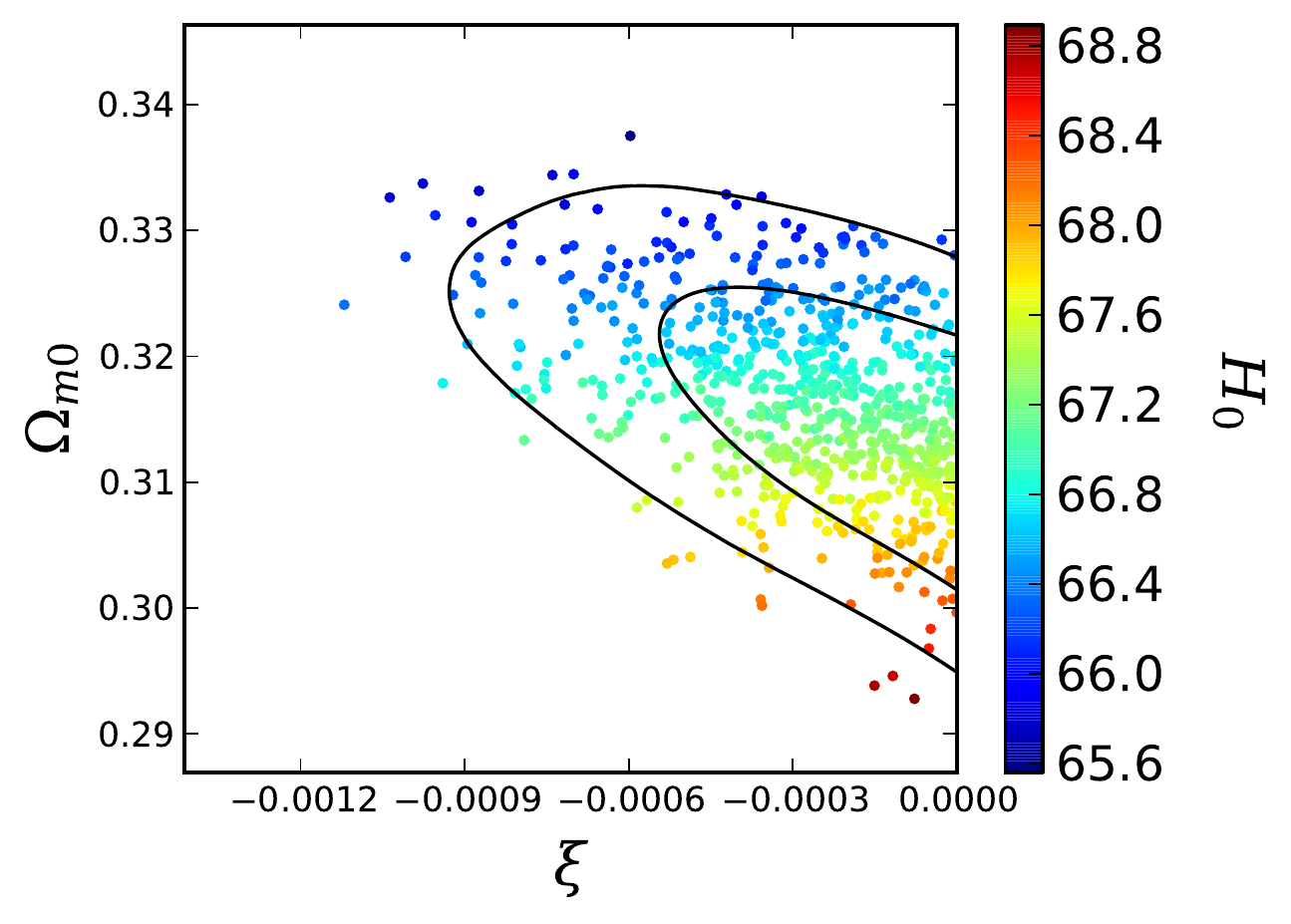}
\begin{center}
\includegraphics[width=0.355\textwidth]{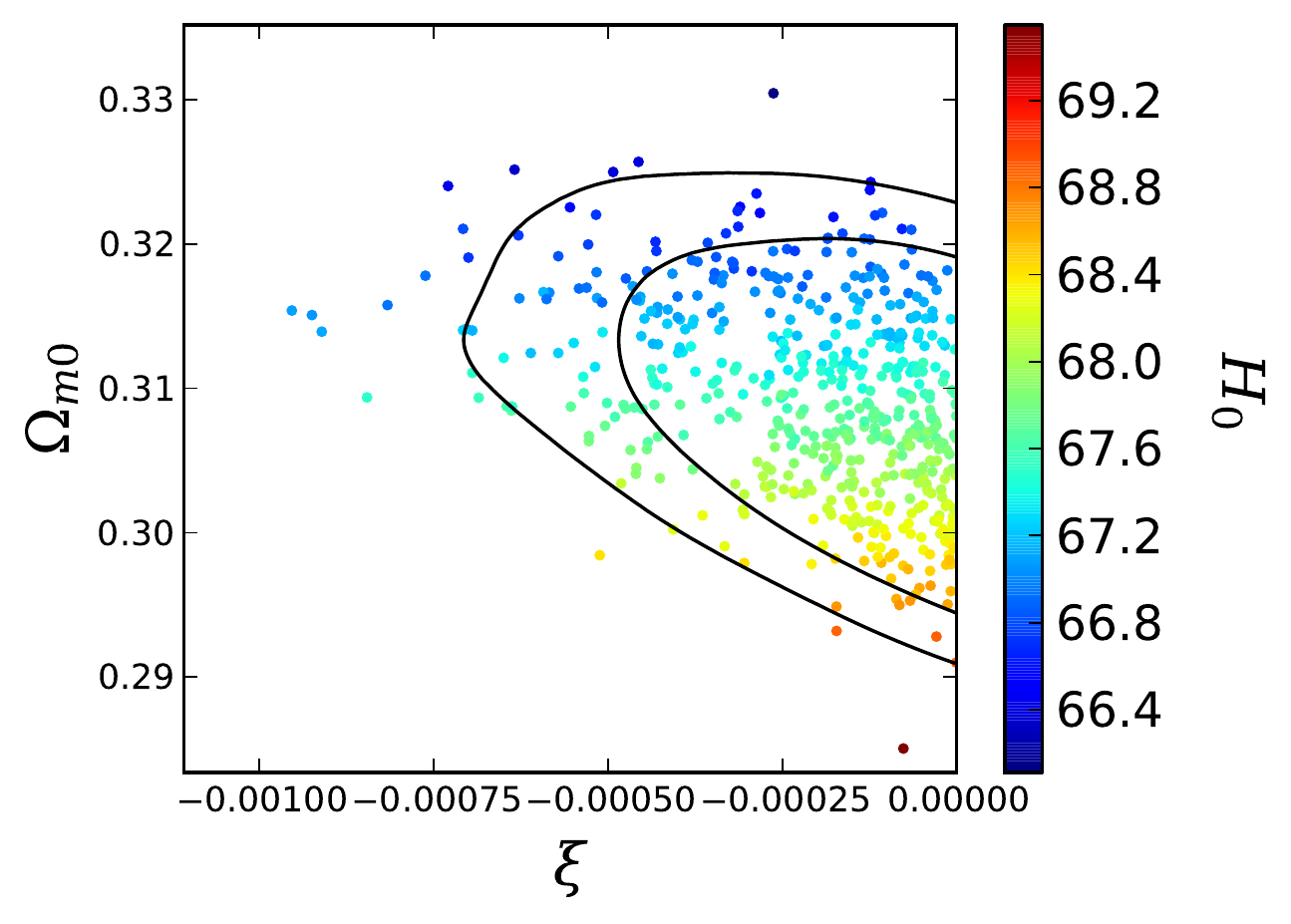}
\end{center}
\includegraphics[width=0.34\textwidth]{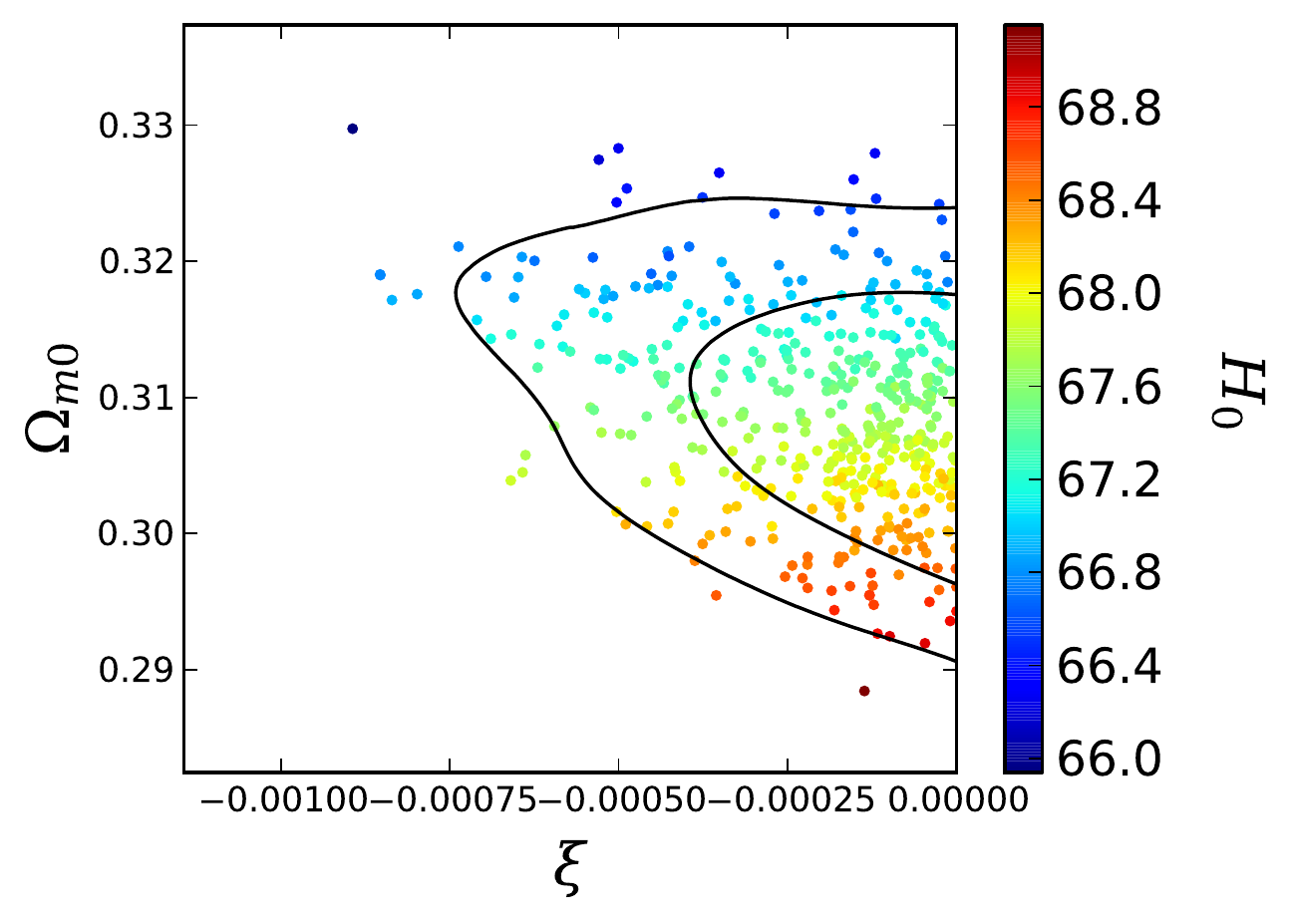}
\includegraphics[width=0.34\textwidth]{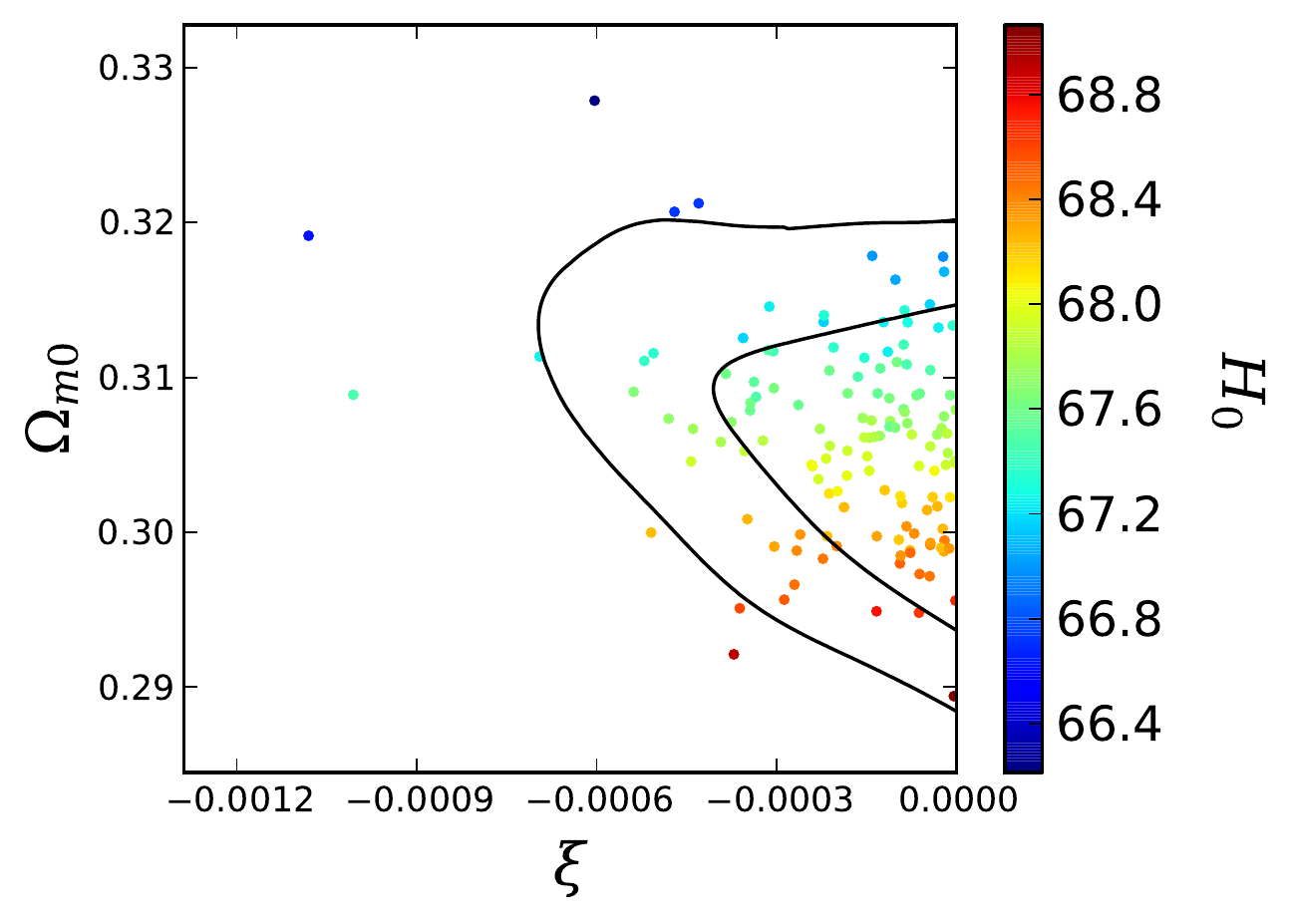}
\caption{Color Online $-$ For the interacting vacuum scenario, we have analyzed the mcmc chains of the combined analysis in the two dimensional $(\xi, \Omega_{m0})$ planes colored by the Hubble parameter values. The upper left and upper right panels respectively represent the analyses CMB+BAO+HST, CMB+BAO+RSD. The center plot stands for the combined analysis CMB+BAO+RSD+HST. Finally, the lower left and right panels respectively represent the analyses CMB+BAO+RSD+HST+JLA+CC and CMB+BAO+RSD+HST+JLA+CC+WL. From all the plots, one thing is clear that the lower values of the Hubble parameter signal for a non-zero interaction in the dark sector while statistically this is consistent with zero and in addition, the density parameter for the matter sector is also allowed to take higher values. }
\label{fig-scattered-ivs}
\end{figure*}

Similar to the IDE model  described in
section \ref{sec-int-general}, here too we have investigated the
three dimensional scattered plots in the $(\xi, \Omega_{m0})$ plane
for all the combined analyses colored by the Hubble parameter values. The
analysis has been presented in Fig. \ref{fig-scattered-ivs} from 
which one can notice that, for lower values of the Hubble parameter, 
the coupling parameter seems to have a tendency to take values away
from $\xi =0$ while for the higher values of $H_0$, the coupling 
parameter takes values very close to zero. 

Following the similar trend as done for IDE model, we compare 
the $\chi^2_{min}$ values for this scenario with respect to
the base cosmological model $\Lambda$CDM. We have similar conclusion for this 
interaction scenario, that means, the $\chi^2_{min}$ values for this model are bigger 
for almost all combined analyses in respect to the $\Lambda$CDM cosmological model.

\begingroup
\squeezetable
\begin{center}
\begin{table*}
\begin{tabular}{cccccccccccc}
\hline\hline
Parameter & CMB & $\begin{array}[c]{c}
\text{CMB+BAO}\\+\text{HST} \end{array}$ & $\begin{array}[c]{c}
\text{CMB+BAO}\\+\text{RSD} \end{array}$ & $\begin{array}[c]{c}
\text{CMB+BAO}\\+\text{RSD+HST} \end{array}$ & $\begin{array}[c]{c}
\text{CMB+BAO}\\+ \mbox{RSD+HST}\\+\text{JLA+CC} \end{array}$ & $\begin{array}[c]{c}
\text{CMB+BAO}\\+ \mbox{RSD+HST}\\+\text{JLA+CC}\\+\text{WL} \end{array}$\\ \hline

$H_0$ (IDE) & $   65.52_{-    3.93-    8.02-   10.22}^{+    4.51+    7.53+    9.45}$ & $   70.77_{-    1.15-    2.32-    2.47}^{+    1.11+    2.31+    2.78}$ & $   67.97_{-    1.02-    1.77-    2.05}^{+    0.83+    1.86+    2.56}$ & $   69.19_{-    0.89-    1.72-    2.22}^{+    0.87+    1.79+    2.46}$ & $   69.64_{-    0.85-    1.75-    2.20}^{+    0.83+    1.69+    2.79}$ & $   68.89_{-    0.82-    1.36-    1.88}^{+    0.68+    1.52+    1.93}$\\

$H_0$ (IVS) & $65.24_{-    1.16-    3.06-    4.14}^{+    1.82+    2.67+    3.21}$ & $   67.81_{-    0.48-    0.98-    1.24}^{+    0.50+    0.97+    1.17}$ & $   67.03_{-    0.54-    1.09-    1.41}^{+    0.54+    1.08+    1.46}$ & $   67.57_{-    0.53-    1.01-    1.35}^{+    0.54+    0.96+    1.31}$ & $   67.61_{-    0.50-    1.03-    1.34}^{+    0.58+    0.98+    1.39}$& $   67.80_{-    0.47-    1.04-    1.54}^{+    0.50+    1.06+    1.23}$\\
\hline\hline
\end{tabular}
\caption{Table displaying the 68\%, 95\% and 99\% confidence level constraints on $H_0$ and $\sigma_8$ for different combined analyses for the interacting scenario with $w_x < -1$ (IDE) and $w_x= -1$ (IVS). We note that the estimation of $H_0$ by latest Planck missions for the base $\Lambda$CDM model yields $H_0= 67.27 \pm 0.66$ km/s/Mpc (Planck TT, TE, EE+lowP) \cite{Ade:2015xua}. }\label{tab:H0}
\end{table*}
\end{center}
\endgroup

\subsection{On the Tension of $H_0$: The role of interaction}
\label{sec-tension}

One of the most talkative issues in  current cosmological research is the tension on the parameter $H_0$. Some recent investigations in the context of interacting dark energy models fueled its further investigations aiming to reach a definite and satisfactory explanation towards this direction, and consequently, people are more focused on how interacting dark energy models may alleviate the tension on $H_0$. The first question that immediately arises is, what exactly such tension is. To illustrate this notion, we need to take into account its distinct measurements from different observational missions.
The estimation of $H_0$ by Planck 2015 missions from the $\Lambda$CDM based cosmological model yields $H_0= 67.27 \pm 0.66$ km/s/Mpc (Planck TT, TE, EE+lowP) \cite{Ade:2015xua} while the local measurement of $H_0$ using the Hubble Space Telescope gives $H_0 = 73.24 \pm 1.74$ km/s/Mpc \cite{Riess:2016jrr}, and puts a huge difference between these estimations.  This effect is generally known as the tension on the Hubble constant. Some recent investigations already shown that interacting dark energy might be able to release such tension on $H_0$ \cite{Kumar:2017dnp, DiValentino:2017iww}. Since the interacting dark energy is purely model dependent, so naturally,  it is quite justified to see how other phenomenological interaction models react with the tension on $H_0$. For a better viewing, in Table \ref{tab:H0}, we summarize the constraints on $H_0$ for both IDE and IVS up to $3\sigma$ confidence level. We see that the addition of one extra degrees of freedom in terms of the coupling parameter significantly increases the error bars on $H_0$ in compared to Planck 2015 \cite{Ade:2015xua}. And the increase of error bars on $H_0$ is prominent for IDE in compared to IVS because the estimated values of $H_0$ for IVS using different combined analyses look similar to Planck 2015 \cite{Ade:2015xua}.  Naturally, for the IDE scenario, one may infer that, due to the large error bars present on $H_0$, the estimated values of $H_0$ are in agreement with the local measurement  ($H_0 = 73.24 \pm 1.74$ km/s/Mpc \cite{Riess:2016jrr}). Thus, one can see that the interaction in the dark sector may provide an explanation towards the reduction of the tension on $H_0$.

\begingroup                                                                                                                     
%\squeezetable                                                                                                                   
\begin{center}                                                                                                                  
\begin{table}[htb]                                                                                                                
\begin{tabular}{ccc}                                                                                                            
\hline\hline                                                                                                                    
$\ln B_{ij}$ & Strength of evidence for model ${M}_i$ \\ \hline
$0 \leq \ln B_{ij} < 1$ & Weak \\
$1 \leq \ln B_{ij} < 3$ & Definite/Positive \\
$3 \leq \ln B_{ij} < 5$ & Strong \\
$\ln B_{ij} \geq 5$ & Very strong \\
\hline\hline                                                                                                                    
\end{tabular}                                                                                                                   
\caption{Revised Jeffreys scale used to test the observational support of any  model $M_i$ with respect to another model $M_j$. }\label{tab:jeffreys}                                                                                                   
\end{table}                                                                                                                     
\end{center}                                                                                                                    
\endgroup 
\begingroup                                                                                                                     
%\squeezetable                                                                                                                   
\begin{center}                                                                                                                  
\begin{table*}[htb]                                                                                                                
\begin{tabular}{cccccc}                                                                                                            
\hline                                                                                                                    
Data set & Model &~~~~$\ln B_{ij}$ &~ Strength of evidence for model $\Lambda$CDM \\ \hline
CMB & IDE & $-2.0$ &  Positive \\
CMB & IVS & $-1.9$ &  Positive\\
CMB+BAO+HST & IDE & $-4.8$ & Strong \\
CMB+BAO+HST & IVS & $-3.5$ & Strong \\
CMB+BAO+RSD & IDE & $-2.9$ & Positive \\
CMB+BAO+RSD & IVS &  $-1.7$ & Positive \\
CMB+BAO+RSD+HST & IDE & $-3.6$ & Strong\\
CMB+BAO+RSD+HST & IVS & $-3.3$ & Strong \\
CMB+BAO+RSD+HST+JLA+CC & IDE & $-1.7$ & Positive \\
CMB+BAO+RSD+HST+JLA+CC & IVS & $-2.2$ & Positive\\
CMB+BAO+RSD+HST+JLA+CC+WL & IDE & $-4.0$ & Strong \\
CMB+BAO+RSD+HST+JLA+CC+WL & IVS & $-3.7$ & Strong \\
\hline                                                                                                                 
\end{tabular}                                                                                                                   
\caption{Summary of $\ln B_{ij}$, for the two interacting scenarios with respect to the reference model  $\Lambda$CDM, for different observational data sets. From the Bayesian evidence point of view, the negative values of $\ln B_{ij}$ mean that the reference model $\Lambda$CDM is preferred over the two interacting scenarios. }\label{tab:bayesian}                                                                                                   
\end{table*}                                                                                                                     
\end{center}                                                                                                                    
\endgroup

\subsection{The Bayesian Evidence}
\label{sec-information}

Model selection \cite{Liddle:2007fy} plays an important role in distinguishing various cosmological models. Keeping the same motivation, in this work we compare both the interacting 
dark energy scenarios with the $\Lambda$CDM cosmological model using the Bayesian analysis. The Bayesian evidence is a powerful statistical technique that quantifies 
the cosmological models based on their performance with the observational data. 
In the following we shortly describe how the Baysian evidence is calculated for
a cosmological model.  
In the Bayesian analysis one needs the posterior probability of the model parameters (denoted by $\theta$), given a particular data set $x$ to test the model, any prior information and a model $M$. Now, recalling the Bayes theorem, one may write 
\begin{eqnarray}\label{BE}
p(\theta|x, M) = \frac{p(x|\theta, M)\,\pi(\theta|M)}{p(x|M)}
\end{eqnarray}
where $p(x|\theta, M)$ is the likelihood function dependent on the model parameters $\theta$ with the data set fixed;  $\pi(\theta|M)$ is the prior used in the analysis. The denominator $p(x|M)$ in the right hand side of eqn. (\ref{BE}) 
is the Bayesian evidence for the model comparison and it is the integral 
over the unnormalised posterior $\tilde{p} (\theta|x, M) \equiv p(x|\theta,M)\,\pi(\theta|M)$ as

\begin{eqnarray}\label{sp-be01}
E \equiv p(x|M) = \int d\theta\, p(x|\theta,M)\,\pi(\theta|M).
\end{eqnarray}
We note that the above equation (\ref{sp-be01}) is also referred to as the marginal likelihood. Now, for 
any particular model $M_i$  and the reference model $M_j$ (which is the base model and it is $\Lambda$CDM here),  
the posterior probability is given by
\begin{eqnarray}
\frac{p(M_i|x)}{p(M_j|x)} = \frac{\pi(M_i)}{\pi(M_j)}\,\frac{p(x| M_i)}{p(x|M_j)} = \frac{\pi(M_i)}{\pi(M_j)}\, B_{ij}.
\end{eqnarray}
where $B_{ij} = \frac{p(x| M_i)}{p(x|M_j)}$, is the Bayes factor of the model $M_i$ relative to the base or reference model $M_j$. For $B_{ij} > 1 $, we refer that the data support the model $M_i$ more strongly than the model $M_j$. The behavior of the models is usually quantified using different values of $B_{ij}$ (or alternatively, $\ln B_{ij}$). Here, we shall use the widely accepted Jeffreys scales \cite{Kass:1995loi} (see Table \ref{tab:jeffreys}) that summarizes the model comparison.

Now, one can calculate the Bayesian evidence using the MCMC chains which directly extract the parameters of the underlying cosmological model. For a detailed explanation on the implementation of the Bayesian evidence for any cosmological model we refer to \cite{Heavens:2017hkr,Heavens:2017afc} where we use the code \texttt{MCEvidence}\footnote{This code is available for free at \href{https://github.com/yabebalFantaye/MCEvidence}{github.com/yabebalFantaye/MCEvidence}.}. 

Thus, using the code \texttt{MCEvidence}, we have calculated the logarithm of the Bayes factor, i.e., $\ln B_{ij}$ where $i$ stands for IDE or IVS and $j$ is the reference model $\Lambda$CDM. In Table \ref{tab:bayesian} we have shown the calculated values of $\ln B_{ij}$ for the two interacting scenarios with respect to the reference model $\Lambda$CDM. From the table, we see that for all the observational data employed in this work, the values of $\ln B_{ij}$ are negative which from the point of view of the Bayesian evidence, one can identify that the reference model $\Lambda$CDM is preferred over the two interacting scenarios. For some combined analysis, the preference of $\Lambda$CDM is strong while for some combined analysis, it is positive. Overall, we see that the present observational data always favor $\Lambda$CDM is favored in respect to the interacting scenarios discussed in this work.

\section{Concluding remarks}
\label{sec-conclu}

An interacting scenario between a pressureless dark matter and a dark energy fluid availing constant barotropic equation of state has been considered. The underlying geometry of the universe is characterized by the spatially flat FLRW line element and the interaction rate $Q = Q (\rho_t^\prime) = Q (\rho_c, \rho_x)$ has been given explicitly in eqn. (\ref{interaction}) or eqn. (\ref{eq-int}).
This interaction rate is very appealing in the sense that the evolution equations for the dark sectors (cold dark matter and dark energy) can be exactly solved, and thus, one can directly measure their deviation from the standard evolution laws of the  dark fluids with no-interaction. We note that initially this kind of interaction
was introduced by Chimento \cite{Chimento:2009hj} where the author proposed a very general interaction rate that recovers the interaction in eqn. (\ref{eq-int}) and discussed its theoretical implications. Later on its observational viability was tested when dark energy is the cosmological constant but at the background level \cite{G:2014mea}, consequently, in a recent article \cite{Sharov:2017iue}, the authors generalized this study for both $w_x = -1$ and $w_x \neq -1$ at the background level with the recent observational data. However, it is quite certain that the dynamics of such interaction models at the large scale of the universe, is promising for a better understanding of the entire scenario. That means, the most important question related with the interaction model is, how the structure formation of the universe  depends when such interaction is included in the cosmological scenario.
Thus, in the present work we discuss the perturbations and the structure formation of the universe when such interaction is present between the dark fluids. Now,
in order to test the resulting cosmological scenarios with the available observational data, we use \texttt{cosmomc}, a markov chain monte carlo package that extracts the model parameters with a sufficient convergence following the Gelman-Rubin statistics \cite{Gelman-Rubin}. The observational data include cosmic microwave background radiation, baryon acoustic oscillations,  redshift space distortions, local Hubble constant, supernovae type Ia from joint light curve analysis, Hubble parameter values at different redshifts from cosmic chronometers and finally the weak gravitational lensing data. For a better analysis, we have considered two distinct interacting scenarios, namely when the dark energy is other than the cosmological constant (i.e., $w_x \neq -1$) and the other one is the cosmological constant itself.

For IDE scenario, the constraints on the model parameters have been summarized in Table \ref{tab:results-I} where we present the 95.4\% limits (lower) on the coupling 
parameter $\xi$. 
And in Fig. \ref{fig-contour1}, we show the contour plots for different combinations of the model parameters at 68.3\% and 95.4\% confidence levels. The right corners of Fig. \ref{fig-contour1} also shows the one-dimensional posterior distributions for some selected model parameters as well. From the observational constraints on the coupling parameter, $\xi$, summarized in the last row of the Table \ref{tab:results-I}, we find that $\xi =0$ is consistent with the observational data. Moreover, from the constraints on the dark energy equation of state, $w_x$, one can see that it is actually very very close to the cosmological constant boundary. Thus, we see that the interaction model is actually equivalent to the non-interacting $\Lambda$CDM background. However, in the large scale  distribution, the interaction model may exhibit some differences even for a very small coupling strength.  
From the imprints on the CMB TT spectra (see the right panel of the Fig. \ref{fig:CMB-ide}) and also from the matter power spectra (see the right panel of the Fig. \ref{fig:Mpower-ide}), it is evident that for a very small coupling strength ($\xi =-0.0001$), the model presents a very minimal deviation from the non-interacting $\Lambda$CDM cosmology.

Now, for the interacting cosmological constant (labeled as IVS), the results have been summarized in Table \ref{tab:results-II}. The corresponding contour plots at 68.3\% and 95.4\% confidence-levels are also shown in Fig. \ref{fig-contour-vacuum1} with the one-dimensional posterior distributions for some selected parameters of this model. From the estimation of the coupling strength shown in Table \ref{tab:results-II}, one can see that $\xi$ is concistent with the non-interaction limit (i.e., $\xi =0$), at least according to the current observational data.  In fact, for this model we have realized a similar trend as in IDE. For instance, from Fig. \ref{fig-scattered-ivs}, similar to IDE model, we find that lower values of the Hubble parameter allow non-zero interaction in the dark sector. The deviation of this interaction scenario from the non-interacting 
$\Lambda$CDM cosmology is also found to be insensitive (see the right panels of Fig. \ref{fig:CMB-ivs} and Fig. \ref{fig:Mpower-ivs}) unlike the IDE scenario where although the deviation is small but they are detectable.

We also raise one interesting point that has become a hot issue at current cosmological research $-$ the observed tension on the $H_0$ parameter from its global \cite{Ade:2015xua} and local measurements \cite{Riess:2016jrr}. We found that the allowance of the interaction increases the error bars on the Hubble parameter measurements, and consequently, the parameters space for $H_0$ is increased. This becomes effective to release the tension partially and is reflected from some combinatons for IDE only. While the interacting vacuum model is not suitable to release the tension. One may argue that the allowance of extra degrees of freedom in the parameters space of the interacting dark energy models (for IDE, the number of parameters is 8 while for IVS this number is 7) might be suitable to alleviate such tension.
Similar results have been reported in some recent works \cite{Kumar:2017dnp, DiValentino:2017iww}, but however, since the theory of interaction is phenomenological and hence its conclusions too, therefore, the analysis with a different interaction model might be perhaps important to see whether the model can avail the same property or not. The relation of the extra degrees of freedom to the tension on $H_0$, in the interacting dark energy models surely needs further attention.  

Finally, we computed the  Bayesian evidence for each interacting scenario with respect to the non-interacting $\Lambda$CDM model (see Table \ref{tab:bayesian}). Our analysis shows that the non-interacting $\Lambda$CDM is preferred over the two interacting dark energy scenarios, at least according to the current 
observational data sets.

\section*{ACKNOWLEDGMENTS}
The authors thank the referee for his/her constructive and illuminating comments that improved the work considerably. W. Yang's work is supported by the National
Natural Science Foundation of China under Grants No.  11705079 
and No. 11647153. RH was supported by Proyecto VRIEA-PUCV N$_{0}$ 039.309/2018. 
SC acknowledges the financial support from the Mathematical Research Impact Centric Support (MATRICS), project reference no. MTR/2017/000407, by the Science and Engineering Research Board, Government of India.
SP thanks Rafael C. Nunes, Burin Gumjudpai and J. A. S. Lima for useful discussions and comments while working on the draft.

%%%%%%%%%%%%%%%%%%%%%%%%%%%%%%%%%%%%%%%%%%%%%%%%%%%%%%%%%%%%%%%%%%%%%%%%%%%%%%%%%%%%%%

\end{document}